\pdfoutput=1
\documentclass{JINST}

\usepackage{wrapfig}
\usepackage{amsmath}
\usepackage{amssymb}
\usepackage{mathptmx}
\usepackage{fontenc}
\usepackage{graphicx}

\newcommand{\unitspace}[0]{\ensuremath{\,}}
\newcommand{\unit}[1]{\ensuremath{\unitspace\text{#1}}}

\newcommand{\mhl}[1]{#1}

\title{The HERMES Recoil Detector}

\author{A.~Airapetian$^{k,n}$,
E.~C.~Aschenauer$^{e,1}$,
S.~Belostotski$^p$,
A.~Borissov$^{d,l}$,
A.~Borisenko$^i$,
J.~Bowles$^l$,
I.~Brodski$^k$,
V.~Bryzgalov$^q$,
J.~Burns$^l$,
G.~P.~Capitani$^i$,
V.~Carassiti$^h$,
G.~Ciullo$^h$,
A.~Clarkson$^l$,
M.~Contalbrigo$^h$,
R.~De~Leo$^a$,
E.~De~Sanctis$^i$,
M.~Diefenthaler$^{m,g}$,
P.~Di~Nezza$^i$,
M.~D\"uren$^k$,
M.~Ehrenfried$^{k,2}$,
H.~Guler$^d$,
I.~M.~Gregor$^d$,
M.~Hartig$^{d}$,
G.~Hill$^l$, 
M.~Hoek$^{l,k}$, 
Y.~Holler$^d$,
I.~Hristova$^e$,
H.~S.~Jo$^j$,
R.~Kaiser$^{l,3}$,
T.~Keri$^{l,k}$,
A.~Kisselev$^p$,
B.~Krause$^d$,
B.~Krauss$^{g,4}$,
L.~Lagamba$^a$,
I.~Lehmann$^{l,5}$,
P.~Lenisa$^h$,
S.~Lu$^k$,
X.-G.~Lu$^e$, 
S.~Lumsden$^l$, 
D.~Mahon$^l$, 
A.~Martinez~de~la~Ossa$^b$,
M.~Murray$^l$,
A.~Mussgiller$^{d,g}$,
W.-D.~Nowak$^e$,
Y.~Naryshkin$^p$,
A.~Osborne$^l$,
L.~L.~Pappalardo$^h$,
R.~Perez-Benito$^k$,
A.~Petrov$^d$,
N.~Pickert$^{g,4}$,
V.~Prahl$^d$,
D.~Protopopescu$^l$,
M.~Reinecke$^d$,
C.~Riedl$^{e,m}$,
K.~Rith$^g$,
G.~Rosner$^l$,
L.~Rubacek$^k$,
D.~Ryckbosch$^j$,
Y.~Salomatin$^{q}$,
G.~Schnell$^{c,j}$,
B.~Seitz$^l$,
C.~Shearer$^l$,
V.~Shutov$^f$,
M.~Statera$^h$,
J.~J.~M.~Steijger$^{o}$,
H.~Stenzel$^k$,
J.~Stewart$^{e,1}$,
F.~Stinzing$^g$,
A.~Trzcinski$^r$,
M.~Tytgat$^j$,
A.~Vandenbroucke$^{j,6}$,
Y.~Van~Haarlem$^{j,7}$,
C.~Van~Hulse$^{c,j}$,
M.~Varanda$^d$,
D.~Veretennikov$^p$,
I.~Vilardi$^{a,8}$,
V.~Vikhrov$^p$,
C.~Vogel$^{g,9}$,
S.~Yaschenko$^{d,g}$,
Z.~Ye$^{e,10}$,
W.~Yu$^k$,
D.~Zeiler$^{g,2}$ and
B.~Zihlmann$^{j,11}$\\
\llap{$^a$}Istituto Nazionale di Fisica Nucleare, Sezione di Bari, 70124 Bari, Italy\\
\llap{$^b$}Nuclear Physics Laboratory, University of Colorado, Boulder, Colorado 80309-0390, USA\\
\llap{$^c$}Department of Theoretical Physics, University of the Basque Country UPV/EHU, 48080 Bilbao, Spain and IKERBASQUE, Basque Foundation for Science, 48011 Bilbao, Spain\\
\llap{$^d$}DESY, 22603 Hamburg, Germany\\
\llap{$^e$}DESY, 15738 Zeuthen, Germany\\
\llap{$^f$}Joint Institute for Nuclear Research, 141980 Dubna, Russia\\
\llap{$^g$}Physikalisches Institut, Universit\"at Erlangen-N\"urnberg, 91058 Erlangen, Germany\\
\llap{$^h$}Istituto~Nazionale~di~Fisica~Nucleare, Sezione~di~Ferrara and Dipartimento~di~Fisica, Universit\`a~di~Ferrara, 44100 Ferrara, Italy\\
\llap{$^i$}Istituto~Nazionale~di~Fisica~Nucleare, Laboratori Nazionali di Frascati, 00044 Frascati, Italy\\
\llap{$^j$}Department of Physics and Astronomy, Ghent University, 9000 Gent, Belgium\\
\llap{$^k$}Physikalisches Institut, Universit\"at Gie\ss en, 35392 Gie\ss en, Germany\\
\llap{$^l$}SUPA, School of Physics and Astronomy, University of Glasgow, Glasgow G12 8QQ, United Kingdom\\
\llap{$^m$}Department of Physics, University of Illinois, Urbana, Illinois 61801-3080, USA\\
\llap{$^n$}Randall Laboratory of Physics, University of Michigan, Ann Arbor, Michigan 48109-1040, USA\\
\llap{$^o$}Nationaal Instituut voor Kernfysica en Hoge-Energiefysica (NIKHEF), 1009 DB Amsterdam, The Netherlands\\
\llap{$^p$}B. P. Konstantinov Petersburg Nuclear Physics Institute, Gatchina, 188300 Leningrad Region, Russia\\
\llap{$^q$}Institue for High Energy Physics, Protvino, 142281 Moscow Region, Russia\\
\llap{$^r$}Andrzej Soltan Institute for Nuclear Studies, 00-689 Warsaw, Poland\\
E-mail: \email{andreas.mussgiller@desy.de}, \email{sergey.yaschenko@desy.de}}
\author{\\
\llap{$^1$}Now at: Brookhaven National Laboratory, Upton, New York 11772-5000, USA\\
\llap{$^2$}Now at: Siemens AG, 91052 Erlangen, Germany\\
\llap{$^3$}Now at: International Atomic Energy Agency, A-1400 Vienna, Austria\\
\llap{$^4$}Now at: Siemens AG, 91301 Forchheim, Germany\\
\llap{$^5$}Now at: FAIR Facility for Antiproton and Ion Research, 64291 Darmstadt\\
\llap{$^6$}Now at: Dept of Radiology, Stanford University, School of Medicine, Stanford, California 94305-5105, USA\\
\llap{$^7$}Now at: Commonwealth Scientific and Industrial Research Organisation (CSIRO), Australia\\
\llap{$^8$}Now at: IRCCS Multimedica Holding S.p.a, 20099 Sesto San Giovanni (MI), Italy\\
\llap{$^9$}Now at: IBA Dosimetry GmbH, 90592 Schwarzenbruck, Germany\\
\llap{$^{10}$}Now at: Fermi National Accelerator Laboratory, Batavia, Illinois 60510, USA\\
\llap{$^{11}$}Now at: Thomas Jefferson National Accelerator Facility, Newport News, Virginia 23606, USA\\
}

\abstract{For the final running period of {\sc Hera}, a recoil detector was
installed at the {\sc Hermes} experiment to improve measurements of hard exclusive
processes in charged-lepton nucleon scattering. Here, deeply virtual Compton scattering
is of particular interest as this process provides constraints on generalised parton
distributions that give access to the total angular momenta of quarks within the
nucleon.

The {\sc Hermes} recoil detector was designed to improve the selection of exclusive
events by a direct measurement of the four-momentum of the recoiling particle.
It consisted of three components: two layers of double-sided silicon strip sensors
inside the {\sc Hera} beam vacuum, a two-barrel scintillating fibre tracker, and a
photon detector. All sub-detectors were located inside a solenoidal magnetic field
with a field strength of $1\unit{T}$.

The recoil detector was installed in late 2005. After the commissioning of all
components was finished in September 2006, it operated stably until the end
of data taking at {\sc Hera} end of June 2007. The present paper gives a brief
overview of the physics processes of interest and the general detector design. The
recoil detector components, their calibration, the momentum reconstruction
of charged particles, and the event selection are described in detail.
The paper closes with a summary of the performance of the detection system.}


\begin{document}


%
\section{Introduction}
%

In the winter shutdown of 2005/2006 the {\sc Hermes} spectrometer~\cite{HermesNIMA:1998} was upgraded
in the target region with a Recoil Detector (RD). The detector surrounded the {\sc Hermes}
target cell and comprised, in a coaxial structure, a set of Silicon Strip Detectors (SSD) situated
inside the {\sc Hera} lepton beam vacuum, a Scintillating-Fibre Tracker (SFT) and a
Photon Detector (PD), \mhl{all surrounded by a superconducting magnet with a field strength
of $1\unit{T}$ in the center of the bore.} The RD was commissioned during the 2006 data taking
and operated in conjunction with the {\sc Hermes} forward spectrometer until the end of {\sc Hera}
data taking in the middle of 2007.

The purpose of the RD was to improve access to hard exclusive electroproduction
of real photons ($\gamma$) or mesons ($m$) off nucleons ($N$), $e + N \rightarrow e + N + \gamma / m$,
at {\sc Hermes}. Hard exclusive processes have come to the forefront of nucleon structure physics because
they provide information on Generalised Parton Distributions (GPDs)~\cite{Mueller:1994, Radyushkin:1996, Ji:1997}.
GPDs can be considered the natural complement to transverse-momentum-dependent parton
distributions, as both are derived from the same parent Wigner distributions~\cite{Ji0:2003, Belitsky:2004}.
In particular, GPDs have quickly risen in importance in hadron physics since it was
shown that they may provide access to the total angular momentum carried by quarks
(and gluons) in the nucleon~\cite{Ji2:1997} and they provide a multi-dimensional picture
of the nucleon structure~\cite{Burkardt:2000}.

Deeply Virtual Compton Scattering (DVCS), i.e., the hard exclusive electroproduction of
a real photon, presently provides the cleanest access to GPDs.
GPDs depend on four kinematic variables: $t$, $x$, $\xi$, and $Q^2$. The 
Mandelstam variable $t=(p-p^\prime)^2$ is the squared four-momentum 
transfer to the target nucleon, with $p$ ($p^\prime$) its initial (final) 
four-momentum. In the `infinite'-target-momentum frame, $x$ and $\xi$ are 
related to the longitudinal momentum of the struck parton, as a fraction 
of the target momentum. The variable $x$ is the average of the initial and 
final momentum fractions carried by the parton, and the variable $\xi$, 
known as the skewness, is half of their difference. 
The evolution of GPDs 
with $Q^2 \equiv -q^2$, where $q = k - k^\prime$ is the difference between 
the four-momenta of the incident ($k$) and scattered ($k^\prime$) lepton, can be calculated 
in the context of perturbative quantum chromodynamics as in the case of 
parton distribution functions.
There exist several GPDs to describe the various possible helicity transitions of
the struck quark and of the nucleon as a whole.
The DVCS process on an unpolarised proton is very well suited to access the GPD $H$,
which describes the dominant transition that conserves the helicities of both
the struck quark and the nucleon.

The DVCS process contributes to the reaction channel $e N \rightarrow e N\gamma$, which
is dominated at {\sc Hermes} kinematics by the Bethe-Heitler (BH) process,
i.e., elastic $e N$ scattering with a bremsstrahlung photon in the initial or final state.
The two processes are experimentally indistinguishable and therefore interfere. The
differential cross section is given by

\begin{center}
  \begin{equation}
\frac{d\sigma}{dQ^2\;dx_{\mathrm{B}}\;d|t|\;d\phi}
=
\frac{x_{\mathrm{B}}e^6}{32(2\pi)^4Q^4\sqrt{1+\epsilon^2}}|\tau_{\textrm{Total}}|^2,
  \label{e:xsec}
  \end{equation}
\end{center}
where
\begin{center}
\begin{equation}
|\tau_{\textrm{Total}}|^2 = |\tau_{\textrm{BH}}|^2 +|\tau_{\textrm{DVCS}}|^2 +
\underbrace{\tau_{\textrm{BH}}\tau_{\textrm{DVCS}}^{\ast}+\tau_{\textrm{DVCS}}\tau_{\textrm{BH}}^{\ast}}_{I}.
\label{e:scam}
\end{equation}
\end{center}

In equation~\ref{e:xsec}, $e$ is the elementary charge of the electron,
$x_{\mathrm{B}}$ is the Bjorken scaling variable $x_{\mathrm{B}} = Q^2/(2pq)$ and
$\epsilon=2x_{\mathrm{B}}\frac{M}{Q}$ with $M$ the proton mass. The angle $\phi$ denotes the
azimuthal orientation of the photon production plane with respect to the lepton scattering plane.
In equation~\ref{e:scam}, the square of the scattering amplitude consists of three parts: one due
to the BH contribution, one due to the DVCS contribution and one due to the interference between
the two, denoted $I$. Although the DVCS contribution to the cross section is small at the kinematic
conditions of {\sc Hermes}, it is `amplified' in the interference term $I$ by the (much) larger BH
contribution. Experimentally, the preferred way to study DVCS is the measurement of cross-section
asymmetries. For an unpolarised hydrogen target, the beam-helicity asymmetry $\mathcal{A}_{\textrm{LU}}$,
where $L$ denotes the longitudinally polarised beam and $U$ the unpolarised target, and the beam-charge
asymmetry $\mathcal{A}_{\textrm{C}}$ can be accessed. These are constructed as

\begin{eqnarray}
\mathcal{A}_{\textrm{LU}}(\phi) &=& \frac{d\sigma^{\rightarrow}(\phi)-d\sigma^{\leftarrow}(\phi)}{d\sigma^{\rightarrow}(\phi)+d\sigma^{\leftarrow}(\phi)}, \\
\mathcal{A}_{\textrm{C}}(\phi) &=& \frac{d\sigma^{+}(\phi)-d\sigma^{-}(\phi)}{d\sigma^{+}(\phi)+d\sigma^{-}(\phi)},
\end{eqnarray}

\noindent
where $d\sigma^{\rightarrow}(\phi)$, $d\sigma^{\leftarrow}(\phi)$, $d\sigma^{+}(\phi)$ and
$d\sigma^{-}(\phi)$ represent cross-sections from positive and negative beam helicity and
positive and negative beam charges, respectively. Various experimental results on these
asymmetries have been published so far by the {\sc Hermes}
collaboration~\cite{PublicationDraft68:2008, PublicationDraft69:2009, PublicationDraft80:2010, PublicationDraft90:2012}.
In these measurements an enriched sample of exclusive events was selected using a missing-mass
technique. An event-by-event selection was not possible as the recoiling proton was outside the
acceptance and the existing spectrometer did not have sufficient resolution. Monte Carlo (MC)
calculations showed that the contribution of events from "associated" production
($e p \rightarrow e N\pi\gamma$, including the resonant production $e p \rightarrow e \Delta^{+} \gamma$)
was expected to be in average $13\%$, while $3\%$ were expected from semi-inclusive
processes. The contribution from the decay products of neutral pions from exclusive reactions
that are misidentified as single-photon events was found to be negligible.

The analysis related to the study of DVCS and {\sc Hermes} employing the RD involves two major
motivations. \mhl{The first is the selection of a DVCS (in the following DVCS corresponds to
DVCS and BH) event sample with a background contamination below $1\%$, and the extraction of
a beam-helicity asymmetry from it.} This allows a cleaner comparison to predictions from the
ongoing theoretical efforts to fit GPD models to {\sc Hermes} data.
The second motivation is the potential to extract an asymmetry in associated production in
the $\Delta$-resonance region.

The present paper is structured as follows. Chapters two and three give an overview of the
general detector design and the individual detector components. The data acquisition system
and the data taking performance are described in chapter four. The energy calibration and
the energy measurement in general are outlined in chapter five, and chapter six explains the
momentum reconstruction. The performance of the RD is summarized in chapter seven. The event
selection with the RD is described in chapter eight. The paper is summarized in chapter nine.

%
\section{Detector Design Overview}
%

\subsection{Design Requirements}
\label{sec:requirements}

\mhl{When using the forward spectrometer alone, exclusivity can only be established
for an event sample on a statistical basis due to its limited energy and position
resolutions. The purpose of the RD was to provide the necessary information to
establish exclusivity on an event-by-event basis and hence reduce the non-exclusive
background to below $1\%$.}

The overall design requirements on the RD were defined by the kinematics and type
of particles involved in the exclusive reactions of interest and by the expected
background processes. The dominant background contribution in the case of DVCS
originates from the associated production. Events with higher-mass resonances can
be removed by an invariant mass constraint. The particle types to be detected were
therefore protons, pions and photons from $\pi^0$ decay.

\begin{figure}[ht]
  \begin{center}
    \includegraphics[width=0.495\textwidth]{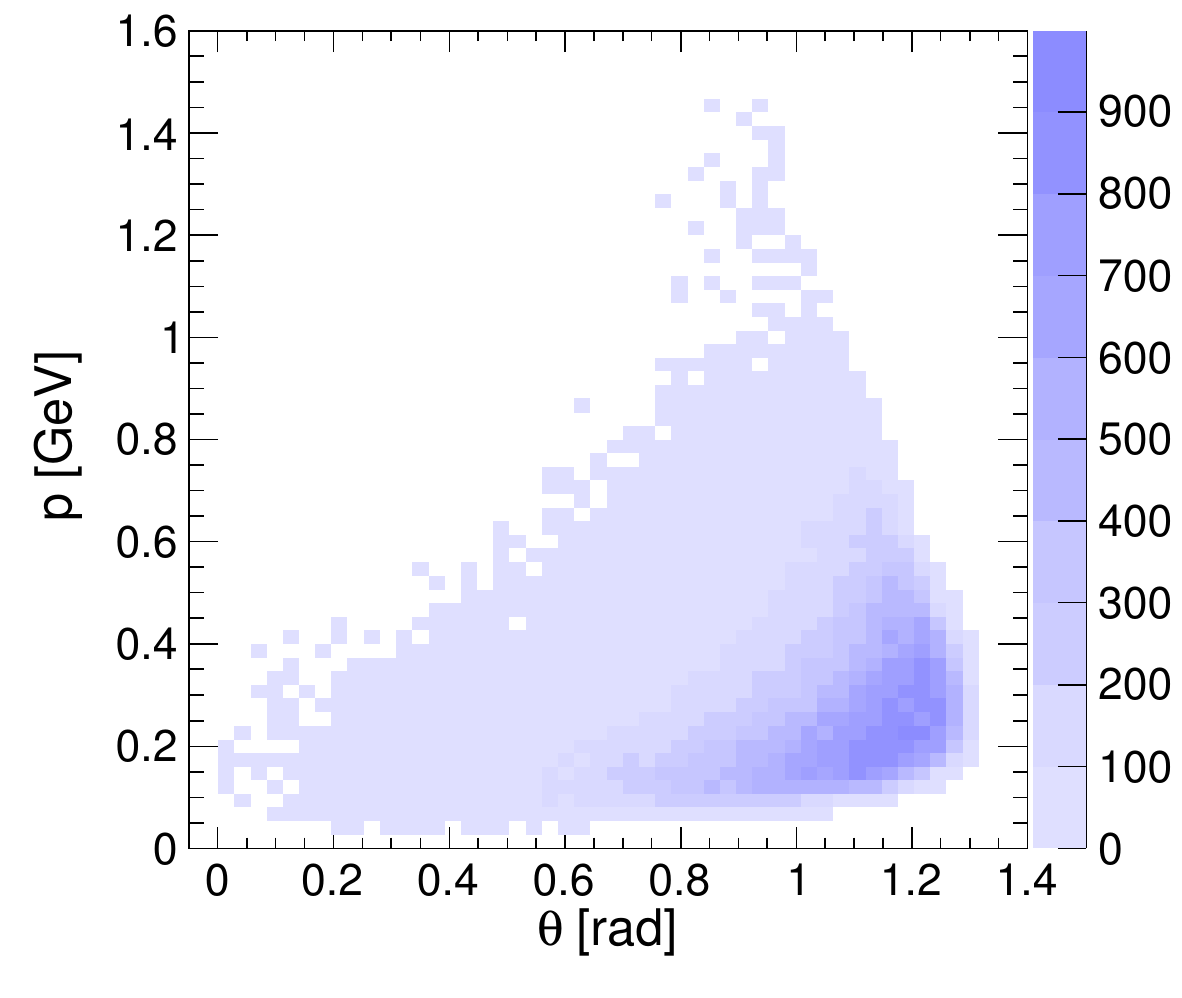}
    \includegraphics[width=0.495\textwidth]{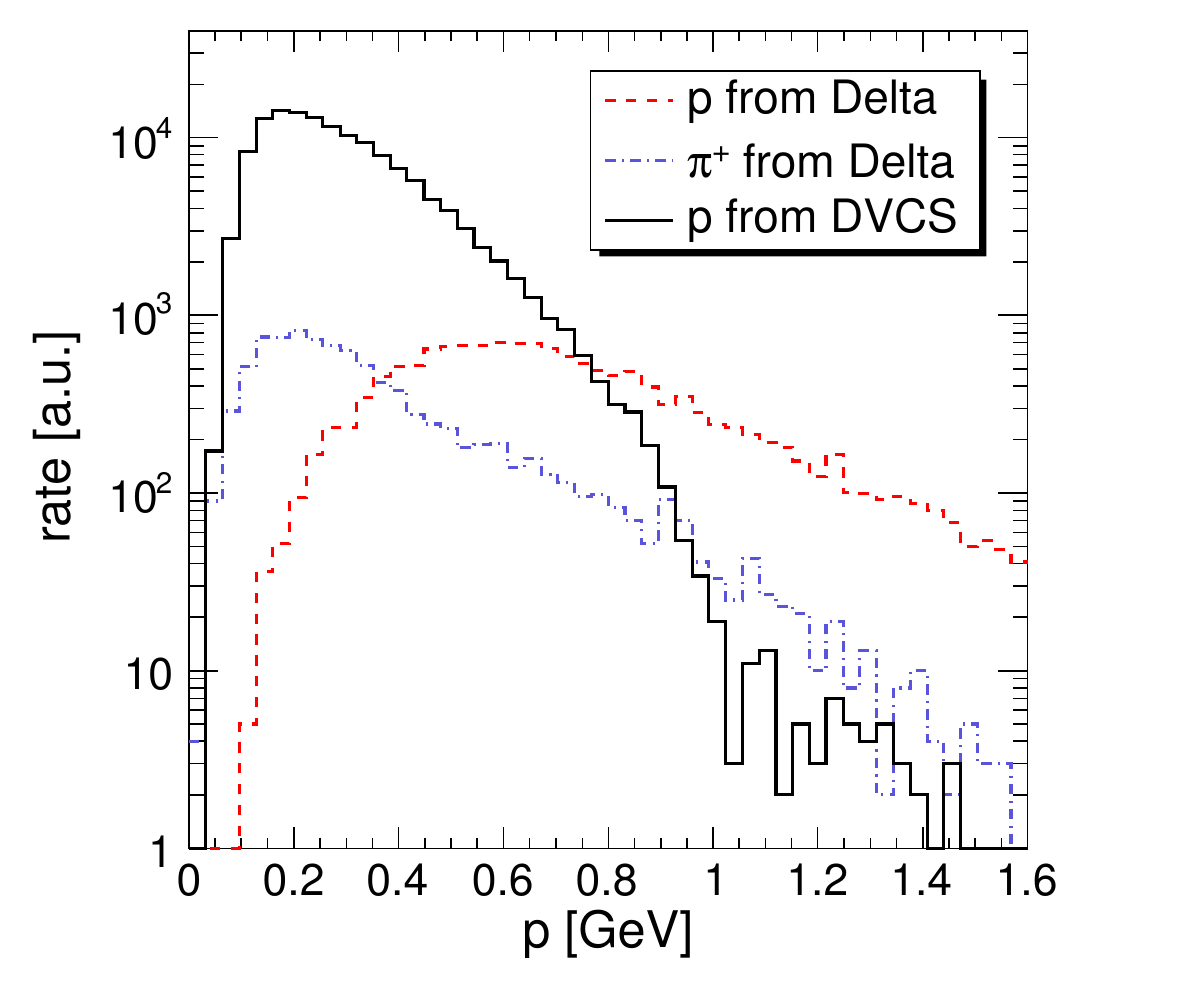}
    \caption{Left: Kinematic distribution of recoiling protons from DVCS in
             momentum $p$ and polar angle $\theta$ in Monte Carlo. Right:
             Monte Carlo momentum distribution for DVCS recoil protons in
             comparison with protons and pions from $\Delta$ decays.}
    \label{fig:DVCS_Kinematics}
  \end{center}
\end{figure}

All exclusive physics processes under investigation at {\sc Hermes} produce a
low-momentum recoil proton at large laboratory polar angles. The left panel of
figure~\ref{fig:DVCS_Kinematics} \mhl{shows the kinematic distribution of recoiling
protons from MC-generated DVCS events in terms of momentum $p$ and polar
angle $\theta$, the angle between the recoiling proton and the beam axis}\footnote{The distributions for exclusive $\rho^0$, $\pi^0$ and
$\eta$ production look similar.}. The right panel shows the MC-generated
momentum distribution of DVCS protons that were left intact, compared to
those of pions and protons from the decay of an intermediate $\Delta$-resonance.

\mhl{The majority of the recoiling proton statistics is located between $0.5\unit{rad}$
and $1.35\unit{rad}$ in $\theta$ and between $50\unit{MeV}$ and $600\unit{MeV}$ in
momentum. The RD was designed to cover the full $\theta$ range and most of the
momentum range down to a lower limit of $125\unit{MeV}$, which is imposed by the SSD.}

The momenta of particles detected in the forward spectrometer were typically much
larger than those of the particle(s) detected by the RD. As a result,
the invariant mass resolution was dominated by the resolution of the forward
spectrometer and could not be improved by the RD. However, the
transverse momenta of all involved particles were of comparable magnitude.
Hence the main exclusivity cuts using the RD can be based on transverse
momentum~\cite{Krauss:PhD}. \mhl{For this to be effective, the momentum resolution of
the RD was required to be better than $10\%$ for momenta below
$500\unit{MeV}$ and $15\%$ for higher momenta.} In the azimuthal angle $\phi$
around the beam axis the design resolution was $0.1\unit{rad}$ below
$500\unit{MeV}$ and $0.05\unit{rad}$ for higher momenta.

\subsection{General Detector Design}

\begin{figure}[ht]
  \begin{center}
    \includegraphics[width=11cm]{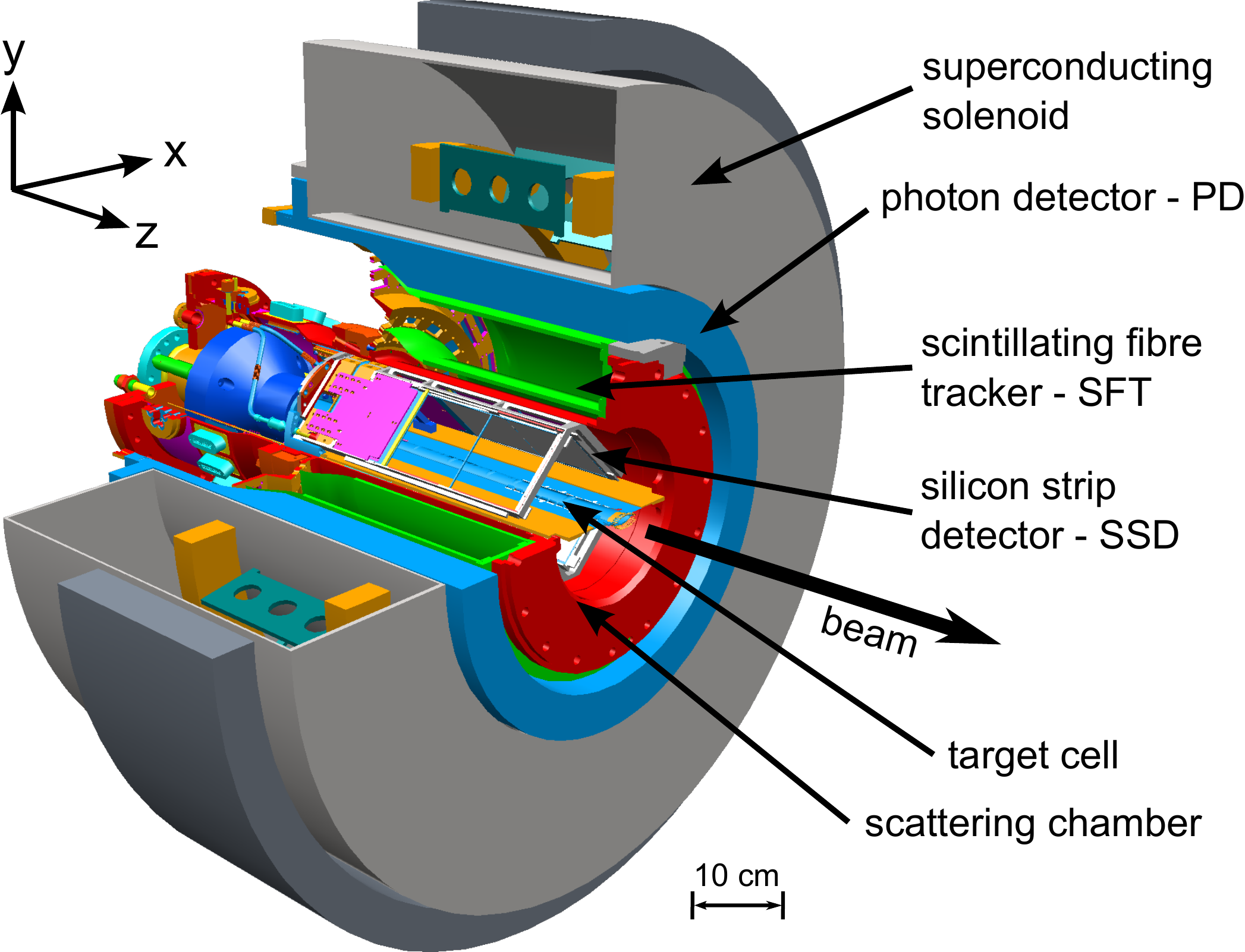}
    \caption{A three-dimensional CAD drawing of the RD.}
    \label{fig:Recoil3D}
  \end{center}
\end{figure}

\begin{figure}[ht]
  \begin{center}
    \includegraphics[width=10.5cm]{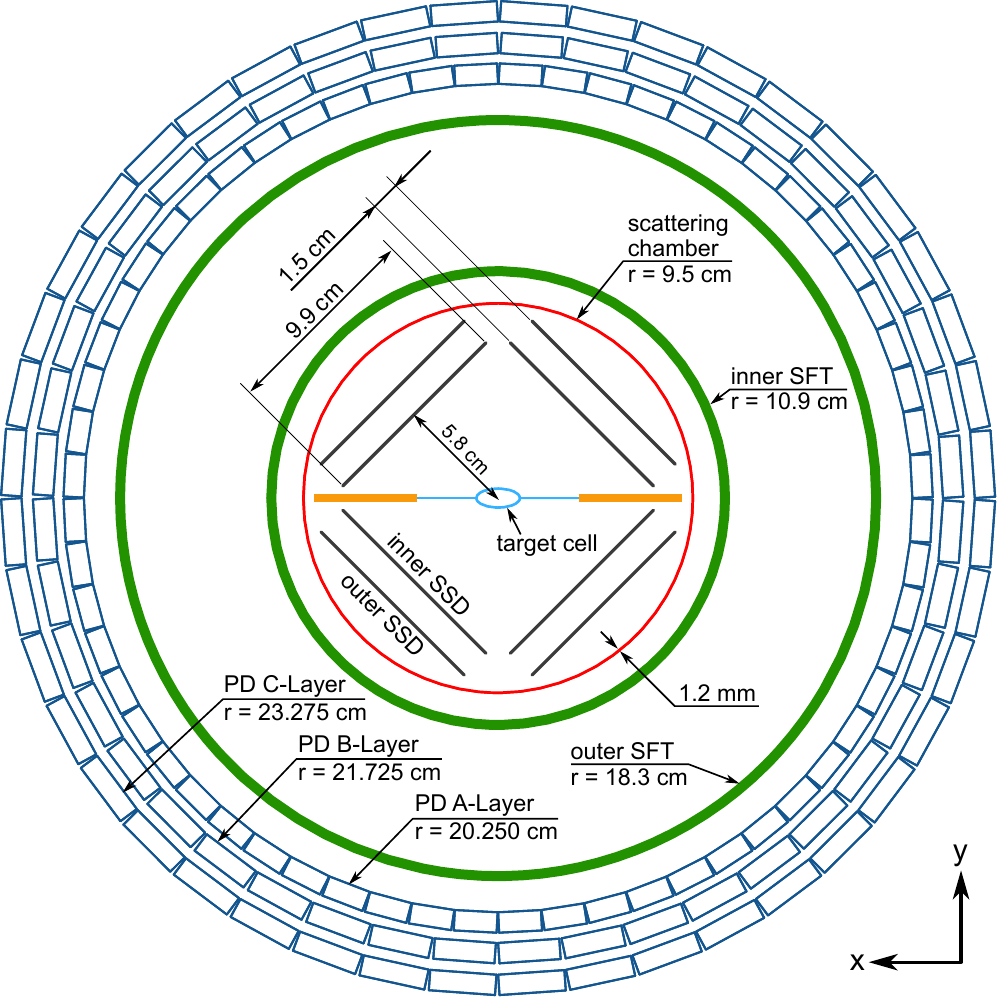}
    \caption{Cross-sectional view of the sensitive detector components of the RD including
             target cell and its support, and the wall of the scattering chamber. The z axis points
             into the plane.}
    \label{fig:RecoilXYView}
  \end{center}
\end{figure}

\begin{figure}[ht]
  \begin{center}
    \includegraphics[width=13.5cm]{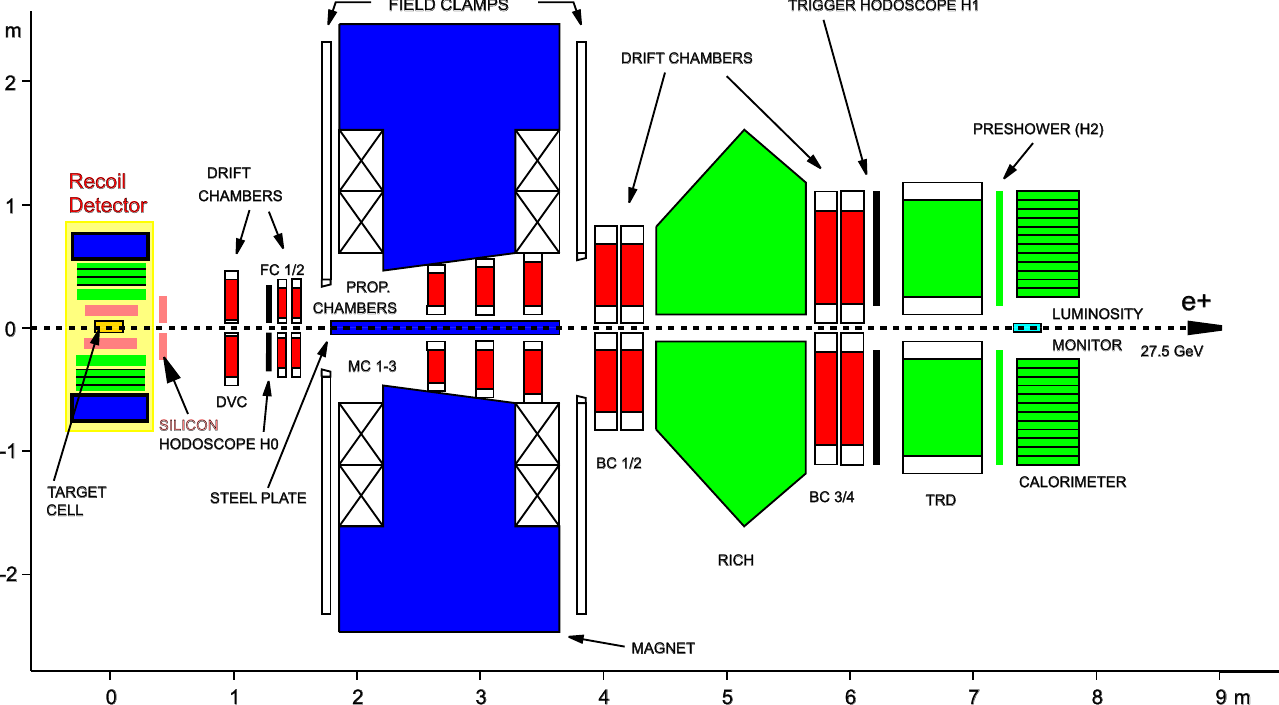}
    \caption{Side view of the {\sc Hermes} spectrometer with the RD installed in
             the target region.}
    \label{fig:Spectrometer0607}
  \end{center}
\end{figure}

The RD consisted of three active detector parts: a silicon strip detector
around the target cell inside the {\sc Hera} lepton beam vacuum, a scintillating-fibre
tracker and a photon detector. They were all surrounded by a magnet providing
a longitudinal field. The thickness of the target-cell wall was chosen to
achieve the lowest possible momentum threshold for the silicon detector.
Figure~\ref{fig:Recoil3D} shows a 3-dimensional sectional view of the CAD model of
the RD including the target cell and the superconducting magnet. The coordinate
system used by {\sc Hermes} and the RD has the z axis aligned along the beam momentum,
the y axis vertical upwards, and the x axis horizontal, pointing towards the outside
of the {\sc Hera} ring. Figure~\ref{fig:RecoilXYView} \mhl{shows a cross-sectional view (in the 
x-y plane) of all sensitive detector components of the RD including target cell and
its support, and the wall of the scattering chamber.} These individual components are described
in detail in the next chapter. The side view of the {\sc Hermes} spectrometer together
with the RD is shown in figure~\ref{fig:Spectrometer0607}.

%
\section{Detector Components}
%

\subsection{Magnet}

The active components of the RD were mounted inside the bore of a super-conducting
solenoid that provided the field for momentum reconstruction and ensured that Moeller
electrons could not reach the SSD. The magnet was built at the D. V. Efremov Scientific
Research Institute in St. Petersburg, Russia. The warm bore of the insulation-vacuum
vessel of the magnet had a minimum diameter of $501\unit{mm}$ to accommodate the detector.
The magnet consisted of two superconducting coils separated by $19.8\unit{cm}$ in a
Helmholtz configuration. The coils were made of NbTi wires with a diameter of $0.85\unit{mm}$
and mounted on massive copper rings inside the cryostat. These copper rings provided
the necessary mechanical stability for the coils as well as cooling because the coils
were by construction only partially immersed in liquid helium. During normal operation,
the liquid-helium level was kept at about $80\unit{\%}$ leaving a small portion of both
coils exposed to helium vapor. The operating current was $166\unit{A}$ resulting in a
$1\unit{T}$ magnetic field at the center of the bore. The inductance was about $10\unit{H}$
with a stored energy of around $136\unit{kJ}$ in the coils. The quench protection was provided
by two large diodes with a critical voltage of $1\unit{V}$. This limited the ramp-up speed of
the magnet to $0.1\unit{A/s}$. In the event of a quench a maximum voltage difference of
$700\unit{V}$ over the coils was observed.

\begin{figure}[b]
  \begin{center}
    \includegraphics[width=0.495\textwidth]{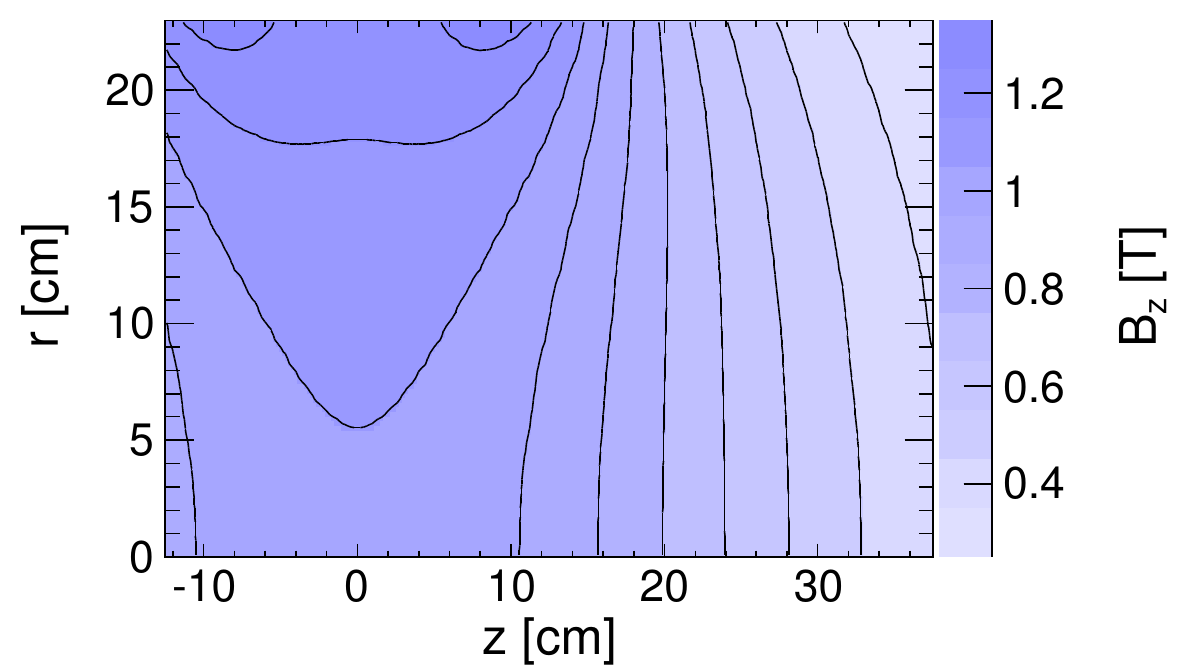}
    \includegraphics[width=0.495\textwidth]{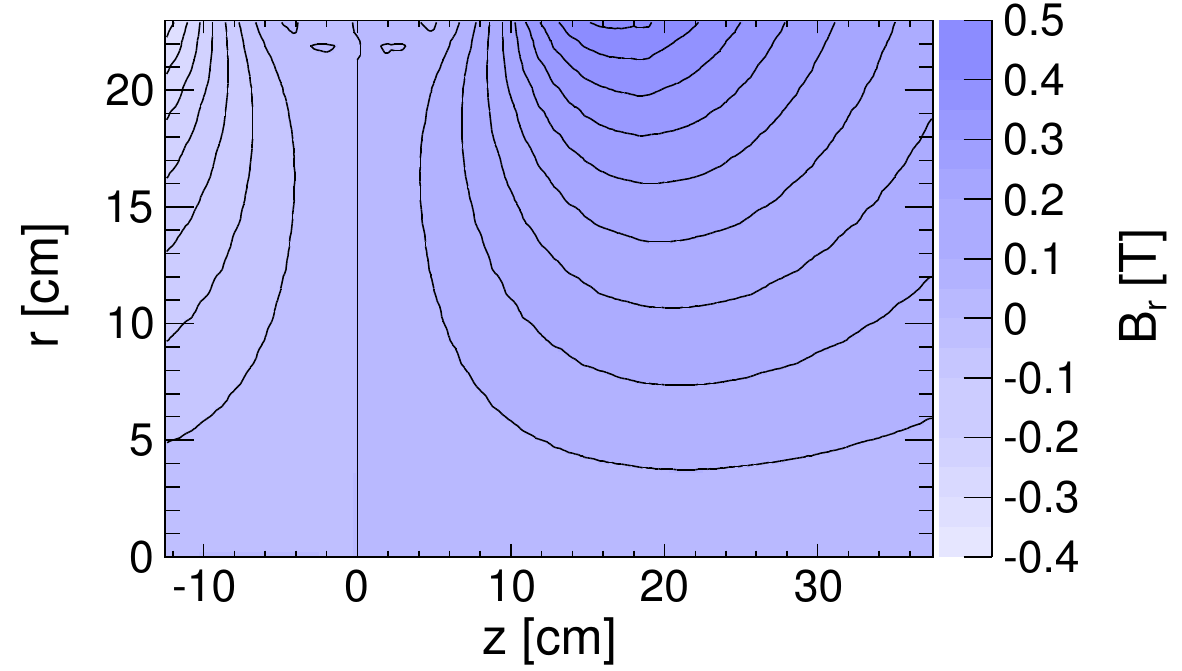}
    \caption{Measured $z$-component (left) and radial component (right) of the magnetic field
             as a function of $z$ and radius $r$.}
    \label{fig:FieldMap}
  \end{center}
\end{figure}

Two concentric heat shields were installed on the inside of the helium vessel as well as on
the outside to protect the cryostat from heat radiated off the vacuum vessel. These heat
shields were cooled by the cold helium boiling off from the cryostat. RhFe thermal resistors were
mounted on both copper rings and on all four heat shields to monitor the temperature of the
whole cryostat. For each temperature sensor a three-point calibration was available to determine
the actual temperature. Two helium-level probes produced by American Magnetics Inc. were
installed in the helium vessel for contingency, each with an active length of $800\unit{mm}$.
The probes ran along the outer surface of the helium vessel in a semi circle. The cryostat was
able to hold a maximum of $55$ liters of liquid helium. The insulation vacuum was better than
$10^{-5}\unit{mbar}$ during operation. In order to further protect the magnet from a potential
quench due to overheating, cold boil-off helium from the magnet was pumped through the hollow
electric leads in order to keep cold the electric terminals, which connected the power supply
to the magnet.

Prior to installing the magnet in the experiment, the magnetic field was measured in the bore
and in the downstream region of the magnet. All three field components were measured 
simultaneously with three calibrated Hall probes. The step size in each dimension was
$1\unit{cm}$ providing an extensive field map for track reconstruction. The $z$ and radial
components of the measured magnetic field map are shown in figure~\ref{fig:FieldMap} with
$z$ pointing in the direction of the lepton beam.

\subsection{Target Cell}

During the period after the RD installation, {\sc Hermes}
exploited both unpolarised hydrogen and deuterium as gaseous targets fed into
a storage cell. This target cell consisted of an elliptical aluminium tube
installed co-axially to the lepton beam. It had a size of $3\unit{cm}$ ($2\unit{cm}$)
in the horizontal (vertical) plane, a wall thickness of $75\unitspace\mu\mathrm{m}$
and an overall length of $25\unit{cm}$. The gas inlet was located in the longitudinal
center of the $15\unit{cm}$ long active length. Up- and downstream of the active
length pumping holes covered by a radio frequency shielding mesh were located.
The overall position of the target cell was shifted by
$12.5\unit{cm}$ downstream with respect to the position of the previously
installed $40\unit{cm}$ long cell in order to achieve optimal acceptance
of both the {\sc Hermes} forward spectrometer and the RD
(see section~\ref{sec:requirements}). The upstream and downstream ends of the
tube were reinforced with elliptical pieces of aluminium. To the sides, the tube
was supported by $4\unit{mm}$ thick aluminium plates. \mhl{At the upstream end the
supports were attached to the scattering chamber, whereas at the downstream end
they were supported by two alignment pins.}

In order to understand the heating of the cell by the beam, which had a bunch
frequency of $10.4\unit{MHz}$ and a bunch length of $28\unit{ps}$, detailed
heating studies on a real target cell were carried out. The studies
revealed that a power of approximately $60\unit{W}$ was generated by the beam during
a typical {\sc Hera} lepton beam injection. The target
cell was therefore cooled by water flowing through pipes soldered to copper
rails running along the support wings on both sides of the cell tube. In order
to monitor the temperature of the target cell construction, especially during injection
of the {\sc Hera} lepton beam, two Pt1000 thermal resistors were mounted on the
support wings. During regular data collection the temperature measured by both
sensors was approximately $16\unitspace^\circ\mathrm{C}$ which was only slightly above the
cooling water temperature, whereas during injection the temperature rose
by about $15\unitspace^\circ\mathrm{C}$.

\subsection{Silicon Strip Detector}

\begin{figure}[b]
  \begin{center}
    \includegraphics[width=0.50\textwidth]{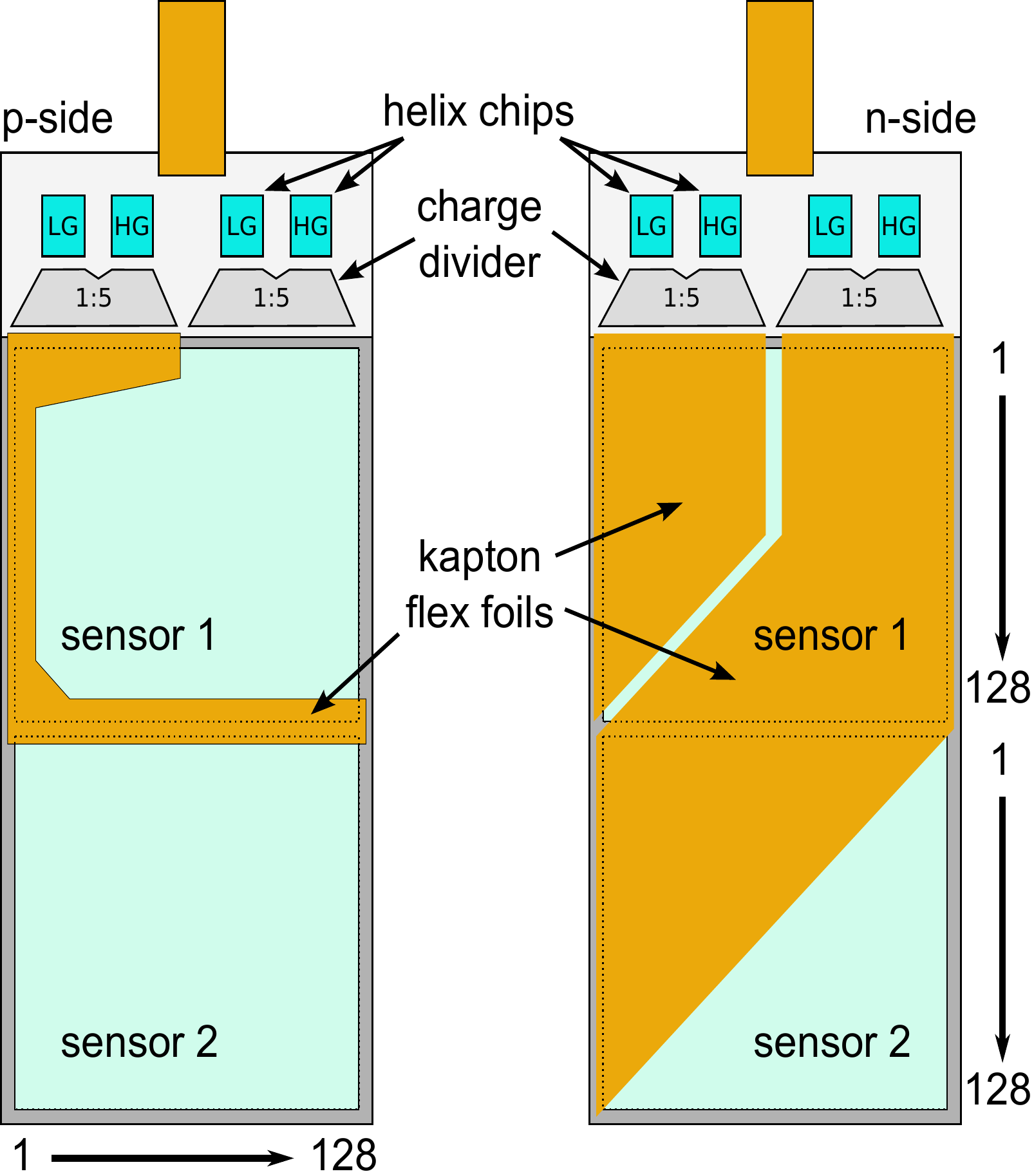}
    \caption{Sketch of the front- and backside of one module of the recoil
    	         silicon strip detector (SSD).}
    \label{fig:module}
  \end{center}
\end{figure}

The innermost active detector component was the SSD
made up of 16 double-sided silicon strip sensors arranged in two layers
around the target cell. In order to allow for a minimum detection threshold for
protons, the amount of passive material between the interaction point and
the first layer of sensors was minimized by placing the silicon sensors and the front-end
read-out electronics inside the {\sc Hera} \mhl{beam vacuum as close as $5.8\unit{cm}$
to the lepton beam.} Each sensor was made of N-type bulk material, and was of the
TTT design by Micron Semiconductors Inc.~\cite{ONeill:2003}. They had a total
area of $99 \times 99\unit{mm}^2$, an active area of $97.3 \times 97.3\unit{mm}^2$
and a pitch of $758.2\unitspace\mu\mathrm{m}$. The individual sensor
thicknesses varied between $295\unitspace\mu\mathrm{m}$ and $315\unitspace\mu\mathrm{m}$.
Two sensors were glued into a holding frame made of Shapal-M~\cite{ShapalM}, which
had a thermal expansion coefficient very similar to that of silicon, and combined with
the front-end read-out electronics to form a module. Figure~\ref{fig:module}
shows a sketch of the front and back sides of one module. The strips on the
p-side (n-side) of each sensor were oriented parallel (perpendicular) to the
{\sc Hera} lepton beam axis. \mhl{The signals from the $128$ strips on each p- and n-side
were routed to the front-end hybrid electronics via $50\unitspace\mu\mathrm{m}$ thick
polyimide foils (flexleads).} On the hybrid the signals were split by a
charge divider network with a ratio of $\approx1:5$ into a high-gain (HG) and a
low-gain~(LG) component to increase the dynamic range. For each sensor side, the HG
and LG signals were fed to two
{\sc Helix}~3.0 front-end amplifier chips~\cite{FallotBurghardt:PhD}, resulting in
eight front-end amplifiers per module. Details on the module and hybrid design
can be found in reference~\cite{Reinecke:2003}. All modules were tested for
functionality in a dedicated laser test stand~\cite{Vandenbroucke:PhD}. A full
system test in combination with all other RD components was performed prior to
the installation at {\sc Hermes} and is described in reference~\cite{Pickert:PhD}.

In total eight modules were mounted on an aluminium holding structure to build
the SSD. In order to minimize the amount of passive material between the
inner and outer SSD layers, the inner and outer modules were mounted such that
the p-sides, which had less areal coverage by flexleads than the n-sides, faced
each other. \mhl{The distance between the inner and outer SSD layers was $1.5\unit{cm}$
and optimized for $\phi$ resolution while maintaining a high $\phi$ acceptance
taking into account geometrical constraints such as the size of the chosen sensors
(including frame), the inner diameter of the scattering chamber, and the thickness of
the support structures of the target cell. Under the assumption of perpendicular
incidence, the material budget of the SSD was $6.45\cdot10^{-3} X_{0}$ in the area
without flexlead coverage and $6.80\cdot10^{-3} X_{0}$ in the area with flexleads.
The material budget of the passive material before the inner SSD sensors (target cell
wall and n-side flexleads) was $1.02\cdot10^{-3} X_{0}$.}

\begin{figure}[ht]
  \begin{center}
    \includegraphics[width=0.37\textwidth]{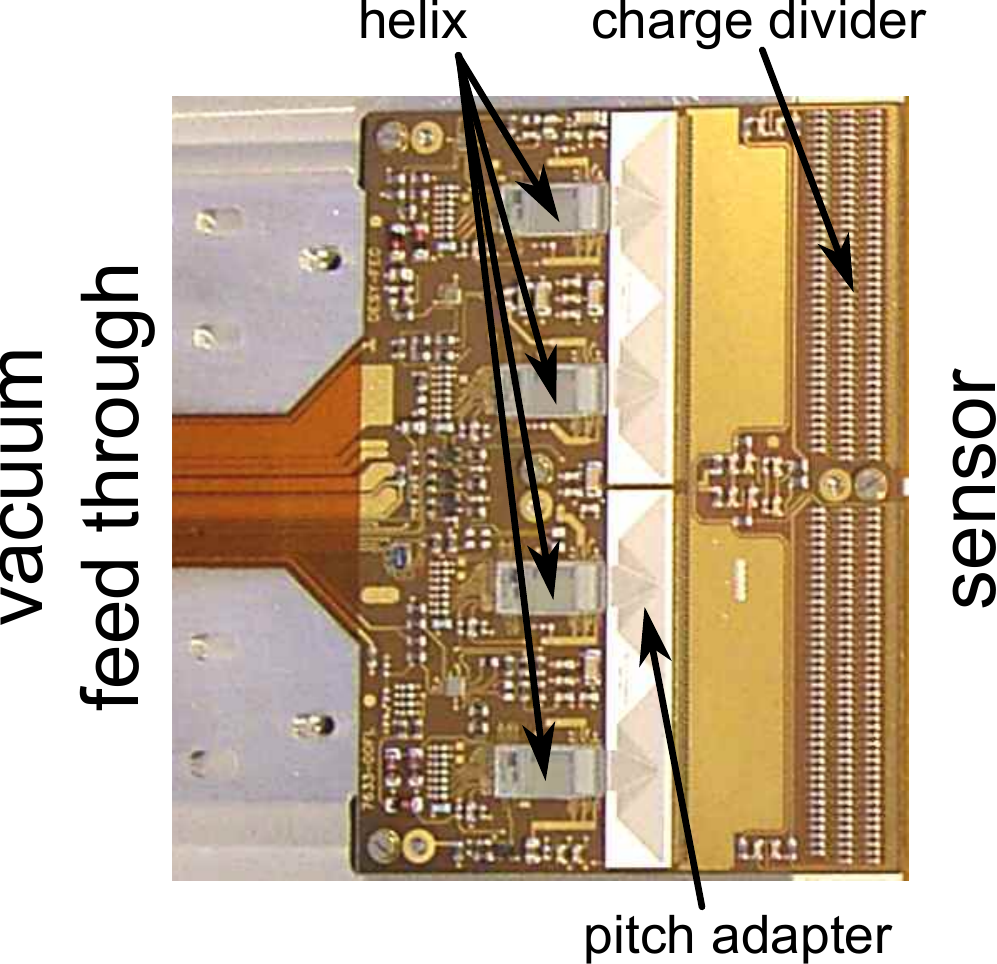}
    \hspace*{1.0cm}
    \includegraphics[width=0.53\textwidth]{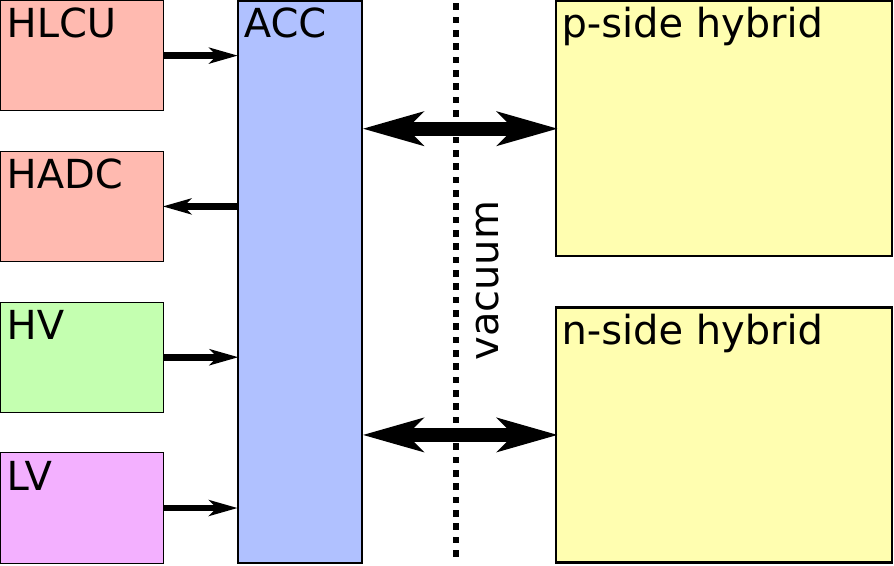}
    \caption{Left: Picture of one of the SSD hybrids. Right: Connection scheme of the
             hybrids to the read-out electronics (see text for details).}
    \label{fig:hybrid}
  \end{center}
\end{figure}

The hybrid was made of a flexible polyimide printed circuit board with a tail that comprised
both the signal and supply lines and provided the connection to the vacuum feed-throughs.
In addition, the p-side (n-side) hybrids were equipped with a temperature (radiation)
sensor. The left panel of figure~\ref{fig:hybrid} shows one of the hybrids with its
four {\sc Helix} chips, the pitch-adapter and charge-divider network on the sensor side,
and the flexible polyimide tail for the connection. With the front-end electronics operated
in vacuum, the hybrids were mounted on a copper heat sink, which was in turn cooled
by ethanol at $-15\unitspace^\circ\mathrm{C}$ resulting in a temperature of typically
$10\unitspace^\circ\mathrm{C}$ on the hybrid.

The sketch on the right side of figure~\ref{fig:hybrid} shows the simplified connection
scheme for the two hybrids of one module (p- and n-sides). The digital control signals
required to download setup sequences and control the {\sc Helix} chips during read-out were
provided by so-called HeLix Control Units (HLCUs). A single HLCU module was able to serve
up to eight daisy chains of {\sc Helix} chips. The analog output was digitised by {\sc Hermes}-ADC
modules (HADCs) of which each could handle the data from up to four chains of {\sc Helix}
chips~\cite{Steijger:2000}. The HLCUs and HADC units were produced by ZOT Integrated
Manufacturing, Musselburgh, UK. A dedicated Analog Control Card (ACC) was used to merge
all signal lines into a single cable per hybrid. The power supply for both the ACC and
the connected hybrids was provided by custom made low-voltage (LV) modules, whereas the
bias voltage for the sensors originated from commercially available high-voltage
(HV) units.

\begin{figure}[ht]
  \begin{center}
    \includegraphics[width=0.65\textwidth]{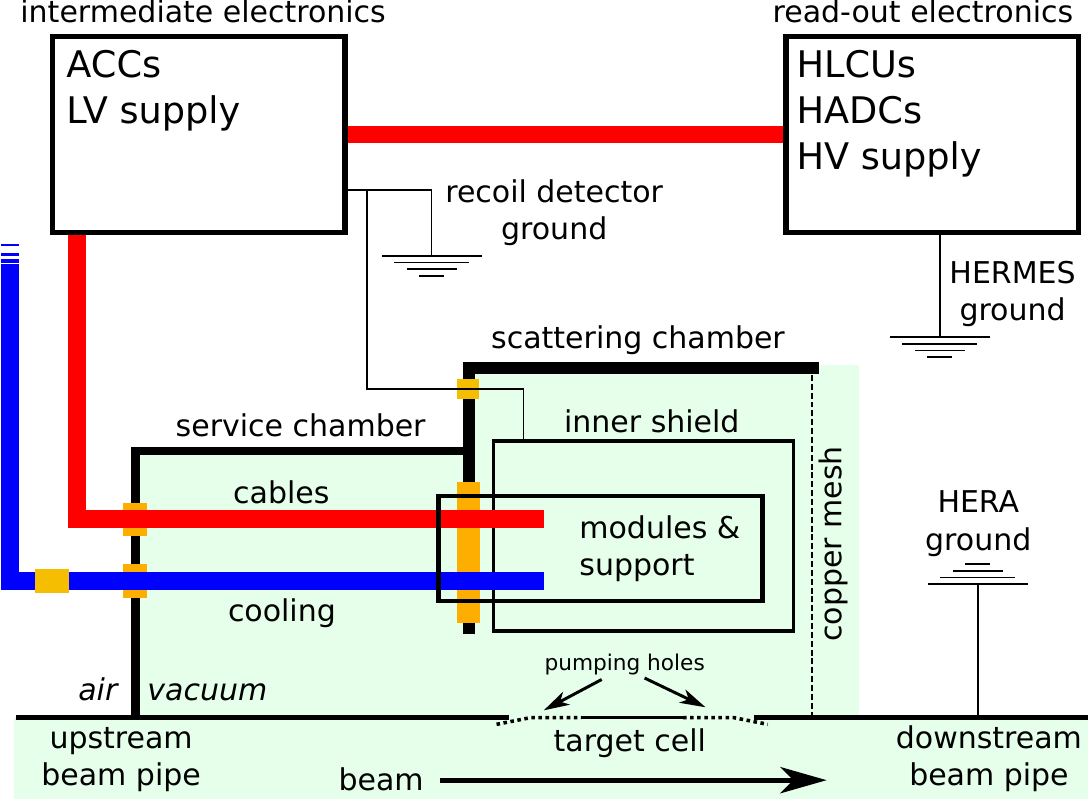}
    \caption{Simplified view of the SSD grounding and shielding scheme.
             The target cell connects the up and downstream beam pipes and acts as
             a first shield to avoid radio frequency pick-up. The modules and the
             holding structures were housed in a cage serving as a secondary shield. The
             in-vacuum installation was electrically decoupled from the scattering
             chamber and beam pipe requiring a dedicated RD ground.}
    \label{fig:ssd_grounding}
  \end{center}
\end{figure}


The silicon sensors and the in-vacuum front-end electronics were operated as close
as $5.8\unit{cm}$ to the {\sc Hera} lepton beam under an environment with high transient
electromagnetic fields. An elaborate grounding and shielding scheme
was therefore developed to avoid radio frequency (RF) pick-up induced by the beam.
Figure~\ref{fig:ssd_grounding} shows a simplified grounding and shielding scheme.
The target cell connected both the up and downstream beam pipes electrically and
served as a primary RF shield. The scattering chamber as well as the so-called service
chamber, which contained the vacuum feed-throughs for power supply, signals and cooling
lines, were connected to the beam pipes. The down-stream end of the scattering
chamber was closed by a fine copper mesh with a connection to the beam pipe that
shielded the downstream detection systems from radio frequency noise generated in the
chamber. \mhl{The beam pipe, target cell, and service and scattering chambers were all kept
at a ground potential common to all {\sc Hera} beam line components. A cage built
around the silicon holding structure acted as a second RF shield (inner shield).}

The holding structure was mounted on the scattering chamber with ceramic insulation
pieces to electrically decouple the in-vacuum installation from the common {\sc Hera}
ground. In order to decouple the cooling lines from common {\sc Hera} ground, the vacuum
feed-throughs were insulated from the service chamber and additional ceramic insulation
pieces were used to break the electrical connection between the in-vacuum cooling lines and
the chiller. The intermediate electronics with the ACCs and the LV power supply, located about
$3\unit{m}$ away from the interaction region, were kept on a dedicated recoil
detector ground, whereas the read-out electronics (HLCUs and HADCs) and the HV supply
were kept at a ground potential common to all {\sc Hermes} read-out electronics components.

\subsection{Scintillating-Fibre Tracker}

\begin{figure}[ht]
  \begin{center}
    \includegraphics[width=0.40\textwidth]{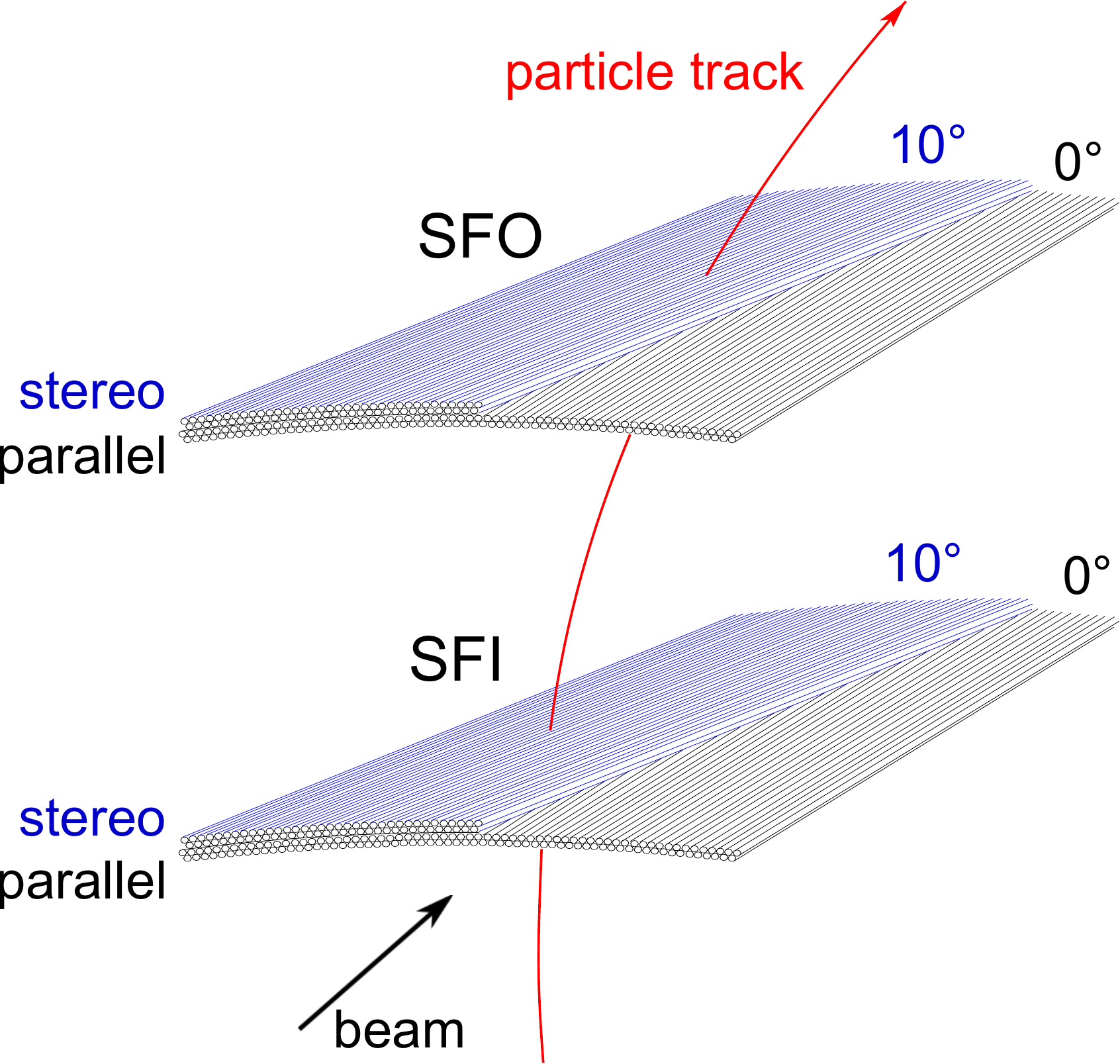}
    \includegraphics[width=0.55\textwidth]{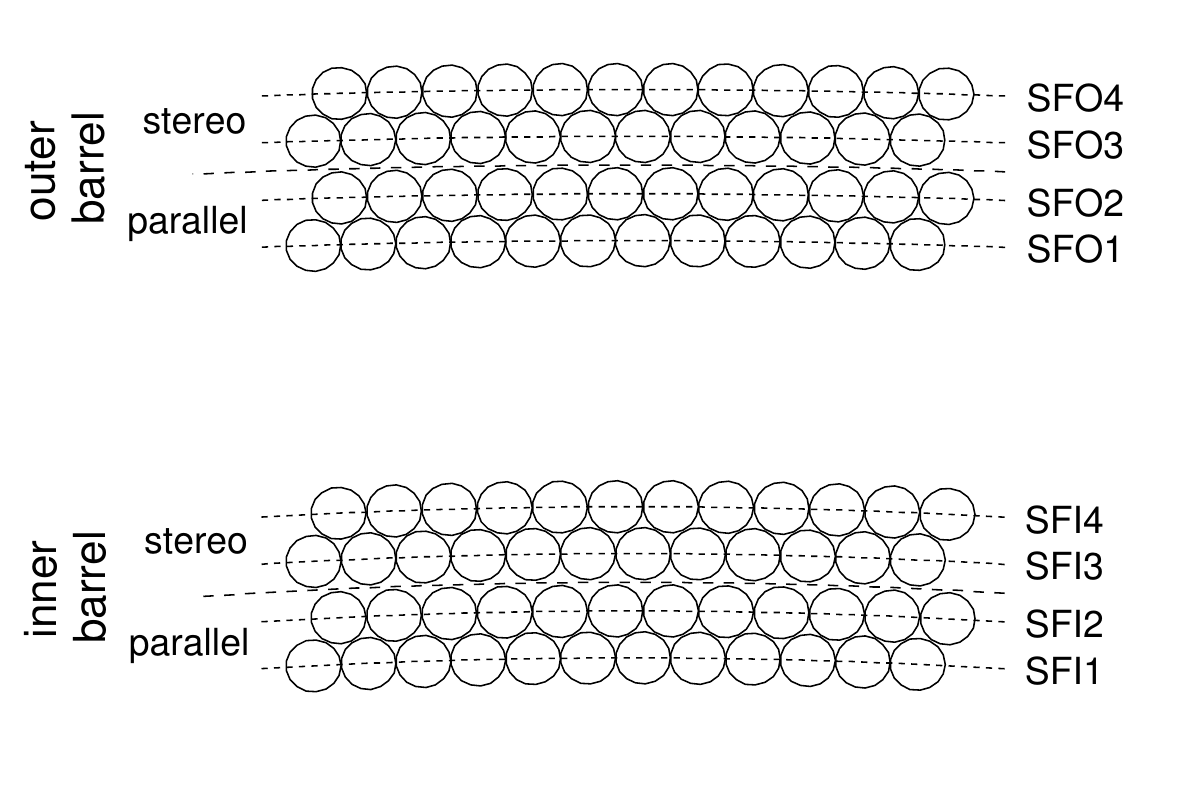}
    \caption{Schematic view of the Scintillating Fibre Tracker (SFT) barrel configuration
             (left) and the arrangement of fibres in the layers (right).}
    \label{fig:SFT_Barrel}
  \end{center}
\end{figure}

Charged particles with sufficiently large momenta escaped the $1\unit{mm}$ thick aluminium
scattering chamber and were detected by the SFT. The SFT consisted of two concentric barrels
of scintillating fibres, the inner barrel (SFI) with a radius of about $11.5\unit{cm}$ and the
outer barrel (SFO) with a radius of about $18.5\unit{cm}$. The active length of both barrels
was $28\unit{cm}$. Each barrel in turn was made of two sub-barrels with the inner sub-barrel
having the fibres oriented parallel to the beam axis, whereas the corresponding outer sub-barrel
had the fibres inclined by $10^\circ$ (stereo layer). The top two panels in
figure~\ref{fig:SFT_Barrel} show a schematic view of the barrel configuration and the
arrangement of fibres in the layers. In both barrels, the two parallel and the two stereo layers
(see figure~\ref{fig:SFT_Barrel} right) can be seen as a single entity, in which for each
sub-barrel the outer layer is shifted with respect to the inner one by half a fibre diameter.

\begin{figure}[ht]
  \begin{center}
    \includegraphics[width=0.60\textwidth]{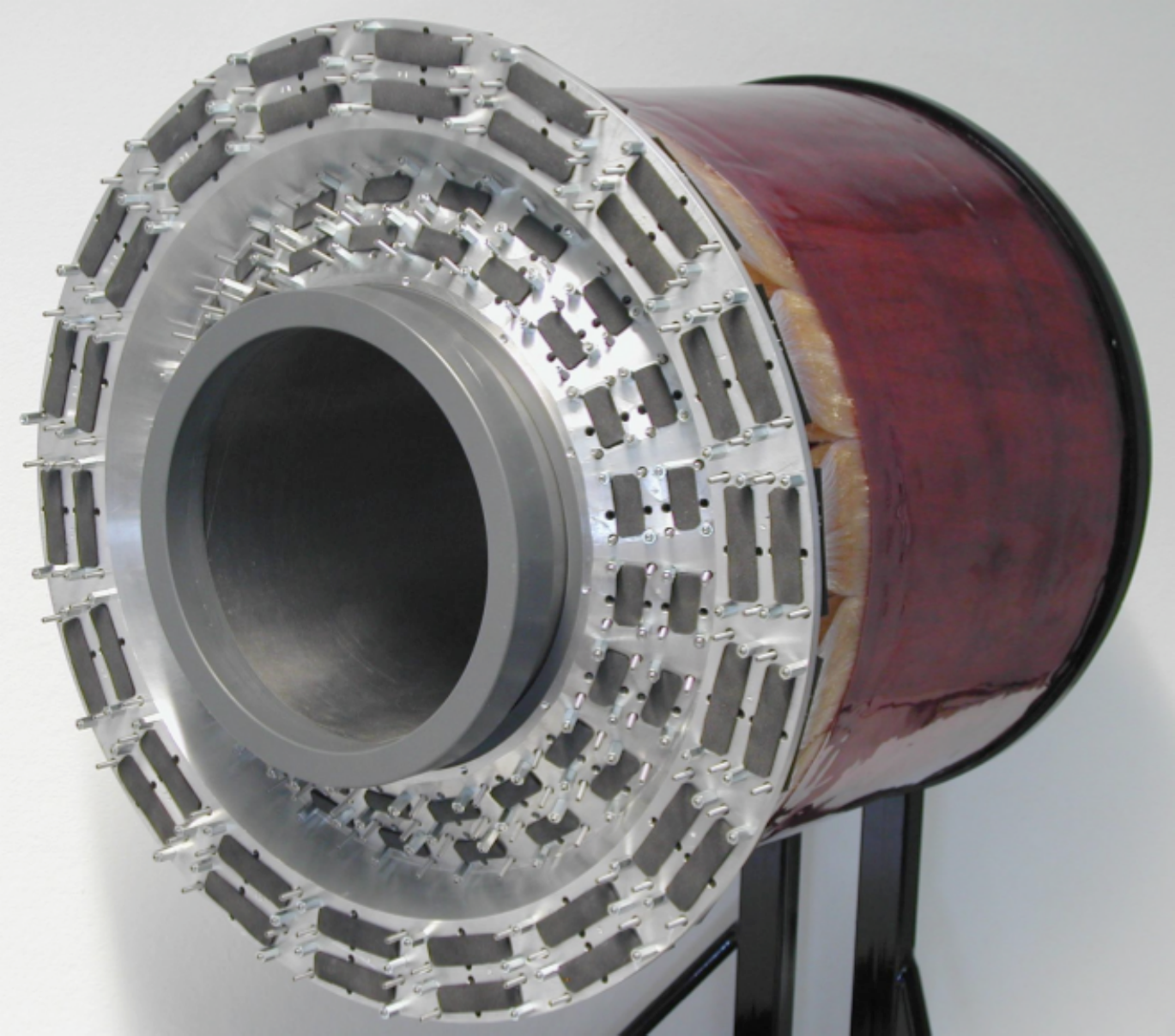}
    \caption{Picture of the assembled SFT mounted over a mockup scattering chamber.}
    \label{fig:SFT_BarrelPicture}
  \end{center}
\end{figure}

The sub-barrels comprised $1318$ inner parallel, $1320$ inner stereo, $2198$
outer parallel and $2180$ outer stereo Kuraray SCSF-78M multiclad scintillating
fibres with a diameter of $1\unit{mm}$~\cite{Hoek:PhD,SFTNIMA:2007}. In order to form a
sub-barrel, modules of 64 (128) fibres for the inner (outer) barrels were joined. Precast moulds
with the corresponding curvature and fibre inclination were used in order to produce
the four different types of modules used in the SFT. The far-end faces of the fibres
were metallised, leading to an increase in light yield by $20\unitspace\%$. The near ends were
glued into custom made connectors mounted on an aluminium holding structure.
Figure~\ref{fig:SFT_BarrelPicture} shows the assembled SFT mounted on a dummy scattering chamber.
The connectors at different radii on the upstream holding ring are visible at the left side of the
picture, indicating the four sub-barrels. The produced scintillation light from each module was
fed to 78 64-channel Hamamatsu H7546B Multi-Anode PhotoMultiplier Tubes (MAPMTs)~\cite{MAPMT:2000}
via $4\unit{m}$ long light guides made of Kuraray clear fibres. Besides the space constraint, also
the sensitivity of the MAPMTs to magnetic field was a reason for placing the MAPMTs at a distance
of $2.5\unit{m}$ from the magnet z-axis. With additional $\mu$-metal and soft-iron shielding the
maximum field at the location of the MAPMTs was reduced to $5\unit{mT}$.

In the inner barrel, 32 fibres from a regular layer and the corresponding shifted layer
shared one MAPMT, with one fibre connected to one MAPMT pixel, whereas for the modules in the
outer SFT barrel two fibres, one from the regular and one from the shifted layer, were
joined in the connector on the MAPMT end of the light guides. In order to allow for a distinction
between cross-talk on the MAPMT surface and cross-talk between neighbouring active fibres,
neighbouring active fibres were not connected to neighbouring pixels on the MAPMT but rather in
a more elaborate scheme (see section~\ref{sec:sftXTalkCorr}).
The MAPMT has its 64~channels arranged in a $8 \times 8$ grid of $2 \times 2\unit{mm}^2$ pixels
and provides a high-speed response while having low internal cross-talk despite the high
integration of electronics components. In addition to the 13~dynode stages for each pixel, the
MAPMT provides a common signal for all pixels after the 12th stage (denoted as dynode--12 signal).
All MAPMTs were tested prior to installation~\cite{VanHaarlem:PhD}.

The readout electronics for the MAPMTs were based on the design of the HADES-RICH
Preprocessing Front-end Module (PFM) \cite{Kastenmueller:1999} with an additional
charge-divider network between the MAPMTs and the PFMs to adjust the signal amplitudes.
This readout system allowed to obtain information on the energy deposit, which
is exploited in the particle-identification procedure (see section~\ref{sec:PID}).
The readout system had an integration time of about $600\unit{ns}$, which meant that
the signals from $7$ {\sc HERA} lepton-beam bunches were summed for each MAPMT channel.
In order to avoid the combination of hits from different bunches reconstructed into a
single track, the dynode--12 signal was read out by a fast commercially available
multi-hit Time-to-Digital Converter (TDC).

\subsection{Photon Detector}

The main purpose of the outermost active detector component, the PD, was the detection
of photons, however, it was also capable of detecting pions and protons with momenta above
approximately $300\unit{MeV}$ and $500\unit{MeV}$, respectively. In order to obtain
good photon-detection efficiency, and taking into account geometric constraints, the PD
was built as a cylindrical volume of three subsequent layers of alternating tungsten as
a converter and plastic scintillator material for the detection of the charged (shower)
particles.

The layer thickness of the photon detector was optimized based on Monte Carlo simulation. The
innermost tungsten layer was $6\unit{mm}$ thick, whereas the two following tungsten layers were
each $3\unit{mm}$ thick. This corresponded to a total thickness of $3.4$ radiation lengths, and
resulted in a $85$\% photon-conversion probability at the characteristic photon energies of
about $150\unit{MeV}$.

\begin{figure}[ht]
  \begin{center}
    \includegraphics[width=1.0\textwidth]{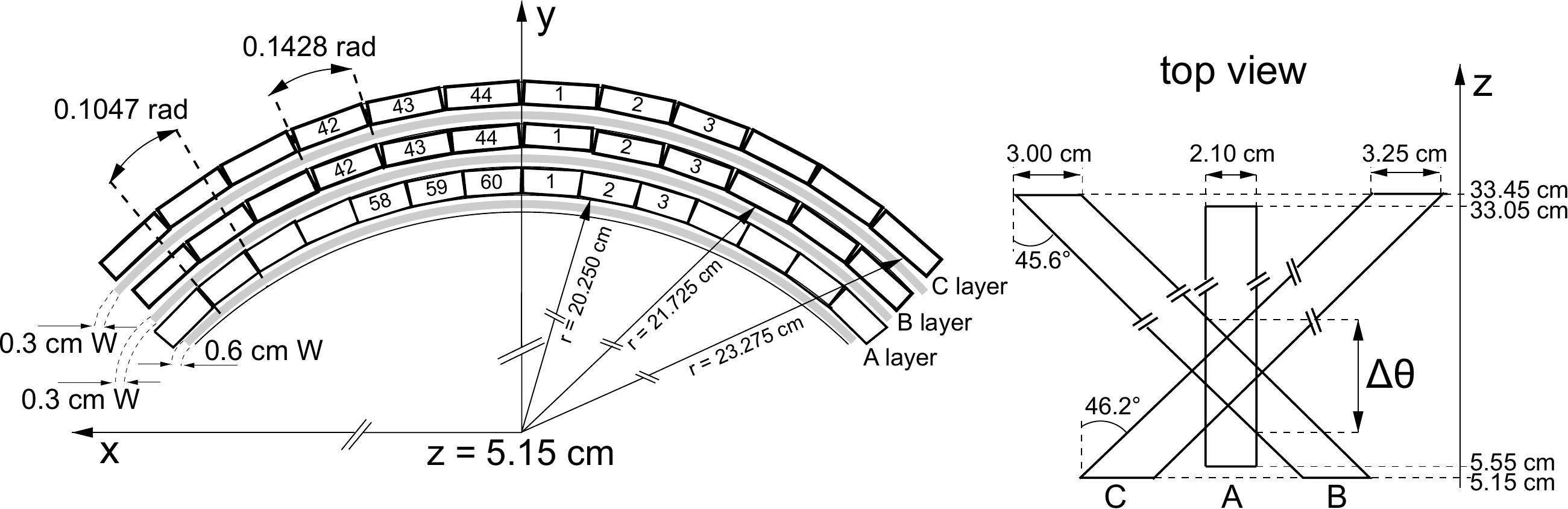}
    \caption{Geometric description of the PD in the {\sc Hermes} coordinate frame. The strip
             numbering is also indicated. For clarity, the picture is not to scale.}
    \label{fig:PD_Geom}
  \end{center}
\end{figure}

The plastic scintillator material was BC-408 from the manufacturer
Saint-Gobain Corporation~\cite{SaintGobain:2008}. Each scintillator layer of the PD was
segmented into strips. \mhl{The inner layer ($A$~layer) had $60$ strips, which had a trapezoidal
cross section for maximal coverage.} The strips were $27.5\unit{cm}$ long, $1\unit{cm}$ thick and
had a mean width of $2.05\unit{cm}$.
They were arranged parallel with respect to the beam line. The
subsequent stereo layers had $44$ strips oriented under an angle of $+45.6^{\circ}$ for the second
layer ($B$~layer) and $-46.2^\circ$ for the third layer ($C$~layer) in order to allow for a
spatial reconstruction of the photons. The overlap between a parallel and a stereo layer resulted
in a polar-angular resolution of about $2.7^{\circ}$. The stereo
strips were fabricated out of rectangular straight blocks of $1.00 \times 2.10\;(1.00 \times 2.25)\unit{cm}^2$
for the second (third) layer, then bent, twisted and cut into their final shape. This resulted
in an effective cross section in the plane orthogonal to the beam axis of
$1.00\times 3.00\;(1.00 \times 3.25)\unit{cm}^2$. The strips covered a length of $28.3\unit{cm}$
along the beam line. A schematic drawing of the PD is depicted in figure~\ref{fig:PD_Geom}.

Two $60\unit{cm}$ long round wavelength-shifting fibres with a diameter of $1.5\unit{mm}$
were glued with BC-$600$ optical cement in grooves along each side of a strip to capture and
redirect the scintillation light from the strips to $2\unit{m}$ long clear light
guides. The fibres (clear light guides) were BCF-$91$A (BCF-$98$) from the same
manufacturer as the scintillator strips. In order to increase the light yield, the strips were
covered with BC-$620$ reflective paint and the wavelength-shifting fibres were mirror
coated at their extremity.

\begin{figure}[t]
  \begin{center}
    \includegraphics[width=0.5\textwidth]{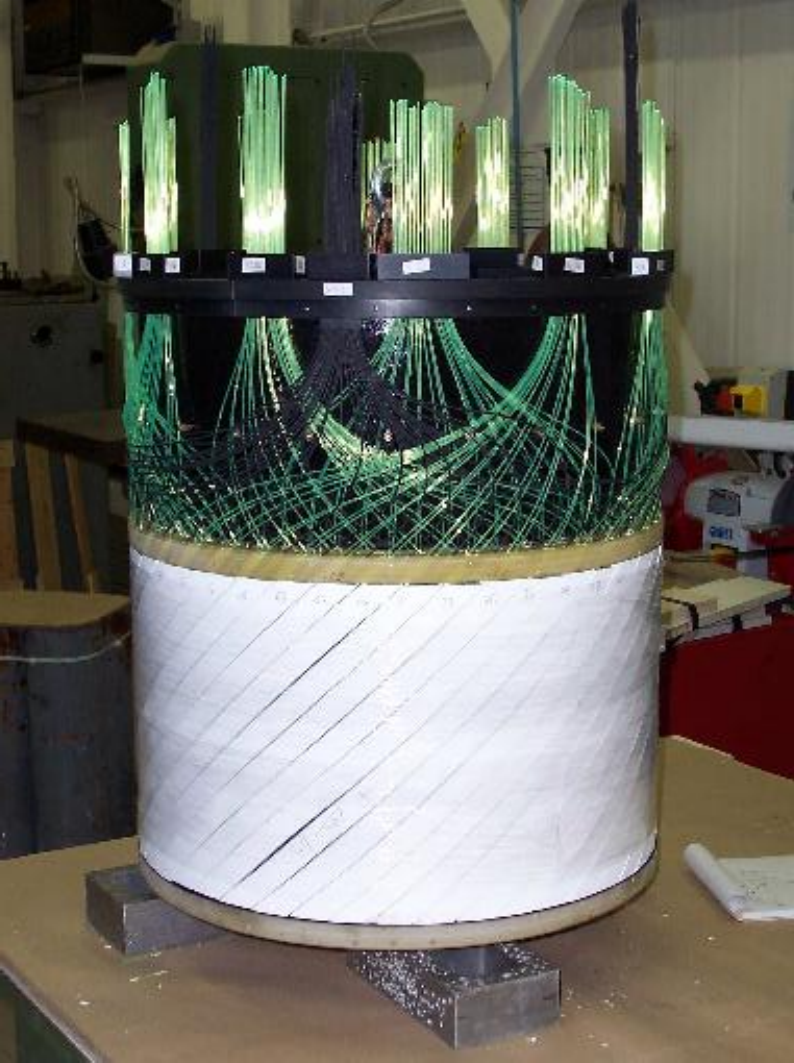}
    \caption{Picture of the PD during its assembly.}
    \label{fig:PD_Picture}
  \end{center}
\end{figure}

Figure~\ref{fig:PD_Picture} shows a picture of the PD during construction. Clearly visible
are the white-painted strips of the outermost layer, which are oriented under a stereo angle.
The wavelength-shifting fibres, with their green light output, were connected to optical
connectors. In the picture the excess ends of the fibres point straight up. Not shown are the
clear light guides that are connected to the depicted optical connectors. In the final assembly,
care was taken to ensure that the detector was completely light tight by covering it with
black foil and black paint.

The readout of the clear light guides was performed by $6$ Hamamatsu 64-channel H7546B multi-anode
PMTs (MAPMTs)~\cite{MAPMT:2000}. The cathode pixel size was $2\times2$~mm$^2$, which allowed
for the connection of one fibre per pixel. \mhl{In order to minimize influence of cross-talk on the MAPMT cathode,
the arrangement of the light guides on the MAPMT pixel matrix followed a scheme similar to
that of the SFT.} The MAPMTs were, each individually, surrounded by two $0.2$~mm thick
$\mu$-metal sheets and placed in a soft-steel case of $14\unit{mm}$ thickness~\cite{VanHaarlem:PhD}.
Per group of three they were then placed in an additional soft-steel box. The boxes were
installed at a distance of $1.5\unit{m}$ from the magnet, where the magnetic field was of
the order of $20\unit{mT}$, which was reduced to $0.2\unit{mT}$ inside the shielding.

The signals from the MAPMTs were transferred to a patch panel, and from there to the transmitter electronics.
In the transmitter, signals originating from the same strip were summed together, amplified,
and transported as a differential signal over $30\unit{m}$ long flat cables to
the receiver electronics. The receiver converted the signals back to non-differential signals,
amplified them, and sent them over $70\unit{m}$ of flat cable, acting as a time delay, to $6$
charge-integrating analog-to-digital converters (ADCs), one per MAPMT. Also here, to allow for
cross-talk correction, care was taken that channels adjacent in the flat cables were neither
neighbouring pixels on the MAPMT surface nor adjacent strips in a PD layer.

Although the time between two {\sc Hera} bunches amounted to $96\unit{ns}$, the gate width for the
readout of the ADCs was set to $250\unit{ns}$ in order to accommodate the full length of the PD
signals. \mhl{According to test measurements performed during commissioning no signal overlay was
observed.}

%
\section{Data Collection}
%

\subsection{Data Acquisition System}

The integration of the RD into the {\sc Hermes} data-taking infrastructure~\cite{HermesNIMA:1998}
involved the additional readout of 8192~channels for the SSD, 5120~channels for the SFT and
192~channels for the PD.

The original {\sc Hermes} Data Acquisition (DAQ) system was built using Fastbus crates with
crate controllers based on the now outdated Motorola M68020 microprocessor and
two Digital Signal Processors (DSPs). Already at the time of the RD upgrade
planning, these modules were not easily available on the market. Programs for microprocessors
and DSPs for each particular crate were written in assembler language and had substantial
volume and complicated structure. All this made extending the existing system by adding
identical readout equipment for the new detectors unreasonable. Instead, another
approach was chosen.

\begin{figure}[t]
  \begin{center}
    \includegraphics[width=0.70\textwidth]{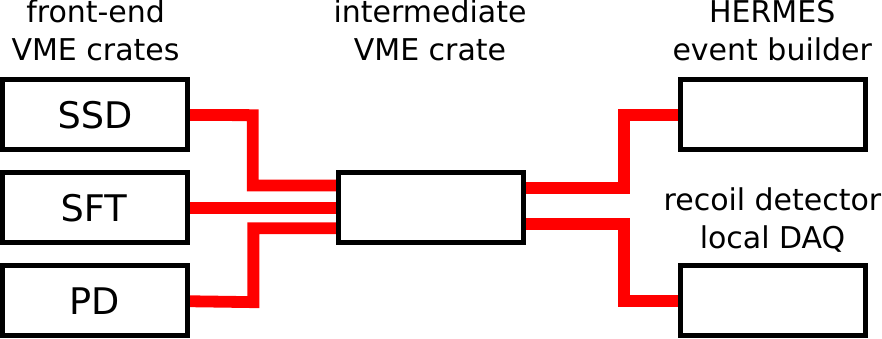}
    \caption{Schematic diagram of the RD data acquisition system. The
             digitisation devices for each sub-detector were housed in separate
             front-end VME crates. These crates were connected via optical links
             to a so-called intermediate VME crate that managed the connection
             of the front-end crates with either the {\sc Hermes} event builder or the
             local RD DAQ.}
    \label{fig:daq_scheme}
  \end{center}
\end{figure}

The basic concept of incorporating the RD readout into the {\sc Hermes} DAQ
was the preservation of the existing hardware and software structure, i.e., building an event
online in contrast to merging two independent data streams offline. The benefits of such an
approach were minimal software changes in the offline data processing and online monitoring.
The only modifications to the software involved the inclusion of new data streams into the online
event reconstruction and the addition of proper decoding and reconstruction in the offline software
chain. Another important aspect of the upgrade was the use of modern and readily available components,
like PCI-VME (Peripheral Component Interconnect - Versa Module Eurocard) bus controllers with onboard
DSPs. The final readout system, schematically shown in figure~\ref{fig:daq_scheme}, consisted of
three VME crates with DSP-based controllers assigned to
each individual sub-detector. The front-end electronics consisted of HADC modules for the SSD, TDCs
and custom-made Readout Controllers (RC) for the SFT, and charge-to-digital converters for the PD. The
controllers were connected via optical links to intermediate VME interfaces located in a fourth
auxiliary crate. This crate also hosted both intermediate VME controllers for the two independent readout
chains. One of these was connected to the main {\sc Hermes} event-builder DSP, responsible for reading
data from all crate controllers and assembling the final event structure. The second VME controller was
connected via an optical link to a PCI controller in a PC dedicated to service tasks on all sub-detectors
of the RD. This control structure allowed the use of the old and new detector components
in parallel and independently of each other. It was, for example, possible to take data with
the {\sc Hermes} forward spectrometer without RD and, at the same time, perform test
and calibration runs using stand-alone programs on the dedicated PC. In order to switch between
this and {\sc Hermes} full data-taking mode, including the RD, the required action was minimized
to reloading the DSPs and changing the trigger dead-time logic~\cite{Keri:PhD}.

In addition to the readout of the main event content, a system for status and control
information was provided. It included the readout of temperature and radiation sensors
for all modules of the SSD, temperature sensors on the target cell and status information
of the SSD cooling system. This so-called slow control information was read out in $10$~second
intervals.

\enlargethispage*{0.5cm}

During data taking, the trigger for the readout of the RD was provided by
the {\sc Hermes} spectrometer. For dedicated pedestal and calibration runs a random trigger
was used. For recording cosmic ray muons, a signal in the lower half or a coincidence
of a signal in the upper and lower halves of the PD was used to trigger the RD
read-out.

\subsection{Performance}
\label{sec:DAQPerformance}

Data were collected in 'runs' which had a size of $500\unit{Megabyte}$ of raw data originating from
all sub-detectors of {\sc Hermes}. Under regular data taking conditions and with all sub-detectors
active (including RD) a single run comprised of the order of $25000$ events. Collection of data
including the RD started in February 2006 with a commissioning phase in order to set up and
tune all RD components.

At the end of February 2006, during a {\sc Hera} lepton beam injection, beam-induced radio
frequency (RF) was the most probable cause of over-heating of the target-cell foil, which
resulted in a hole in the foil of about $1\unit{cm}^2$ in size. The opening in the primary
SSD radio frequency shield allowed RF to escape. This caused in turn the SSD to stop
operating properly and meant also a substantial delay in the commissioning.
Only after a second RF incident during injection and the repair, exchange and
repositioning of several modules of the SSD, the commissioning could continue and was
successfully finished end of September 2006. During the final installation, the flat
cable connecting the n-side hybrid of one module to the vacuum feed-through broke and
caused this module side to become unusable. Of the $91000$ runs collected in the years
2006 and 2007, $69\unit{\%}$ were collected with fully commissioned SSD of which
$4.7\unit{\%}$ runs were marked as having one or more non-working {\sc Helix} chips.
This state was mainly caused by read-out chips losing their programming, which was
typically quickly recognized by the shift crew and resolved by reprogramming. The
overall data-taking efficiency of the SSD in the time after final commissioning was
$90\unit{\%}$.

In the case of the SFT, the overall data taking performance was $96\unit{\%}$. Runs were
labeled as bad due to either desynchronisation of the SFT readout system with respect
to the {\sc Hermes} DAQ or due to data corruption. Both effects were automatically cured
by reinitialisation of the SFT readout system at the beginning of each run.

The PD had a data-taking efficiency of $99.6\unit{\%}$ during the data taking period with
all sub-detectors commissioned and fully operational.

%
\section{Energy Measurement and Calibration}
%

\subsection{Silicon Strip Detector}

In order to improve the reconstruction of the momentum of protons, the energy deposits
in both SSD layers are used as additional measurements. For good performance a
precise energy measurement is thus essential.

An initial energy calibration was performed at the Tandem Accelerator of the
University of Erlangen as described in reference~\cite{Vogel:PhD}. Each module was
mounted on a movable carrier inside a vacuum chamber and each strip was irradiated
with protons of $3.5$, $4.0$ and $9.0\unit{MeV}$ kinetic energy. This allowed
for a determination of the ratio between high and low gain channels, the
cross-talk behavior as well as an absolute energy calibration. However, as the
environmental conditions at the final installation at {\sc Hera} were extremely different
compared to the ones present during the calibration, the obtained coefficients
could not be used during data taking and the analysis. The subsequent
sections describe the steps necessary for a precise determination of the energy
deposit and energy calibration based on data collected during detector operation.

\subsubsection*{Raw-Data Corrections}

In order to achieve the required resolution and absolute precision, several
corrections on the raw data must be applied. They are described in the following
paragraphs.

\paragraph{Pedestal Stability}

\begin{figure}[b]
  \begin{center}
    \includegraphics[width=0.495\textwidth]{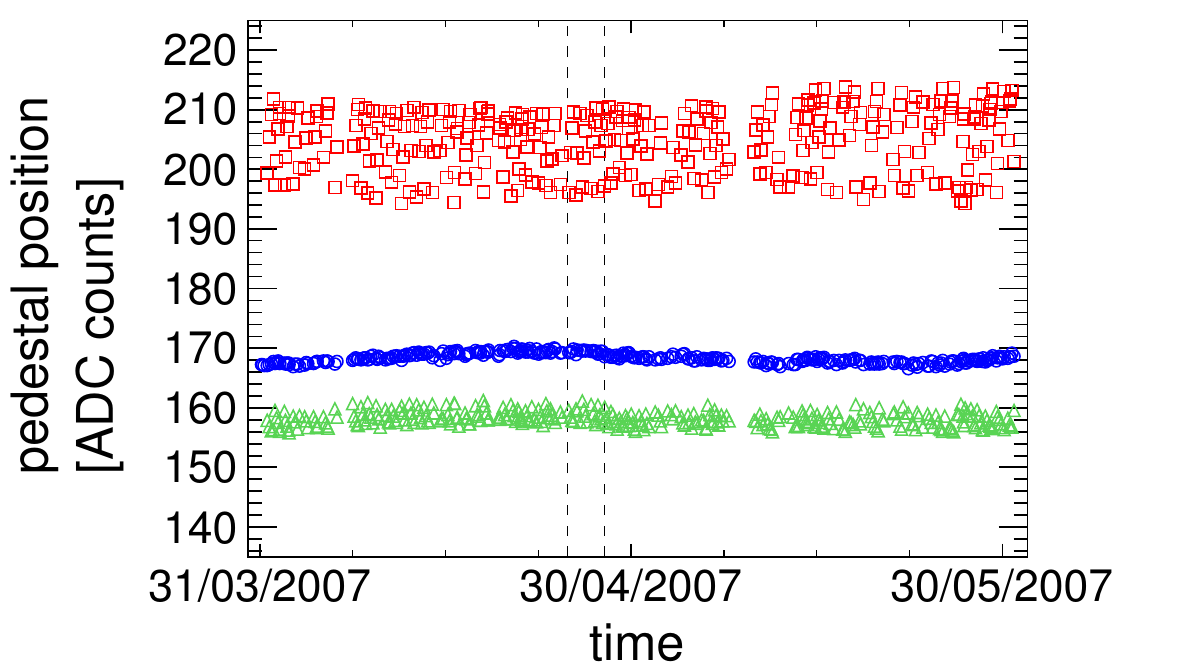}
    \includegraphics[width=0.495\textwidth]{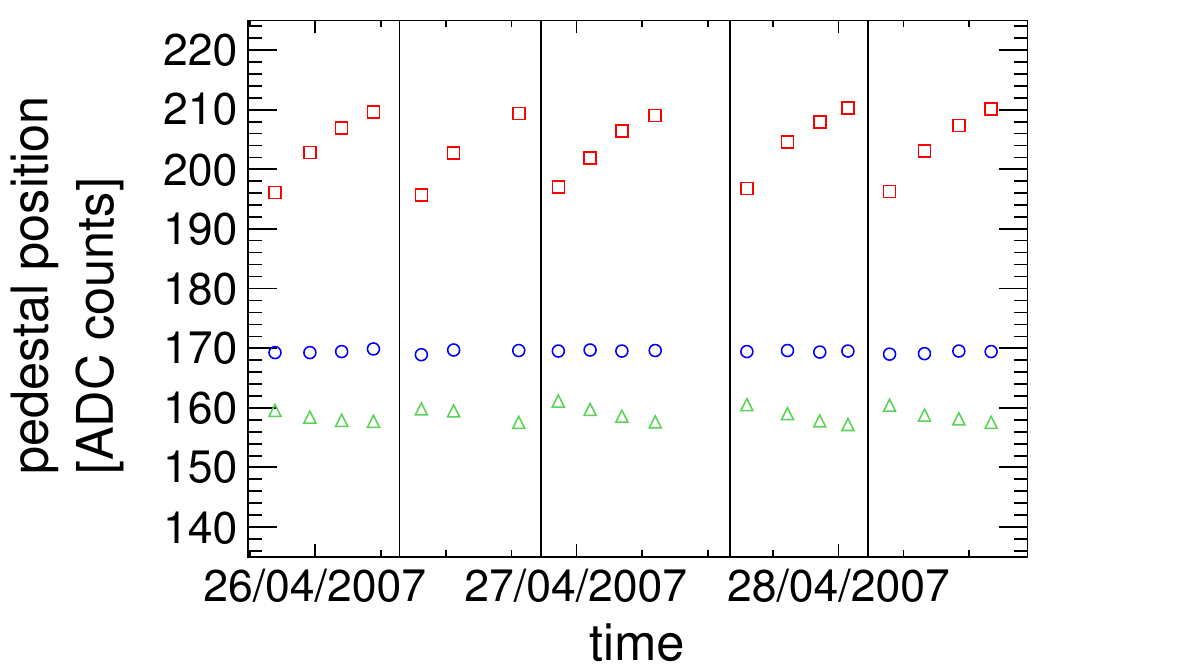}
    \caption{Left: Measured pedestal position for three selected channels
     		 versus time for about two months of collected pedestal data. Right: A
		     zoom into the period indicated by the dashed lines in the panel on the left
		     showing the pedestal data for five consecutive {\sc Hera} fills.}
    \label{fig:SSD_PedStabi}
  \end{center}
\end{figure}

\begin{figure}[t]
  \begin{center}
    \includegraphics[width=0.495\textwidth]{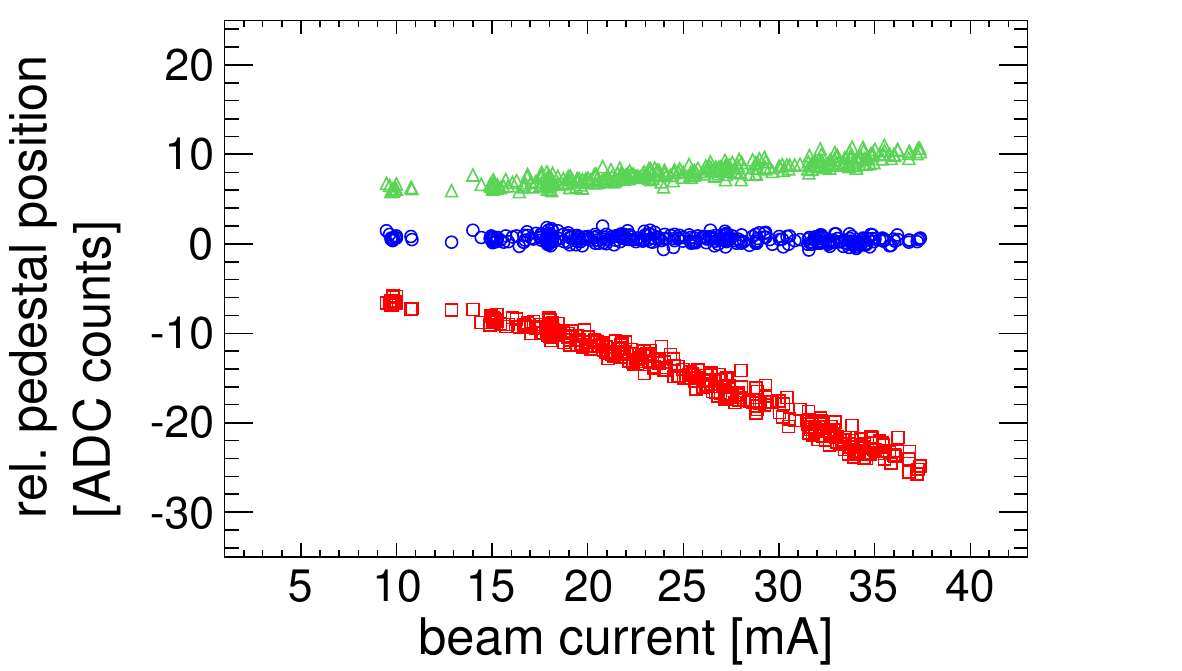}
    \caption{The same pedestal data as in the previous figure displayed versus the beam
             current at which the data were collected. The pedestal data collected without
             beam is used as an absolute reference. The graphs are
             shifted by $5$ ADC counts with respect to each other for better visibility.}
    \label{fig:SSD_PedVsBeam}
  \end{center}
\end{figure}

During data taking, every $2$ -- $3$ hours dedicated pedestal data were collected.
The panels of figure~\ref{fig:SSD_PedStabi} show the mean value of the pedestal
for three selected channels for about two months worth of collected
pedestal data. The channel indicated by the triangles (boxes) shows medium (large)
fluctuations, whereas the channel indicated by the circles only shows a slow drift
over time. In figure~\ref{fig:SSD_PedVsBeam} the same pedestal data are displayed
versus the beam current at which the pedestal run was taken, using the position of
the pedestal obtained from the most recent pedestal run without beam as an absolute
reference. For the channels indicated by triangles and boxes a clear dependence of
the pedestal position on the beam current is visible. A behavior that resulted in a
drift of more than one ADC channel during a typical {\sc Hera} fill with a beam current
of $\ge 25\unit{mA}$ was shown by $25\%$ of the low gain and $79\%$ of the high gain
channels, randomly distributed throughout the SSD.

For both the high and the low gain channel of each strip a quadratic parameterisation of the
beam-current dependence was used to correct for the drift of the pedestals between two
pedestal runs. The parameterisations gave an offset based on the change of the beam current
between the last pedestal run and the analysed event. This offset was added to the raw
data to correct for the drift.

\paragraph{Correlated Noise}
The HADC used to digitise the data delivered by the {\sc Helix} chips performed an online
common-mode correction, i.e., the correction for a common shift of the signal base-line,
based on the data from the first $16$ non-hit channels out of the first $32$ channels of
each chip. For pure common-mode noise this implementation is sufficient. However, the data
contain additional correlated noise that cannot be corrected for by this common-mode
correction algorithm. 

\begin{figure}[b]
  \begin{center}
    \includegraphics[width=0.495\textwidth]{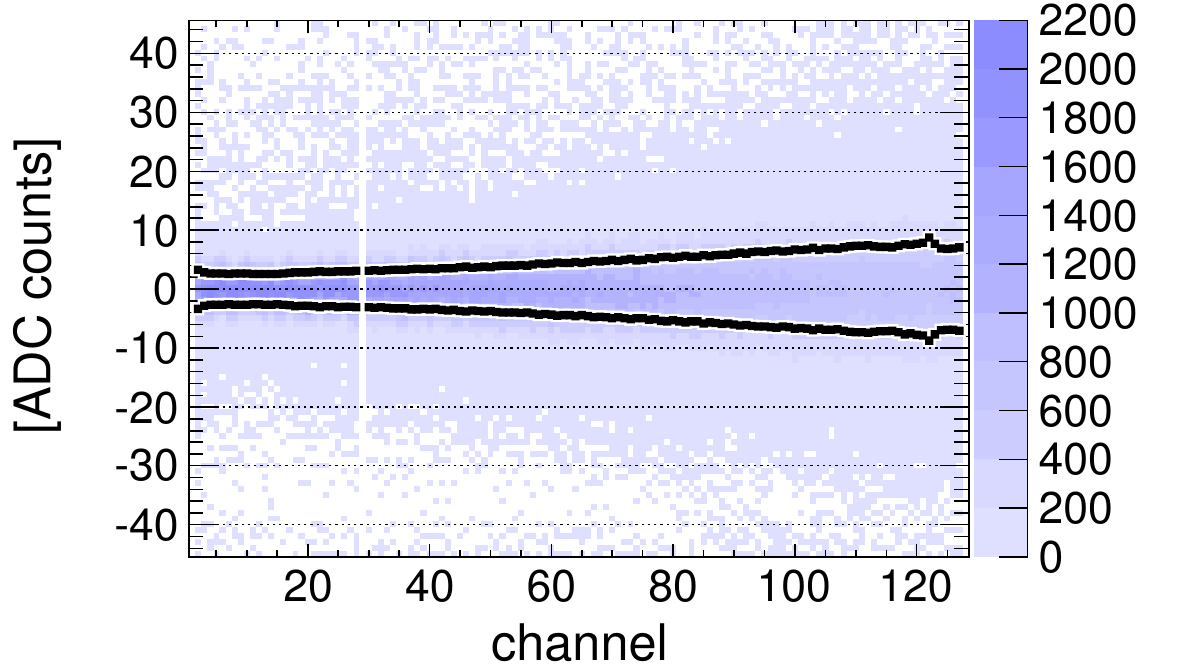}
    \includegraphics[width=0.495\textwidth]{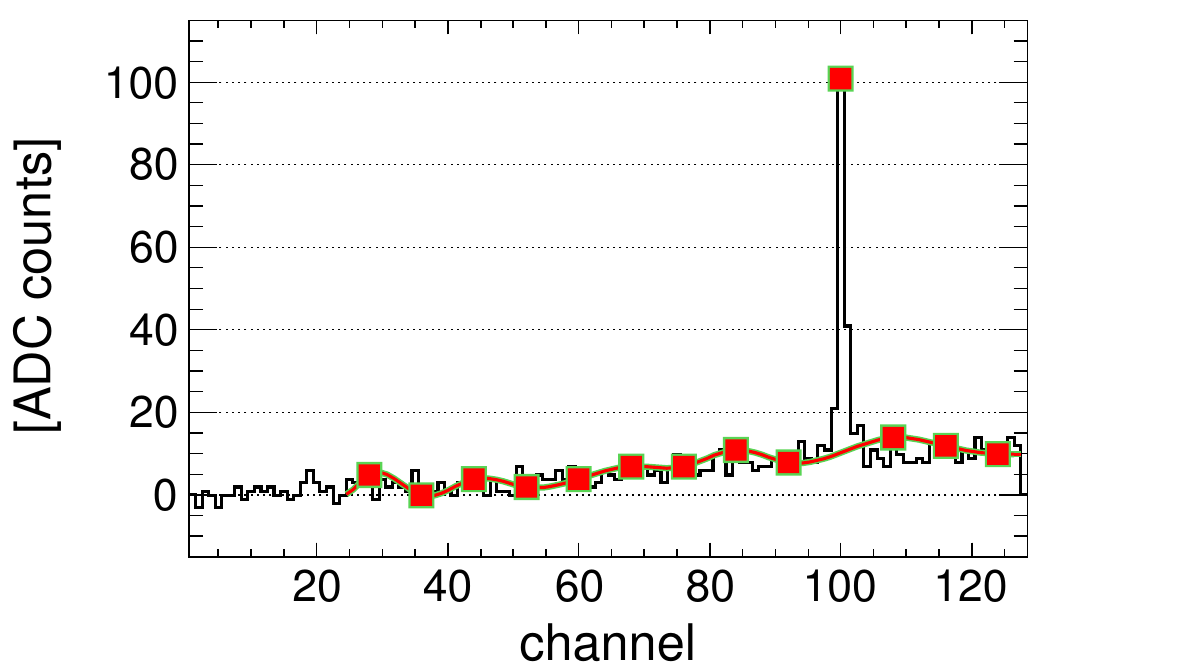}
    \caption{Left: Pedestal ADC spectrum for all channels of a chip versus channel number
             after pedestal subtraction and common-mode correction in the ADC. The lines in
             the figure indicate the $\pm 1 \sigma$ band of the pedestal width. The blank
             vertical band close to channel $30$ is caused by a non-working channel. Right:
             Data after pedestal subtraction and common-mode correction for a single event.
             The markers indicate the channels used for the correlated noise correction with
             a spline interpolation shown as dashed line. The data was taken without
             zero-suppression.}
    \label{fig:SSD_SplineNoCorr}
  \end{center}
\end{figure}

The left panel of figure~\ref{fig:SSD_SplineNoCorr} shows the ADC value for the 128 channels
of a chip after pedestal subtraction and common-mode correction performed by the ADC. The
common-mode correction performed well in the beginning of a chip (up to approximately channel 20).
Beyond this region, the width of the pedestal increased with increasing channel number. In
the right panel of figure~\ref{fig:SSD_SplineNoCorr} the ADC values of all channels of a chip
versus channel number are shown for a single event. At around channel 100 one can
observe an actual charge signal with an amplitude of $100$ ADC counts. However,
a substantial offset is present, as the baseline for this particular event drifts
towards positive ADC values for increasing channel numbers.

\begin{figure}[t]
  \begin{center}
    \includegraphics[width=0.495\textwidth]{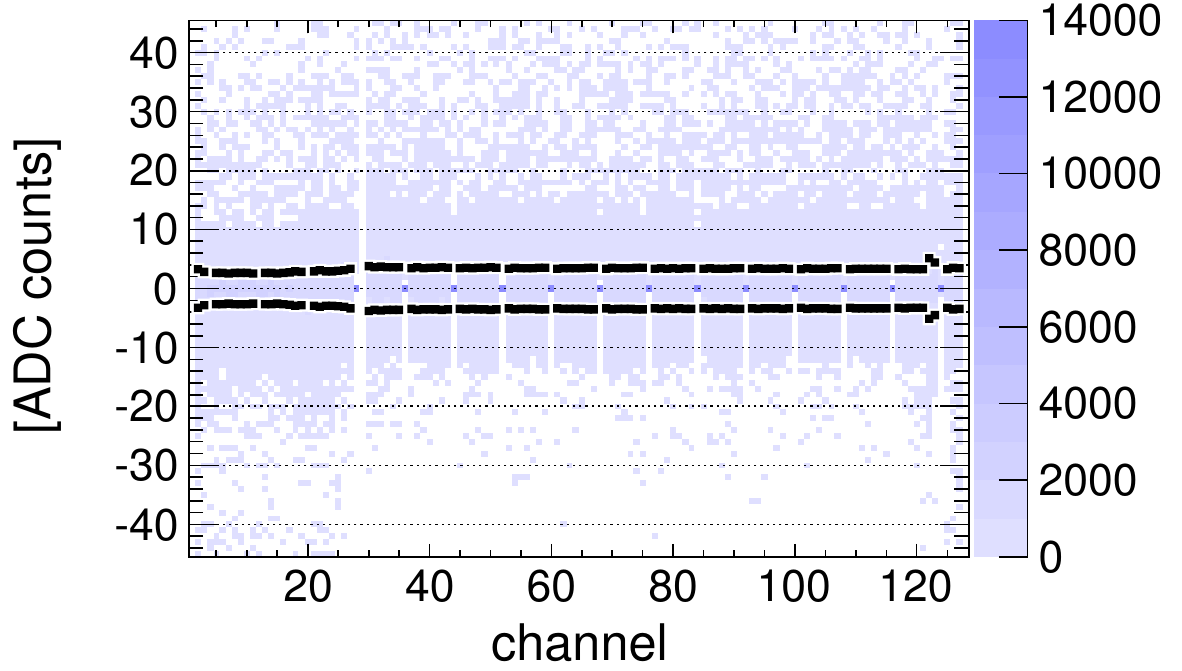}
    \includegraphics[width=0.495\textwidth]{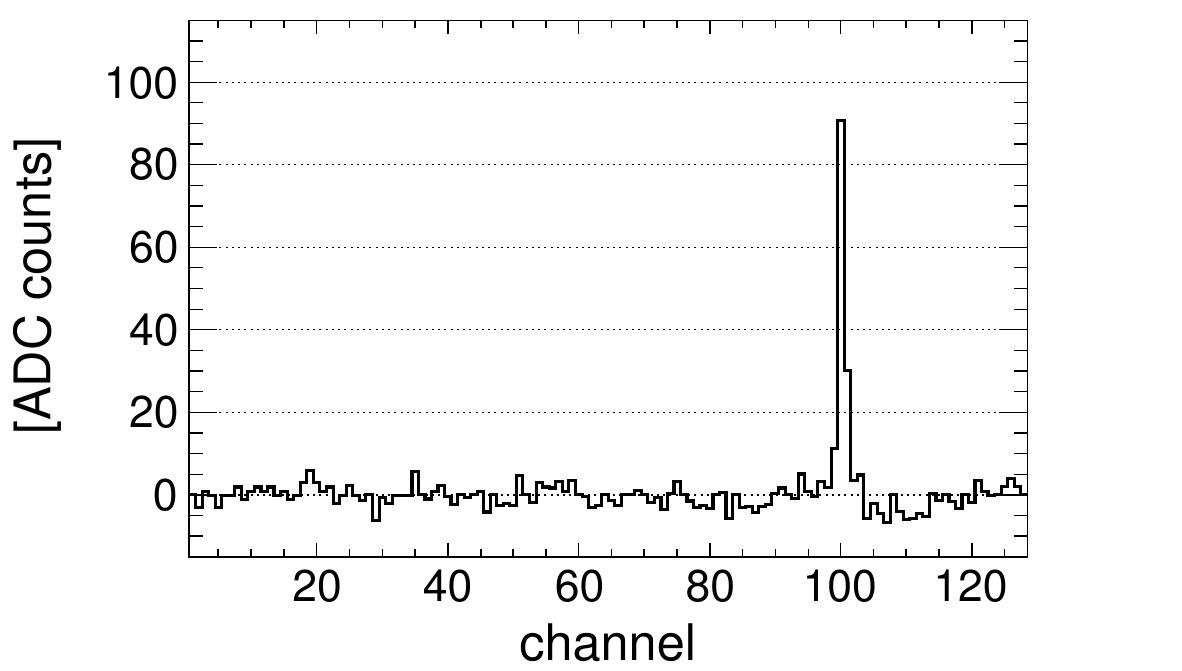}
    \caption{Left: Pedestal ADC spectrum for all channels of a chip~versus~channel number
             after the correction for the correlated noise. The lines in the figure indicate
             the $\pm 1 \sigma$ band of the pedestal width. The blank vertical band close
             to channel $30$ is caused by a non-working channel. Right: Data for the
             sample event from the previous figure after correction for the
             correlated noise.}
    \label{fig:SSD_SplineCorr}
  \end{center}
\end{figure}

In order to correct for the drift, the data from every eighth channel (starting from channel $28$)
of each chip were always read out and used as base points for a spline interpolation. In the right
panel of figure~\ref{fig:SSD_SplineNoCorr}, the base points are indicated by square markers. The
actual spline is shown as a line through the base points. The result from the interpolation
was used to correct for the drift. In the case that the signal of a spline-interpolation base point
after common mode correction is above a certain threshold and hence part of an actual cluster,
the average signal of the two neighbouring base points is used in the interpolation.

In the right panel of figure~\ref{fig:SSD_SplineCorr}, the same event is shown after the
spline-interpolation correction. The baseline over the full chip is flat and the amplitude of
the hit at channel 100 is reduced to approximately 90 ADC counts. For a larger data sample, the
spline-interpolation corrected ADC values versus channel number are shown in the left panel
of figure~\ref{fig:SSD_SplineCorr}. Again, the lines indicate the $\pm 1 \sigma$ band of the pedestal width.
After the correction, the pedestal width is below $3$ to $4$ ADC counts over the full sensor
side. A detailed description of the method including results from systematic studies can be found
in reference~\cite{Vilardi:PhD}.

\begin{figure}[t]
  \begin{center}
    \includegraphics[width=0.495\textwidth]{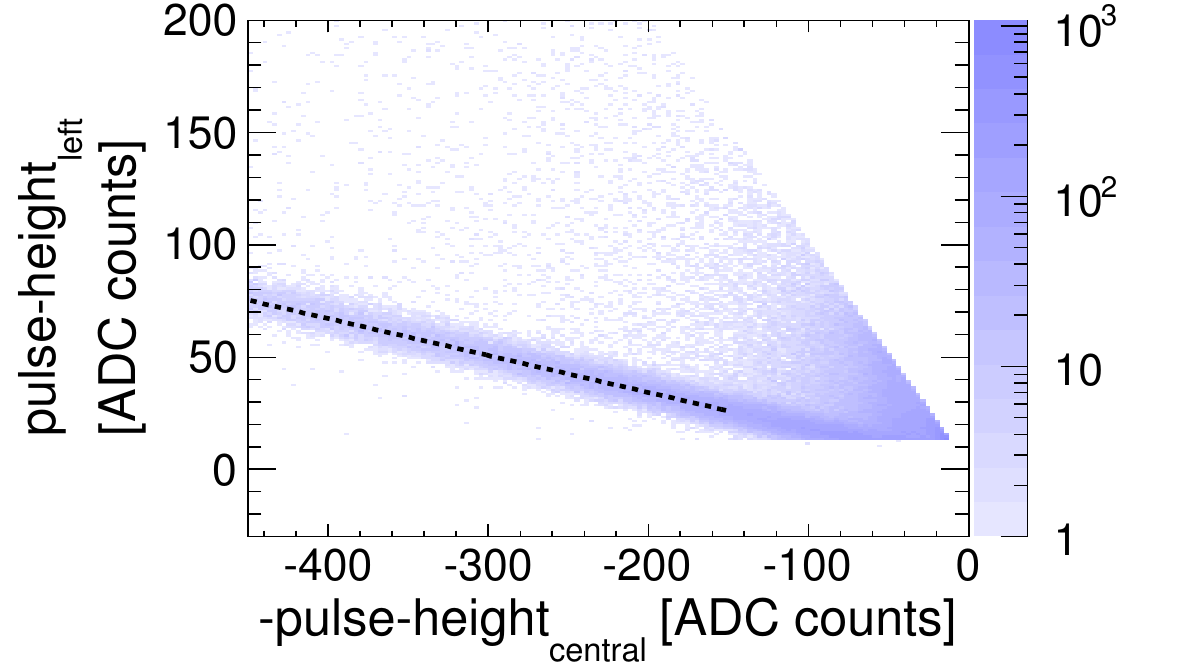}
    \includegraphics[width=0.495\textwidth]{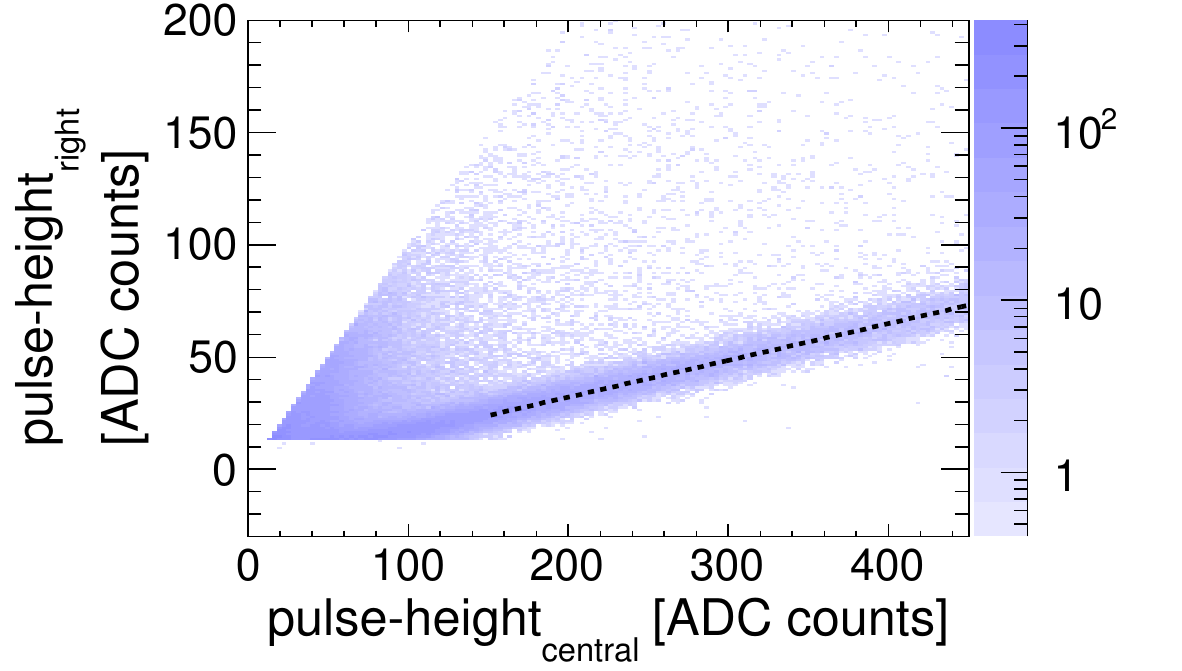}
    \caption{ADC value of the channel to the left/right of the channel with the highest
    	         ADC value in a cluster (central channel) versus the ADC value of the central
	         channel before cross-talk correction.}
    \label{fig:SSD_XtNoCorr}
  \end{center}
\end{figure}

\paragraph{Cross-talk}
For each actual signal in a channel a significant fraction of that signal migrates into
the neighbouring channels. This cross-talk is due to coupling between strips on the sensor,
the traces on the flex foils and the chip itself. Figure~\ref{fig:SSD_XtNoCorr} shows the
ADC value of neighbouring channels as a function of the ADC value of the central channel
of a cluster with central channel denoting the channel with the highest ADC value. Both to
the left and to the right a band indicated by dashed lines is seen. The slopes of the bands
correspond to the constant fraction of signal that migrates to the neighbours. The
cross-talk to the left neighbour was between $11\unit{\%}$ and $16\unit{\%}$, whereas to the
right it was between $15\unit{\%}$ and $21\unit{\%}$. In addition, the cross-talk is different
for even and odd numbered strips which can be explained by the internal design of the
{\sc Helix} chip.

\begin{figure}[t]
  \begin{center}
    \includegraphics[width=0.495\textwidth]{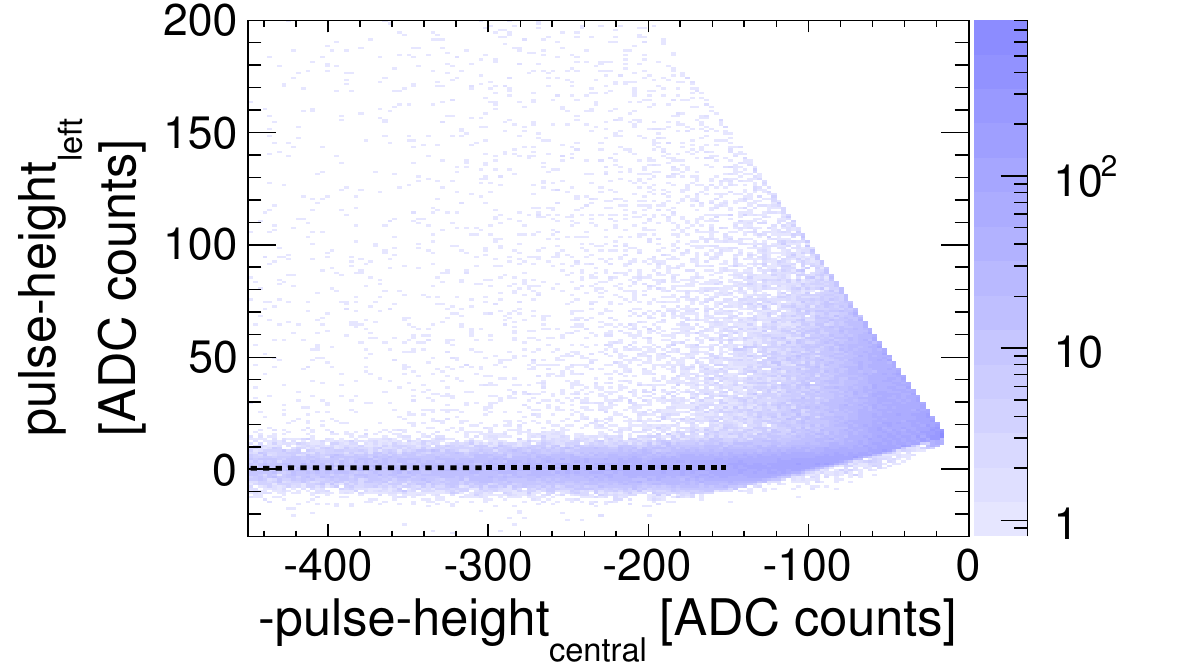}
    \includegraphics[width=0.495\textwidth]{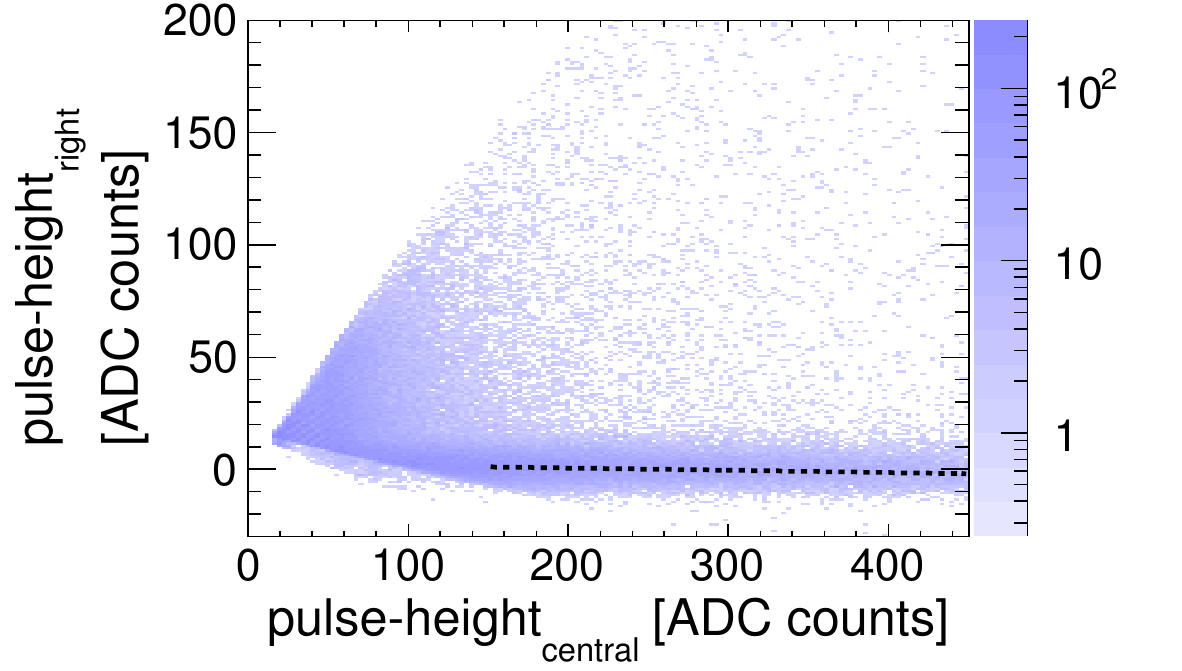}
    \caption{ADC value of the channel to the left/right of the channel with the highest
    	         ADC value in a cluster (central channel) versus the ADC value of the central
	         channel after cross-talk correction.}
    \label{fig:SSD_XtCorr}
  \end{center}
\end{figure}

The left-right asymmetry in the cross-talk would introduce a bias in the position
measurement in the silicon sensor and therefore needs to be corrected for.
In addition, the cross-talk correction is capable of recovering signals below
the hardware threshold. The algorithm
assumes that the $128$ uncorrected ADC values originate from a $128 \times 128$ matrix
multiplied by the $128$ actual ADC values. The matrix elements represent the fractions
of signal migrating into other channels, i.e., the fraction of signal remaining in a
channel is represented by the diagonal elements and the cross-talk to neighbouring channels
by the off-diagonal elements. The cross-talk to the second neighbour is of the order of
$2\unit{\%}$. Cross-talk effects to channels farther away from the central channel
are insignificant and therefore neglected in the correction. By matrix inversion\footnote{
The correction algorithm assumes that cross-talk affects only the immediate and second 
neighbours to the left and right of the central channel. It is therefore sufficient to invert
a $5 \times 5$ submatrix to perform the correction.}
it is possible to reconstruct the actual ADC values. Figure~\ref{fig:SSD_XtCorr} shows the
ADC values of the left and the right neighbour versus the ADC value in the central channel
after the cross-talk correction. After the correction the bands are horizontal and have a mean
around zero. A detailed description of the SSD cross-talk correction can be found in
reference~\cite{Zeiler:PhD}.

\subsubsection*{High-Low Gain Relation}

\begin{figure}[t]
  \begin{center}
    \includegraphics[width=0.495\textwidth]{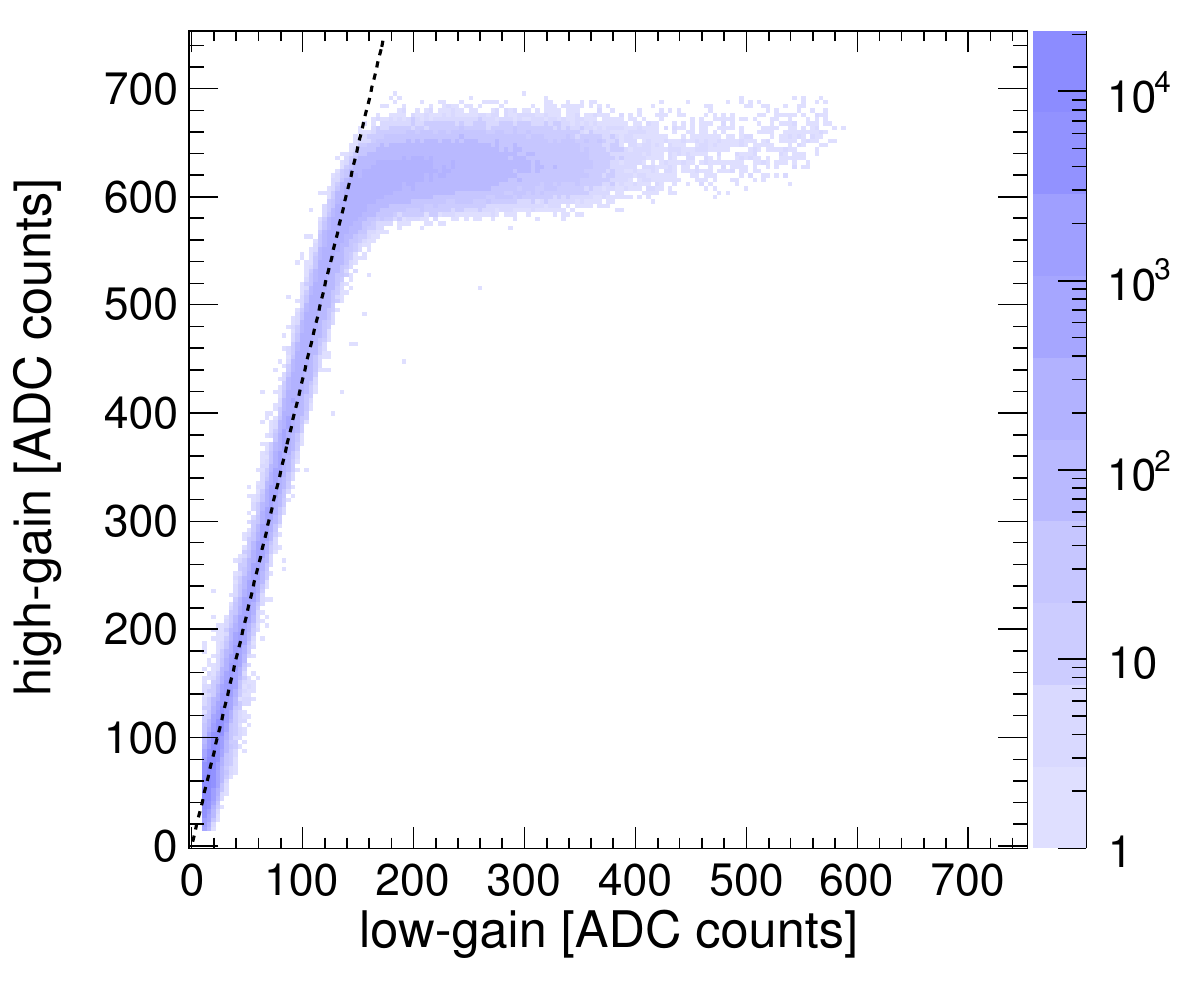}
    \caption{ADC values of the high-gain versus those of the low-gain channel for all 128 strips
             of a sensor side.}
    \label{fig:SSD_HiLo}
  \end{center}
\end{figure}

For each strip the cross-talk corrected ADC values for both high-gain and low-gain channels are
combined to an effective low-gain ADC value. In the left panel of figure~\ref{fig:SSD_HiLo}, the
ADC values of the high-gain channel are plotted versus the ADC values of the low-gain channel,
for all 128 strips of one sensor side. The figure shows a linear behavior with a slope of
$\approx4.3$ up to $\approx500$ ADC counts for all 128 high-gain channels after which they
reach their individual saturation levels. For each individual strip the relation between
both channels is extracted by a fit of a second-order polynomial to the data. A quadratic
polynomial with the offset fixed at zero is chosen to account for non-linearities of individual
strips not visible in the integral distribution shown in the left panel of figure~\ref{fig:SSD_HiLo}.
For ADC values below the saturation level in the high-gain channel, the effective ADC value
is calculated from the high-gain value, whereas for ADC values above the saturation level in
the high-gain channel it is simply the low-gain ADC value.

\subsubsection*{Clustering}

A cluster is formed by combining adjacent strips that have a real signal. A new cluster is
created in the case that the signal of a strip is above a threshold corresponding to $60\unit{keV}$
energy deposit. Neighboring strips are added to the cluster if their signal is above
a secondary threshold corresponding to $50\unit{keV}$. The amplitude (position) of a cluster
is calculated as the sum (barycenter) of energy deposit values of all strips in the cluster.

\subsubsection*{Absolute Energy Calibration}

\begin{figure}[t]
  \begin{center}
    \includegraphics[width=0.495\textwidth]{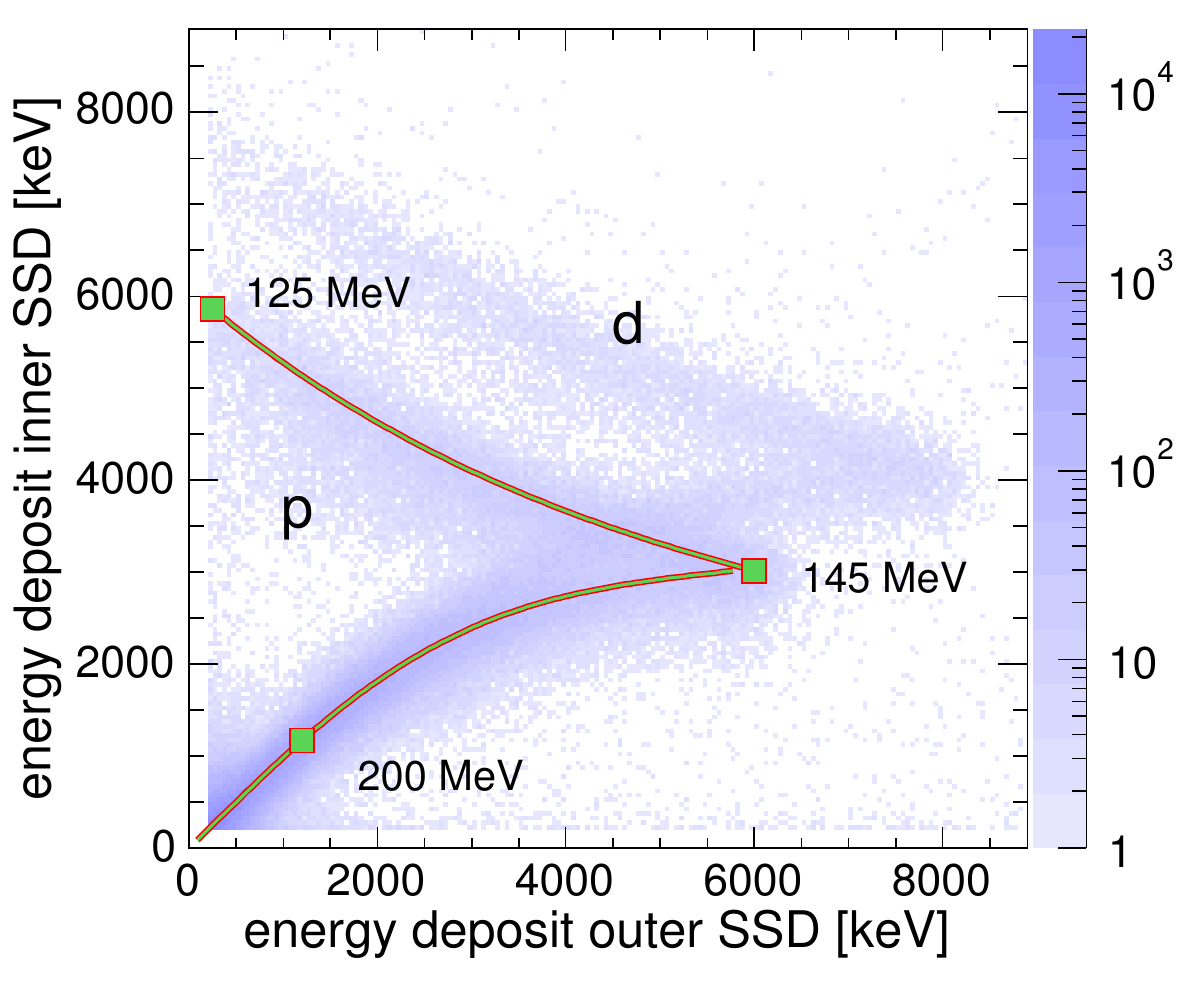}
    \includegraphics[width=0.495\textwidth]{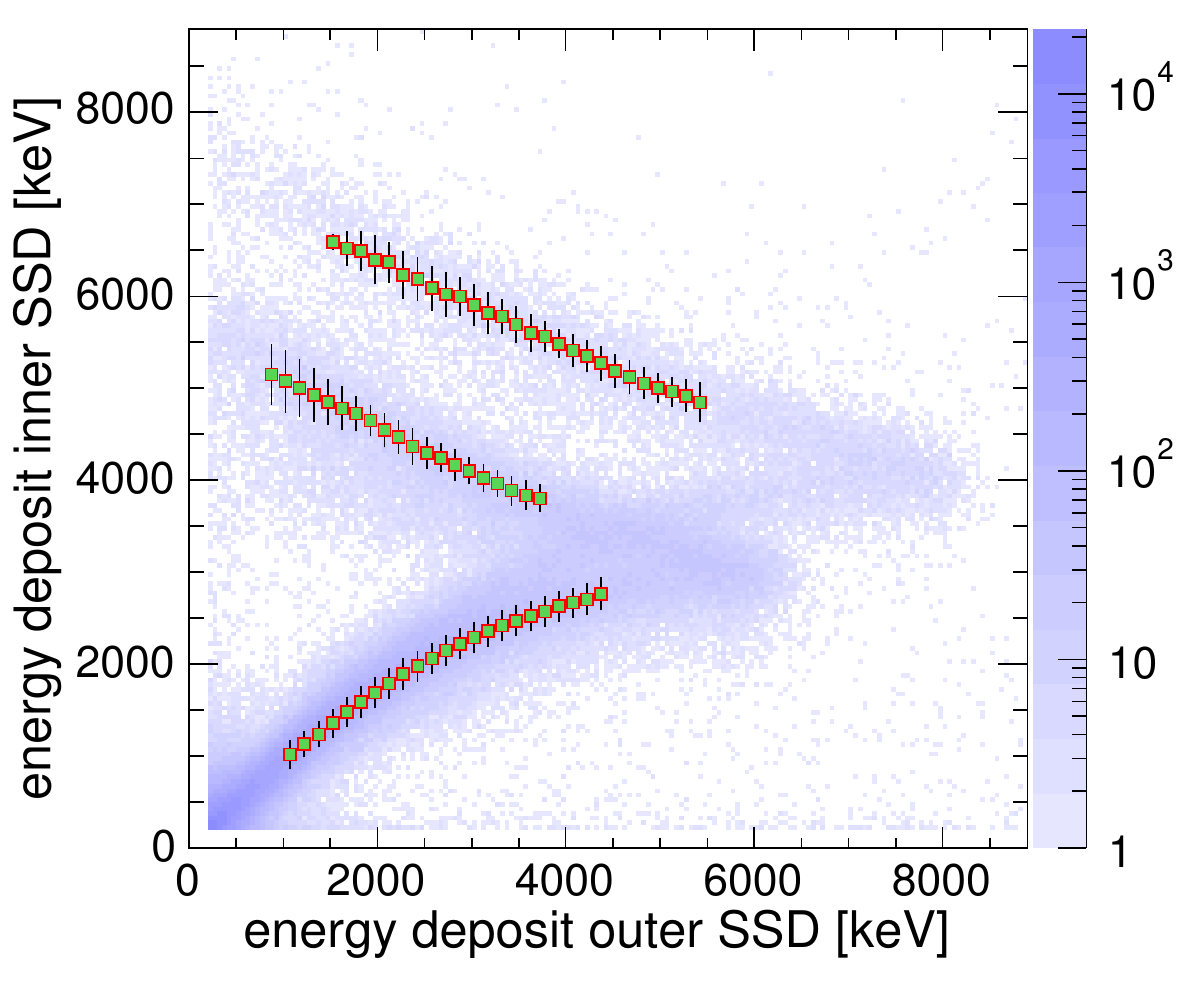}
    \caption{\mhl{Left: Energy deposit in an inner SSD sensor versus the
             energy deposit in the corresponding outer SSD sensor obtained from data taken
             with the deuterium target using preliminary calibration constants. The solid
             curve represents the most probable energy deposit and the markers the
             corresponding momentum of protons. At a momentum of $125\unit{MeV}$ protons
             have sufficiently high momentum to reach the second
             layer of the SSD where they are stopped up to a momentum of $145\unit{MeV}$. The
             second band originates from deuterons being stopped in the second SSD layer.
             Right: The same data with markers indicating
             the most probable energy deposits of protons and deuterons stopped in the outer
             layer and protons that punch through the outer layer as obtained from fits to the
             distribution.}}
    \label{fig:SSD_dEE}
  \end{center}
\end{figure}

For protons stopped in the outer SSD layer as well as for protons with too low momentum
to reach the SFT layers, the momentum is reconstructed using just the energy deposits in
both SSD layers. In this momentum region the quality of the momentum reconstruction
is therefore directly related to the precision of the energy calibration. In the first
iteration, the energy calibration is obtained via the energy deposits for protons and deuterons
that are stopped in the outer SSD layer and for protons that punch through this layer. The left
panel of figure~\ref{fig:SSD_dEE} shows the energy deposit in the inner SSD layer versus
the energy deposit in the corresponding outer SSD layer after a correction for the incident
angle. The curves in the panels show the results of a Geant4 simulation~\cite{GEANT4:2003,GEANT4:2006}
with protons that penetrate the sensors under an angle of $90^{\circ}$, i. e., they represent
the theoretical energy deposit. At a momentum of about $125\unit{MeV}$, a proton has
sufficient energy to punch through the inner SSD layer and reach the outer layer where it
is stopped. For momenta above approximately $145\unit{MeV}$ protons also punch through
the outer layer.


\begin{figure}[t]
  \begin{center}
    \includegraphics[width=0.92\textwidth]{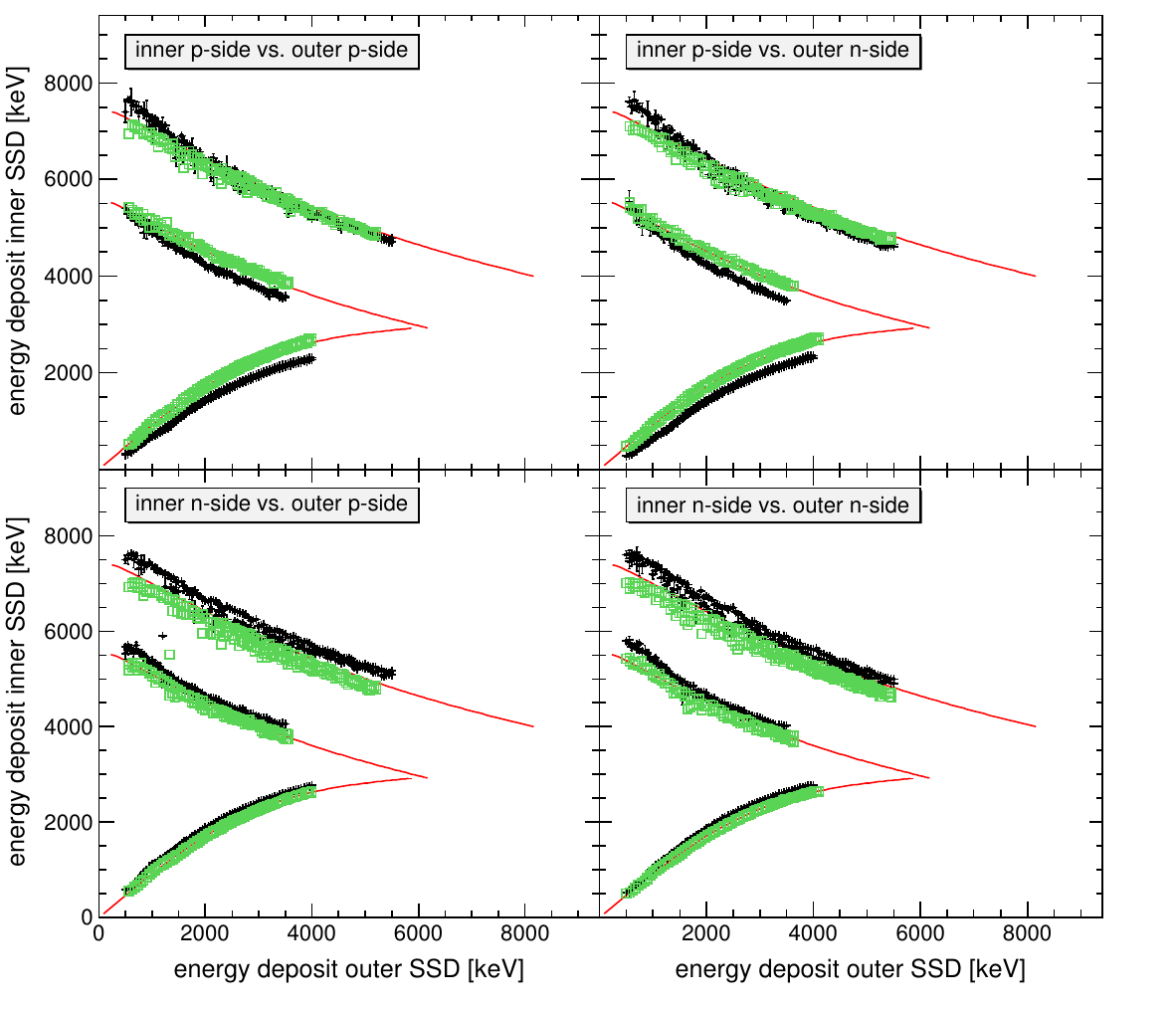}
    \includegraphics[width=0.92\textwidth]{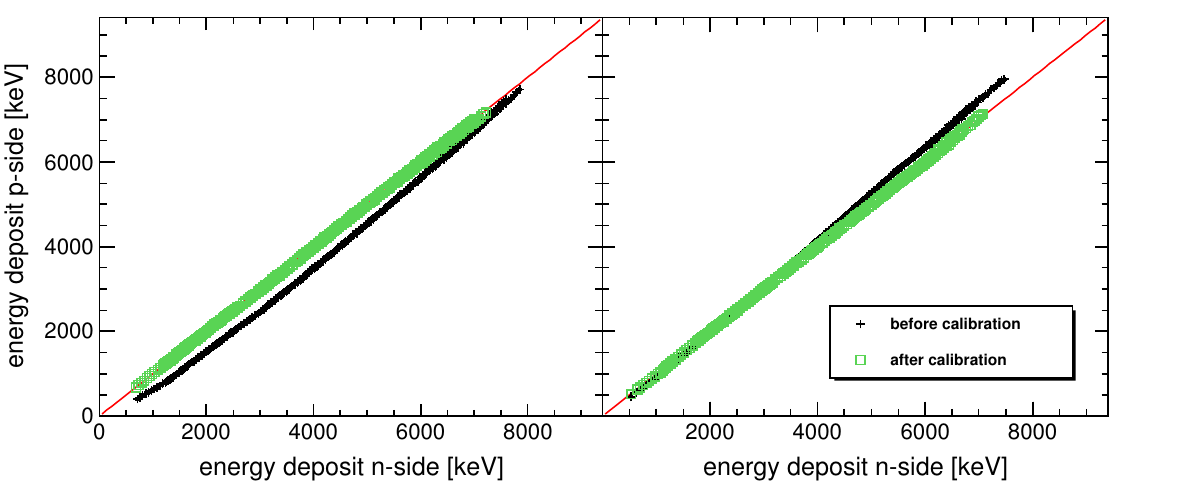}
    \caption{Top: Energy deposit in an inner sensor versus the energy deposit in the
             corresponding outer sensor for all combinations of p- and n-side energy
             deposit measurements. Bottom: Energy deposit measured by the p-side versus
             energy deposit measured by the n-side for an inner and the corresponding
             outer SSD sensor. In all six panels, the curves represent the theoretical
             values taking the individual sensor thicknesses into account. The crosses
             show the data before calibration whereas the open boxes depict the same data
             after calibration.}
    \label{fig:SSD_CalFit}
  \end{center}
\end{figure}

The comparison of the distribution shown in the left panel of figure~\ref{fig:SSD_dEE}
with the theoretical values is used to obtain the absolute energy calibration for all
sensor sides. This is shown in the right panel of figure~\ref{fig:SSD_dEE} in which the
most probable values of a Landau distribution convoluted with a Gaussian, obtained from
fits to the distribution of energy deposits in the inner SSD layer versus the energy
deposit in the outer SSD layer, are presented. The fit results for each
combination of an energy deposit measured on one side of an inner sensor with an energy deposit
measured on one side of an outer SSD sensor are then used in a
combined fit to the theoretical curves. \mhl{The overlap between the proton and deuteron bands
for particles passing through both SSD layers could lead to a bias in the absolute
energy calibration, which is taken into account by assigning lower weights to these data points in the
combined fitting procedure. The procedure} is shown in figure~\ref{fig:SSD_CalFit} for
one specific combination of inner and outer sensors. The deviation between the measured data
and the theoretical energy deposition clearly indicates a non-linear detector response. In order
to compensate for this non-linearity the energy deposits of clusters in the inner and outer
layers are expressed by

\begin{equation}
  \Delta E_{\mathrm{inner}} = f\left(ADC_{\mathrm{inner}}^{\mathrm{cluster}}\right) 
  ~~~ ~~~  \mathrm{and} ~~~  ~~~ 
  \Delta E_{\mathrm{outer}} = f\left(ADC_{\mathrm{outer}}^{\mathrm{cluster}}\right),
\end{equation}

\noindent
respectively, where $ADC^{\mathrm{cluster}}$ are the sums of ADC values in a cluster, $\Delta E$
are the calibrated energy deposits and $f$ is a function of four parameters describing two
connected polynomials. The function is constructed such that there is a smooth transition at
the connection point and that the behavior is quadratic (linear) for $ADC^{\mathrm{cluster}}$
values below (above) the connection point. In this way, non-linearities in the region of low
$ADC^{\mathrm{cluster}}$ values can be corrected, while there is still stable behavior for
large values of $ADC^{\mathrm{cluster}}$.

\enlargethispage*{0.5cm}

The final $16$ calibration coefficients for the combination of both p- and n-sides of one
inner and one outer sensor are obtained by a combined minimization of the differences between
the measured band positions (crosses) and the theoretical curves shown in all panels
of figure~\ref{fig:SSD_CalFit}. The results of the calibration procedure applied to the
uncalibrated data points is shown as open boxes in the panels demonstrating a good agreement
of the calibrated data points with the expected curves. 

\subsection{Scintillating-Fibre Tracker}

\begin{figure}[t]
  \begin{center}
    \includegraphics[width=0.495\textwidth]{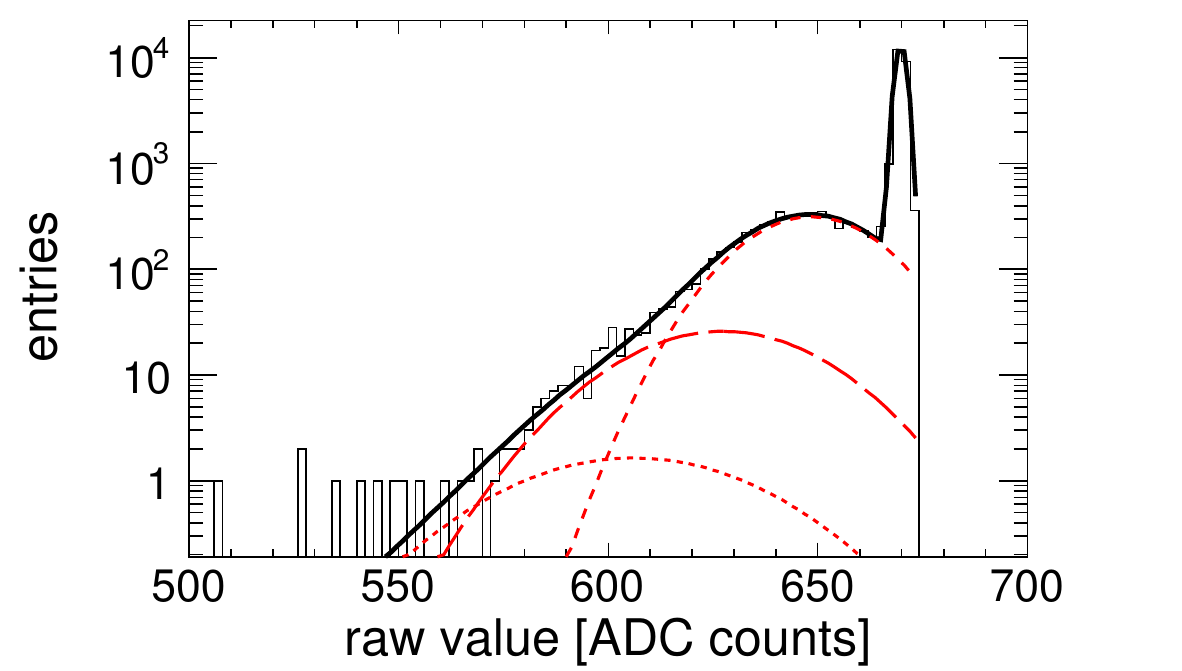}
    \includegraphics[width=0.495\textwidth]{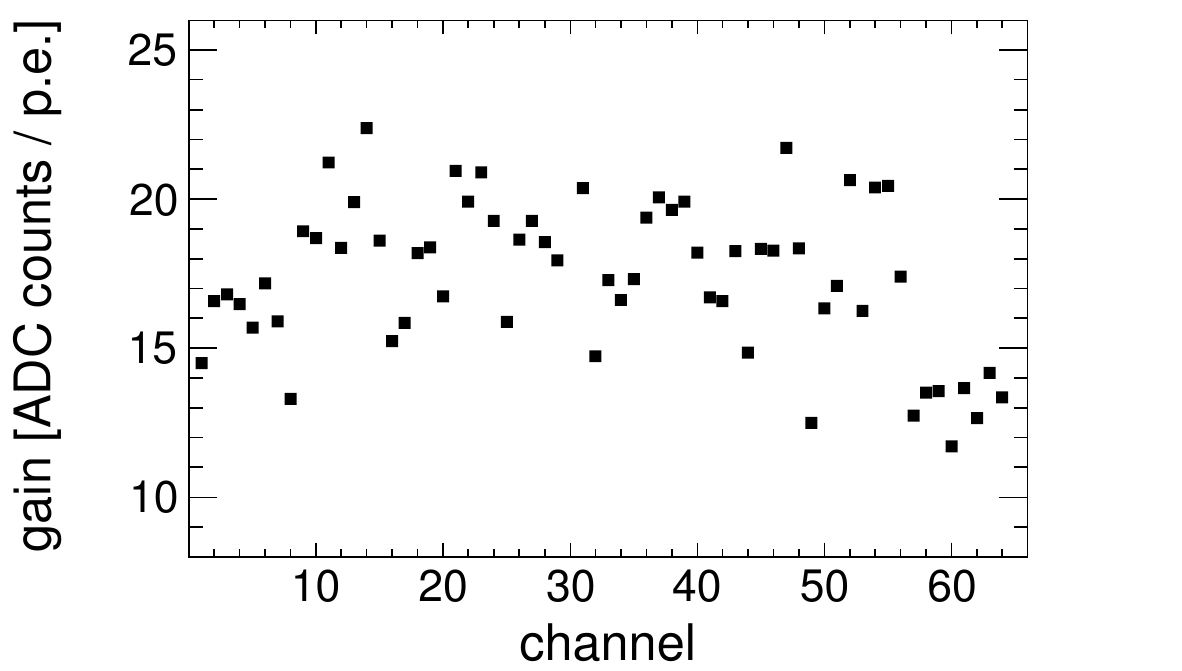}
    \caption{Left: Raw ADC amplitude spectrum for a single MAPMT pixel obtained by flashing
             the surface of the MAPMT with blue light. The position of the pedestal peak
             (at the right of the spectrum) is obtained from a fit to the distribution.
             \mhl{The spectrum is inverted due to the readout logic of the used PFM.}
             Right: Gain expressed as ADC counts per photoelectron for all $64$ channels
             of one MAPMT.}
    \label{fig:SFT_PMTgain}
  \end{center}
\end{figure}

Although the main purpose of the SFT was to provide position information for the reconstruction
of charged tracks, the measured energy deposits are also valuable quantities that are used
in the particle-identification procedure. In contrast to the SSD, this requires only a
relative energy calibration. For the SFT, the conversion
from ADC values to energy deposits is performed in two stages. In the first stage, the raw
ADC values are converted to the number of photoelectrons seen by a MAPMT pixel. The second
stage is the conversion from the number of photoelectrons to an actual energy deposit.
As mentioned before, cross-talk is assumed to originate from neighbouring fibres or
neighbouring pixels on the MAPMT. The correction is therefore performed in units of
photoelectrons.

\begin{figure}[b]
  \begin{center}
    \includegraphics[width=0.495\textwidth]{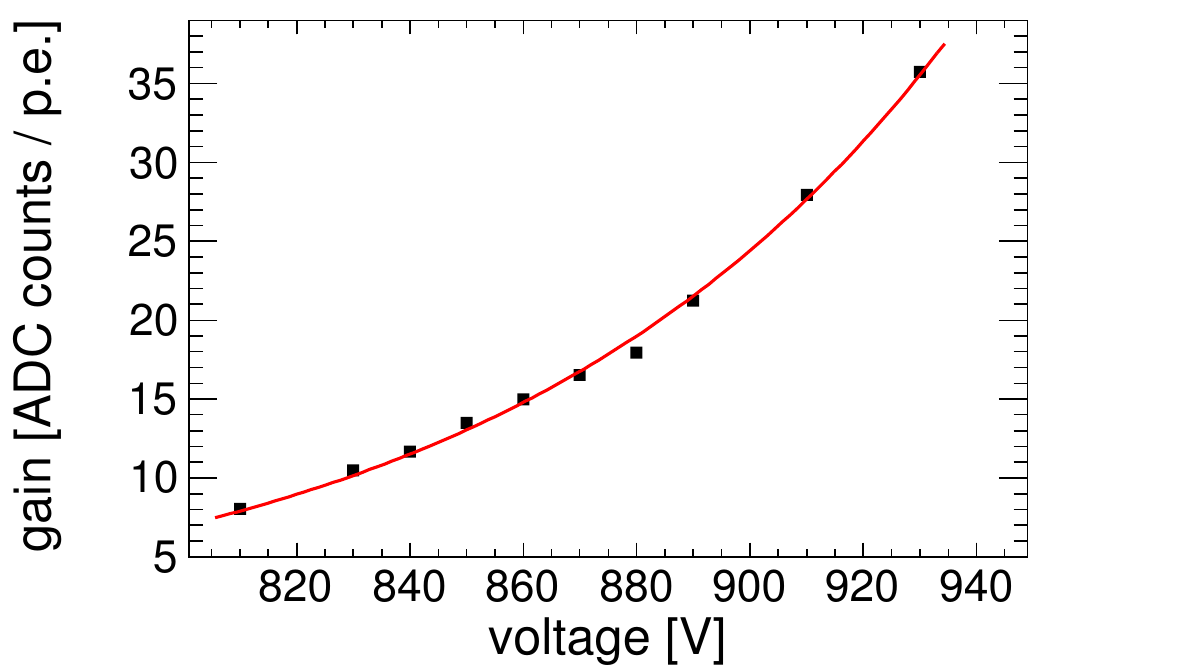}
    \caption{Dependence of the MAPMT gain on the applied supply voltage.}
    \label{fig:SFT_PMTgainVsU}
  \end{center}
\end{figure}

\subsubsection*{PMT Gain Calibration}

The gain of the Hamamatsu MAPMT is of the order of $10^5$ -- $10^6$ for supply voltages
between $800\unit{V}$ and $1000\unit{V}$. In order to calibrate the response of the MAPMT
pixels in terms of photoelectrons, each pixel was illuminated with blue light originating from a
LED~\cite{Yu:PhD} and the data were acquired with the same read-out system as used in the
final SFT installation. The obtained distributions were fitted with a function describing
both the pedestal peak and the signal contributions from $1$, $2$, and $3$ photoelectrons.
The left panel in figure~\ref{fig:SFT_PMTgain} shows the ADC spectrum for a single pixel
including the pedestal peak at high ADC values and the signal contributions. The MAPMT gain
is then the distance between the single photoelectron and the pedestal peak. The right
panel of figure~\ref{fig:SFT_PMTgain} shows the variation in gain $g$ for all 64 channels
of the same MAPMT.

The position of the single-photoelectron peak and hence the MAPMT gain varies exponentially
with the applied voltage according to

\begin{equation}
g = b + e^{K\cdot U},
\label{eq:SFT_PMTgain}
\end{equation}

\noindent
where $U$ is the applied high voltage and $b$ and $K$ are coefficients determined
from the LED data. It was found that the coefficients $K$ are the same for all channels
of one MAPMT, whereas the coefficients $b$ reflect the variation in gain shown in the
right panel of figure~\ref{fig:SFT_PMTgain}. Figure~\ref{fig:SFT_PMTgainVsU} shows the
exponential behavior of the measured gain as a function of the applied high voltage
together with a fit of the function described in equation~\ref{eq:SFT_PMTgain}.

\begin{figure}[b]
  \begin{center}
    \includegraphics[width=0.495\textwidth]{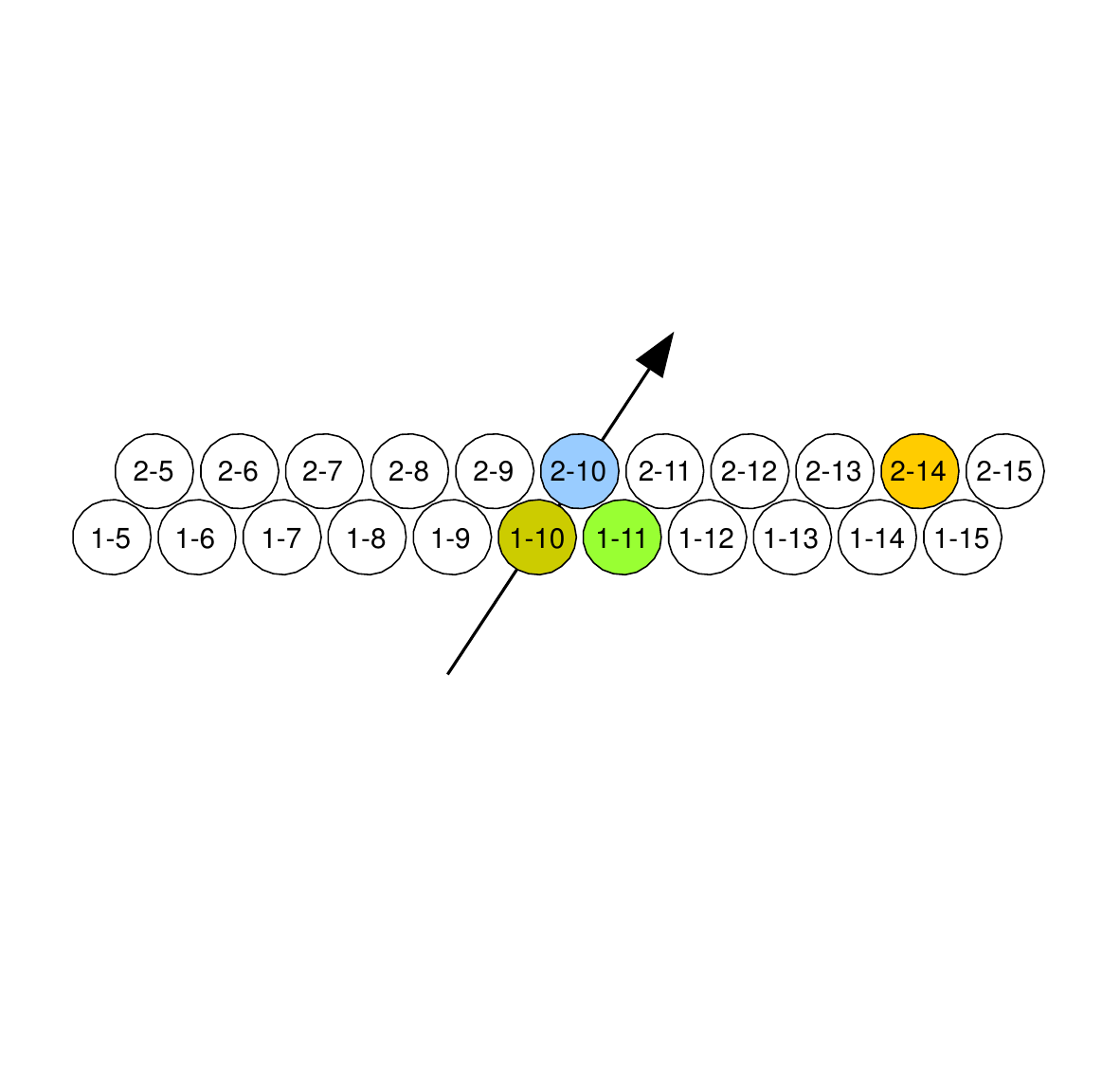}
    \includegraphics[width=0.495\textwidth]{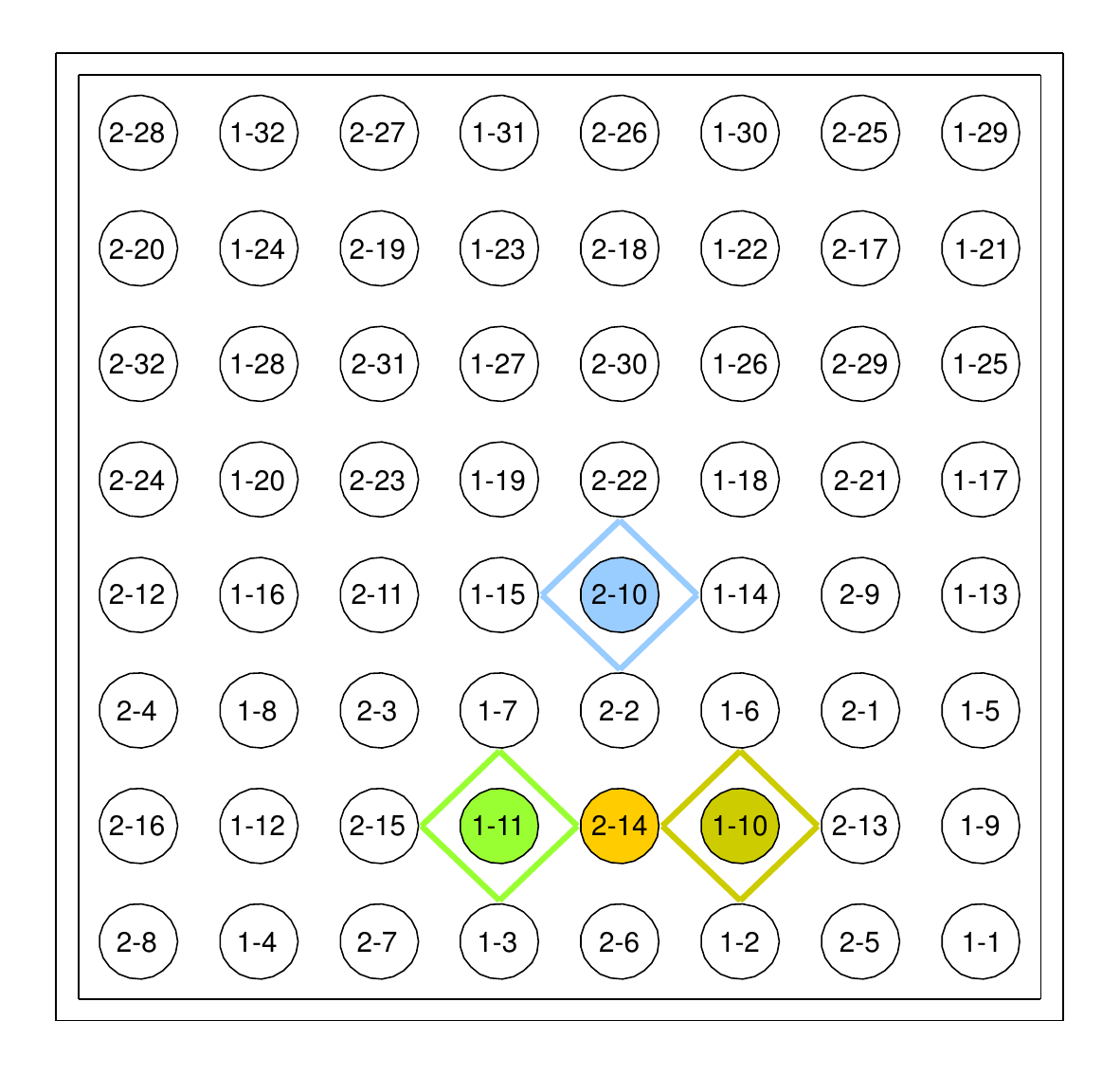}
    \caption{\mhl{The left panel shows the schematic view of eleven consecutive fibres}
             in a regular and the corresponding shifted layer of a given sub-barrel,
             indicated by the prefix $1-$ and $2-$, that are connected to the same MAPMT.
             The mapping on the MAPMT surface is shown in the right panel. A particle traverses
             fibre $10$ in both layers resulting in a direct signal with an additional
             signal in fibre $1-11$ due to cross-talk between neighbouring fibres. The
             signal seen in fibre $2-14$ originates from cross-talk on the MAPMT surface.}
    \label{fig:SFT_XTalk}
  \end{center}
\end{figure}

\subsubsection*{Cross-Talk Correction}
\label{sec:sftXTalkCorr}

It is necessary to discriminate cross-talk between neighbouring fibres in the sensitive
volume of the SFT and cross-talk on the surface of the MAPMTs. Cross-talk between
neighbouring fibres does neither influence the position nor the energy-deposit information
provided by the SFT. However, cross-talk on the MAPMT surface leads to wrong hit-position
information. In order to disentangle the two sources of cross-talk a
sophisticated mapping between the active fibres and the pixels on the MAPMT surface
was developed. In this scheme, neighbouring sensitive fibres are not connected to
neighbouring pixels on a MAPMT. The left panel of figure~\ref{fig:SFT_XTalk} shows \mhl{eleven}
consecutive fibres in a regular and the corresponding shifted layer of a given sub-barrel.
The mapping of the fibres on the MAPMT surface is shown in the right panel of the figure.
The pictures also show the signals originating from an example event, in which a track
passes through fibres 1-10 and 2-10, and in the corresponding pixels a signal is observed.
In addition, also fibre 1-11 has a signal which is caused by cross-talk between
neighbouring sensitive fibres. The signal in fibre 2-14 originates from cross-talk on the
PMT surface as can be seen from the right panel of figure~\ref{fig:SFT_XTalk}.

For all pixels with a signal above a threshold of $1.5$~photoelectrons, the correction
algorithm considers the signals in the direct neighbours as shown in the right panel of
figure~\ref{fig:SFT_XTalk}. In the example the signal in pixel 2-14 is below threshold,
so only pixels 1-10, 1-11 and 2-10 are corrected for cross-talk. The signal in
pixel 2-14 is then split and distributed according to the amplitudes of the signals
in pixels 1-10 and 1-11. After the correction pixel 2-14 no longer has a signal,
leaving pixel 2-10 with the original signal and pixels 1-10 and 1-11 with an increased
signal. For details see reference~\cite{Yu:PhD}.

\subsubsection*{Clustering}

Similar to the case of SSD, consecutive fibres from a SFT-layer containing signals after
the cross-talk correction are combined to clusters. The threshold for combining fibres
is imposed by the threshold used in the cross-talk correction algorithm and thus corresponds
to $1.5$~photoelectrons, as well. In the example, the clustering algorithm produces a cluster
containing fibres 1-10 and 1-11 for the lower layer and a cluster with fibre 2-10
for the upper layer. For clusters with more than one fibre, the amplitude-weighted
mean fibre number of the cluster is assigned as a virtual fibre number that is later
used to determine the precise position of the cluster. For the inner SFT barrel the
mean cluster size is $1.2\unitspace\unit{fibres}$ for all layers, while it is
$1.6\unitspace\unit{fibres}$ for the outer barrel layers.

\begin{figure}[h]
  \begin{center}
    \includegraphics[width=0.495\textwidth]{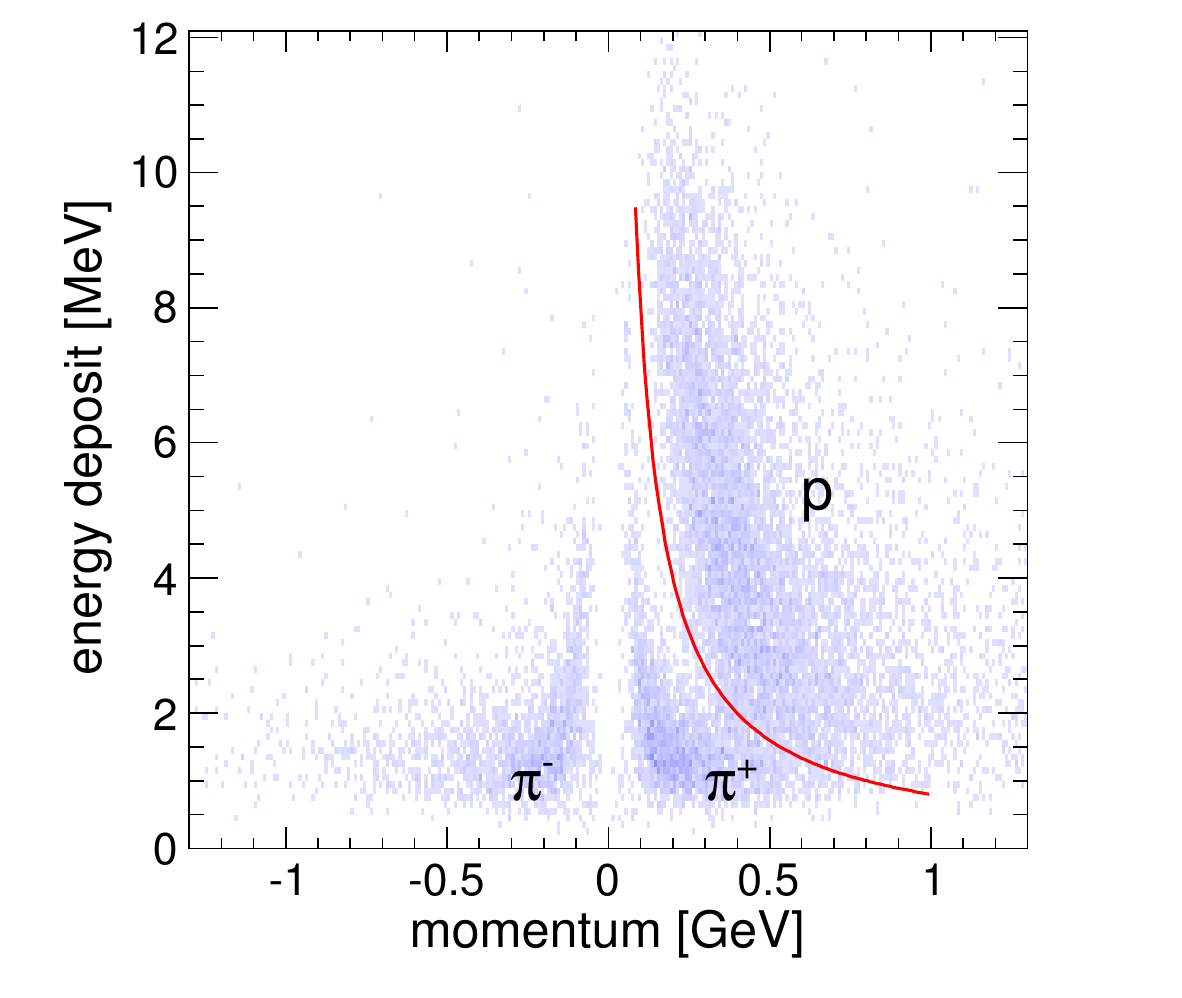}
    \includegraphics[width=0.495\textwidth]{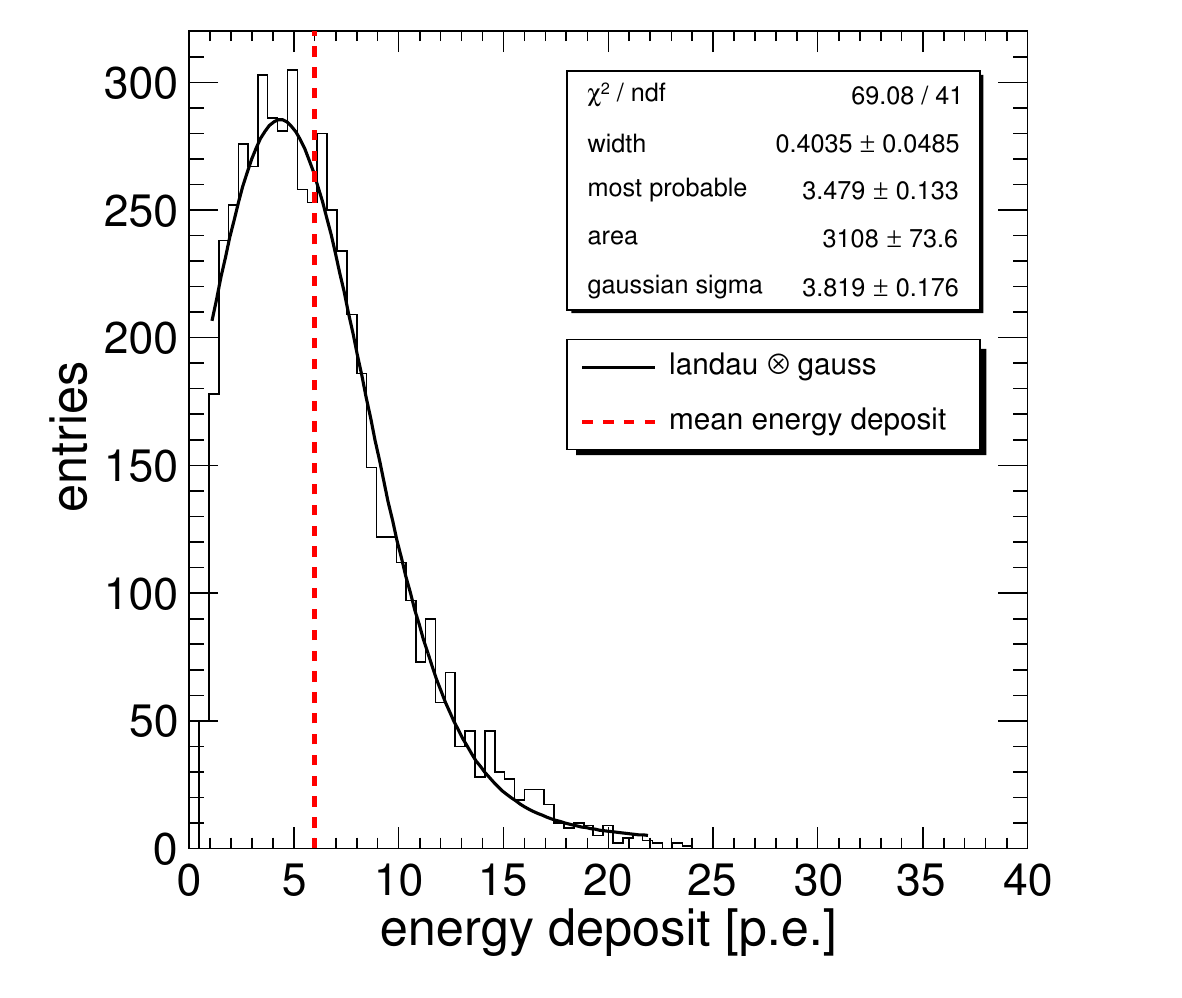}
    \caption{Left: Distribution of the total energy deposit in all SFT layers
             versus reconstructed momentum. The curve indicates a possible cut to separate
             protons from positively charged pions. In the calibration pions of both
             charges were used. Right: Distribution of energy deposits in a single fibre
             together with a fit of a Landau distribution convoluted with a Gaussian. \mhl{The
             fit is used to obtain the numeric mean value of energy deposits for MIPs as
             indicated by the dashed vertical line.}}
    \label{fig:SFT_dEvsMom}
  \end{center}
\end{figure}

\subsubsection*{Energy Calibration}

For the purpose of particle identification it is sufficient to extract a relative
calibration and equalize the response from all fibres.

The left panel in figure~\ref{fig:SFT_dEvsMom} shows the sum of energy deposits in
all SFT layers versus the reconstructed momentum, where negative values for the
reconstructed momentum correspond to negatively charged particles. The two bands
on the positive momentum side originate from positively charged pions and protons and the single
band on the negative momentum side corresponds to negatively charged pions. Negatively and
positively charged pions with a momentum larger than $0.2\unit{GeV}$ are considered
to be minimum-ionizing particles (MIPs) and used in the calibration. Positively charged pions
were distinguished from protons by requiring the total energy deposit to lie below
the curve shown in the left panel of figure~\ref{fig:SFT_dEvsMom}.

The right panel in figure~\ref{fig:SFT_dEvsMom} displays the energy-deposit
distribution after correction for incident angle in units of number of photoelectrons
for a single fibre, in the case that in the cluster the fibre was the leading fibre,
i.e., the one with the highest signal. A fit of a Landau distribution
convoluted with a Gaussian is used to extract the mean energy deposit for MIPs
for each fibre. The fit is indicated by the solid line, whereas the mean energy
deposit obtained by the fit is shown by the vertical dashed line. This mean energy
deposit is used for the relative calibration to equalize the response throughout the
layers.

\begin{figure}[b]
  \begin{center}
    \includegraphics[width=0.495\textwidth]{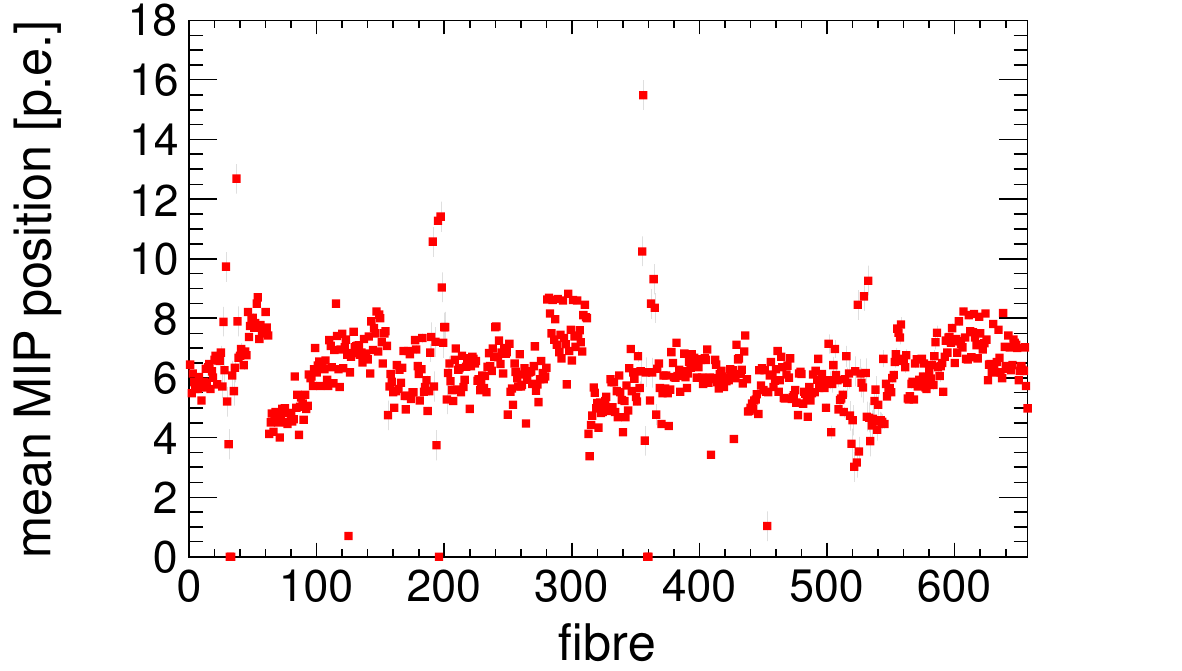}
    \includegraphics[width=0.495\textwidth]{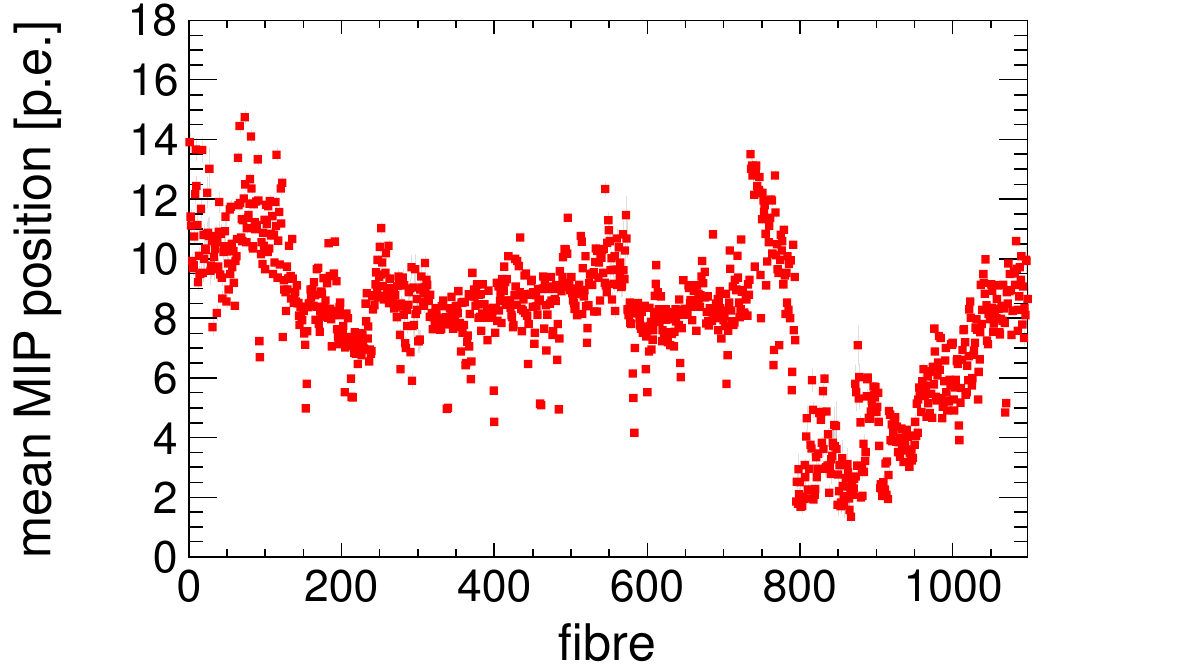}
    \caption{Mean position of the MIP peak for all fibres of the inner parallel
             layer of the inner barrel (left) and the parallel layer
             of the outer barrel (right) of the SFT.}
    \label{fig:SFT_MIPvsLayer}
  \end{center}
\end{figure}

Figure~\ref{fig:SFT_MIPvsLayer} shows the mean amplitude of the reconstructed
MIPs for all fibres of the inner parallel layer of the inner SFT barrel
(left panel) and the parallel layer of the outer barrel (right panel). In the region of
fibres $18$, $193$, $360$ and $520$ in the inner barrel layer large fluctuations
and larger error bars are visible. This is related to the holding structure
of the SSD, \mhl{which limits the acceptance and hence causes a drop
in statistics and possibly introduces systematics in the determination of the mean value.}
For the outer barrel this effect is not as strong, because in
the analysis both negatively and positively charged pions are used and the shadowing
of the SSD holding structure is less pronounced due to the bending of the particles in the
magnetic field. There is also a substructure of $32$ consecutive fibres visible
in the left panel. Each of these groups corresponds to one bundle of lightguide
connections between the active fibres and the MAPMTs. Due to the various bending angles
of the lightguides in a bundle, a different amount of light is lost, which results
in the observed substructure. In the outer barrel a similar,
yet less prominent effect with a substructure of $64\unitspace\unit{fibres}$ is observed.
Most probably the fluctuations are smaller here due to the larger bending radii
of the lightguides used for the outer SFT barrel.

In addition, the mean MIP amplitude of the fibres with numbers between $800$
and $960$ in the outer barrel are smaller compared to the remaining
fibres in the layer. With a threshold of $1.5$~photoelectrons, this has
clearly an influence on the MIP detection efficiency as it will be discussed
in section~\ref{sec:DetEffy}.

\subsection{Photon Detector}

By construction, the limited energy-deposition resolution of the PD does not allow 
the reconstruction of photon energies, due to the given photon-energy sampling fraction
between $6$\% and $10$\%. However, measured energy deposits are quantities valuable 
for the selection of photon signals and the identification of charged particles.
For signals observed in the PD, the efficiency for photon identification is about $99$\%, whereas
charged-particle identification has an efficiency of around $95$\% in the case the corresponding
track is reconstructed by the SSD and SFT.


\subsubsection*{Clustering}

For the processing of the PD signals, in a first step all strips with an ADC value
below threshold ($2\sigma$ above pedestal) are
discarded as well as strips with signals induced by cross-talk in the flat
cables. The latter signals are identified as signals with an energy amplitude smaller
than half the size of signal amplitudes observed 
in directly neighbouring flat-cable channels.

In the second step, a MAPMT clustering algorithm is applied that closely follows the one
used for the SFT. Initially, the local maxima on the MAPMT are determined.
A pair of pixels (associated to the same PD strip) is considered a local maximum,
if no higher signal exists in any of its neighbouring pixels, i.e. $8$ neighbouring
pixel pairs for pixel pairs not located at the edge of the cathode.
\mhl{Then the remaining pixel pairs are examined. Such a pixel pair is considered to
be a MAPMT cross-talk hit if its summed energy signal lies below $1.5\unit{MeV}$ and
if it corresponds to a PD strip that has no adjacent strip for which a signal is recorded.  
If any of these two conditions is not satisfied, the pixel pair is promoted to a local maximum.}
Finally, the energies of the cross-talk hits are added to that of the local maximum. If
a cross-talk hit is associated with several local maxima, its energy is shared
between them, proportionally to the energy value of each of the associated
local maxima. However, since a calibration in terms of photoelectrons was not
available for the PD, signal amplitudes in terms of energy units were considered.

In the third step, neighbouring strips in each of the PD layers are combined to
clusters. Similar to SSD and SFT, also here the energy contribution
from each of the individual strips is appropriately taken into account for the
reconstruction of the energy and the energy-weighted position of the clusters.
The cluster width amounts in most of the cases to one strip for pions and
protons and to one or two strips for photons.
From the reconstructed clusters only those with energy values above $1\unit{MeV}$
are stored. The justification for this threshold value lies in the rejection of
noise signals, which have been shown to be induced by high electron-beam currents and 
the presence of gas in the target cell.

In the fourth, and last step clusters are associated with tracks reconstructed by
the SSD and the SFT, if possible.

\subsubsection*{Calibration}

The calibration of the PD is based on signals from charged pions reconstructed
by SSD and SFT with momenta above $0.5\unit{GeV}$. In this momentum
range pions behave as MIPs in the PD. Collisional losses in the scintillating
layers result in an energy deposition of $2.1\unit{MeV/cm}$ per active detector
layer. The selection of positively charged pions is based on particle-identification
information from SSD and SFT.

\begin{figure}[t]
  \begin{center}
    \includegraphics[width=0.495\textwidth]{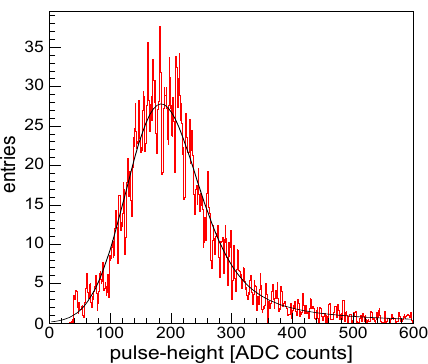}
    \caption{ADC spectrum for energy depositions in the PD by
             minimum-ionizing particles in units of ADC counts.}
    \label{fig:PD_ADCFit}
  \end{center}
\end{figure}

The calibration follows an iterative procedure: a first calibration is performed
on the ADC signals; in a second step, the calibration is based on cross-talk-corrected
energy signals. 

The signal spectra is fitted with a Landau distribution convoluted with a Gaussian
distribution. The numeric mean of this function, determined within certain
boundaries, is taken as the calibration point and converted into $\unit{MeV/cm}$ units
based on MC studies. A typical example of such a signal spectrum is shown in
figure~\ref{fig:PD_ADCFit} together with the fit function for signals generated in a
PD strip.

\begin{figure}[t]
  \begin{center}
    \includegraphics[width=0.95\textwidth]{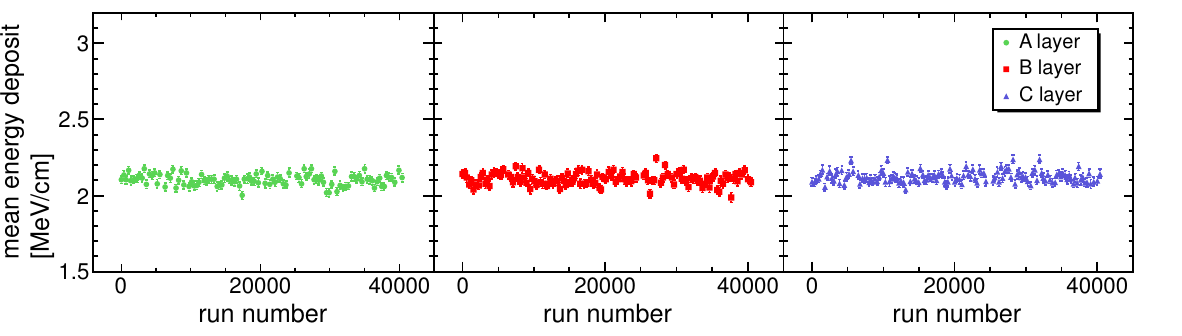}
    \caption{Mean energy deposition of minimum-ionizing particles in a cluster in
             the $A$ layer (circle), $B$ layer (square), and $C$ layer (triangle) of
             the PD as a function of run number.}
    \label{fig:PD_CalibTime}
  \end{center}
\end{figure}

The result of the calibration procedure is shown in figure~\ref{fig:PD_CalibTime}
for signals from MIPs. In the figure the mean energy deposition normalized to path
length in a cluster of the $A$ layer (circle), $B$ layer (square), and $C$ layer
(triangle) is shown as a function of {\sc Hermes} run number, with the run range
corresponding to the entire data collected in the year $2007$.
As can be seen, the calibrated data show a mean energy-deposition 
signal of $2.1$~MeV/cm, stable over time within the systematic 
uncertainty of $0.03$~MeV/cm.

\enlargethispage*{1.0cm}

%
\section{Momentum Reconstruction}
%

The momenta of protons, which can be reconstructed in the RD vary
from very low values of about $125\unit{MeV}$ to momenta of up to $1.4\unit{GeV}$.
For protons with momenta above $250\unit{MeV}$, which reach the SFT, momenta
can be reconstructed using their deflection in the field of the recoil-detector magnet.
Protons with lower momenta are stopped in the outer layer of the SSD. For these protons,
the accuracy of the momentum reconstruction using the deflection in the magnetic field
is degraded due to the rather small distance of only $1\unit{cm}$ between the two SSD
layers. However, the kinetic energy of these protons can be directly reconstructed from
the energy deposits in the two SSD layers. The energy deposits can also be employed for
protons in the intermediate momentum range, i.e., for protons with high enough momentum
to already reach the SFT, in order to improve the momentum resolution.

The track reconstruction algorithm is therefore divided into two stages. In the first
stage, a search for tracks with coordinate information in the SSD and SFT layers is
performed, and the momentum reconstruction is based solely on the coordinates of the hits.
In the second stage, the energy-deposit information is used in addition to the coordinate
information assuming that the particle is a proton. Both stages require the coordinate
information that is provided by SSD and SFT. This hit-position reconstruction is
performed separately prior to the track search.

\subsection{Spacepoint Reconstruction}

In the following, `spacepoint' refers to a combination of clusters reconstructed in
one sub-detector layer, which contains information on the 3D position as well as the
energy deposition of the hit.

\subsubsection*{SSD}

With the strips of the p-sides oriented perpendicular to those of the
n-sides, the hit position in the sensor plane is obtained directly from
the cluster positions given in terms of strip numbers. A constraint on the correlation
between measured energy deposits from both sides is applied in order to reduce the
number of ghost space points, i.e., the combination of two clusters not belonging to
the same track. The energy deposition of a reconstructed spacepoint is calculated as
the mean value of the energy deposits given by the two clusters forming the spacepoint.
In the case that one sensor side exhibits significantly higher noise, the energy
deposit of the formed spacepoint corresponds to the energy deposit of the cluster
from the other side.

\subsubsection*{SFT}

In the case of the SFT, the angle between parallel and stereo fibres was
$10\unitspace^{\circ}$. Spacepoints are formed considering geometrically allowed
combinations of clusters in the parallel and stereo layers. As the energy deposits
from the different layers are only weakly correlated, no hard restrictions can be applied,
and unavoidably ghost spacepoints can be created. This has to be taken into account in
the subsequent track search. In addition, due to the orientation of the stereo fibres, the
resolution in the $z$-coordinate reconstruction is worse by a factor of
$1/\sin(10\unitspace^{\circ})\approx 5.76$ in comparison to the resolution in the $xy$-plane.

\subsection{Track Finding}

The track search is performed by combining spacepoints from the SSD and SFT layers.
In the first stage of the track search, 4-spacepoint track candidates are considered,
i.e., tracks with one spacepoint in each sub-detector layer (inner SSD, outer SSD,
inner SFT, and outer SFT). Each track candidate is fitted with a helix hypothesis and
accepted if the $\chi^2$ value lies below a certain value. This value is set to $20$
according to MC studies, which allows the rejection of most of the ghost
tracks with four spacepoints with negligible influence on the
efficiency of the track search. From now on, spacepoints belonging to already accepted
4-spacepoint tracks are not considered any longer. During the following stage of the
track search, all possible 3-spacepoint combinations are fitted and again tracks with
$\chi^2$ below the threshold value are accepted. Finally, all possible 2-spacepoint
tracks with hits in the inner and outer SSD layer (including space points belonging
to accepted 3-spacepoint tracks) are considered. In the case of 2-spacepoint tracks
a $\chi^2$ restriction can not be employed due to the lack of degrees of freedom.

\subsection{Momentum Reconstruction by Coordinate Information}

The magnetic field of the RD magnet was precisely
measured~\cite{Statera:PhD} and can be used directly during the momentum
reconstruction. The procedure of momentum reconstruction in the magnetic field of
the RD is described in the framework
of references~\cite{Osborne:PhD, Hill:PhD,Mahon:PhD}. MC momentum-resolution
studies show that a simple assumption of a homogeneous magnetic field can be used
without essential degradation of the accuracy of the momentum reconstruction. Such a
simple approach is faster and can be used during track search.

During track fitting, the following function is minimized in the case of independent
measurement errors and neglecting multiple scattering:

\begin{equation}
	\chi^{2} = \sum_{i=1}^{N} \frac{(S^{\mathrm{fit}}_{i}(P) - S^{\mathrm{meas}}_{i})^{2}}{(\sigma^{\mathrm{meas}}_{i})^{2}}, 
\label{eq:ChiNoCor}
\end{equation}

\noindent
where $N$ is the number of measurements, $S^{\mathrm{meas}}_i$ and $\sigma^{\mathrm{meas}}_i$ are the measured coordinates
and their uncertainties, respectively, and $S^{\mathrm{fit}}_i(P)$ are coordinates as
functions of the kinematic parameters $P$ to be fitted to the measured ones. In the
momentum reconstruction, the vertex position in the $xy$-plane is assumed to be the beam
position. The beam position is determined by a separate beam-finder computer program
for each individual run using 4-spacepoint tracks reconstructed in the recoil detector. The
kinematic parameters are chosen to be convenient for a solenoidal magnetic field geometry:
$\lambda = 1/p_{t} = 1/(p*\sin(\theta_{v}))$, $\phi_{v}$, $\cot(\theta_{v})$, and $z_{v}$,
where $p_t$ is the transverse momentum, and $\theta_v$, $\phi_v$ and $z_v$ are respectively
the polar angle, azimuthal angle and $z$ coordinate at the vertex position of the track.
During minimisation of equation~\ref{eq:ChiNoCor}, the coordinates are expressed
as functions of the kinematic parameters in the following way:

\begin{equation}
	\phi_{det} = \phi_v - C \cdot R \cdot B \cdot \lambda \cdot \frac{0.3\unit{GeV}}{200\unit{T cm}},
\end{equation}
\begin{equation}
	z_{det} = z_{v} + R\cot(\theta_{v}),
\end{equation}

\noindent
where $C$ is the particle charge, \mhl{$B$ is the magnetic field} and $R$ is the distance from a
detection layer to the vertex. In order to take multiple scattering into account,
equation~\ref{eq:ChiNoCor} is modified by inclusion of correlation coefficients in the
error covariance matrix:

\begin{equation}
	\chi^{2} = \sum_{i,j=1}^{N} (S_{i}^{\mathrm{fit}}(P) - S_{i}^{\mathrm{meas}})V_{i,j}(S_{j}^{\mathrm{fit}}(P) - S_{j}^{\mathrm{meas}}), 
\label{eq:ChiCor}
\end{equation}

\noindent
where $V_{i,j}$ are elements of the covariance matrix $V$. The covariance matrix is
determined from a MC simulation for protons. For pions, the effect of multiple
scattering is found to be negligible.

\subsection{Momentum Reconstruction Using Energy Deposits in the SSD}

For tracks containing hits in the SSD layers, the measured energy deposit in the
sensors can be used to improve the momentum reconstruction based on the hit
coordinates only. This is achieved by adding an additional term to
equation~\ref{eq:ChiCor}:

\begin{equation}
	\chi^{2} = \sum_{i,j=1}^{N} (S_{i}^{\mathrm{fit}}(P) - S_{i}^{\mathrm{meas}})V_{i,j}(S_{j}^{\mathrm{fit}}(P) - S_{j}^{\mathrm{meas}}) +
	\sum_{i=1}^{M} \frac{(E_i^{\mathrm{fit}} (P) - E_i^{\mathrm{meas}})}{(\sigma^{\mathrm{meas}}_{i})^{2}}.
\end{equation}

\noindent
Here $M$ is the number of energy deposit measurements, 
$E_i^{\mathrm{meas}}$ is the measured energy deposit in a silicon sensor and
$E_i^{\mathrm{fit}} (P)$ is the fit value of the energy deposit. The latter is a function
(lookup table) of kinematic parameters and is determined from detailed MC
studies taking also into account the passive materials~\cite{Burns:PhD}.

\subsection{Alignment}

\subsubsection*{Alignment of SSD and SFT}

Knowledge of the detector alignment usually includes internal alignment of
individual sub-detectors and relative alignment of sub-detectors. For the
alignment procedure several assumptions are made:

\begin{itemize}
	\item The positions and spacing of strips in each SSD sensor are assumed to be
	      known precisely, as the accuracy of these parameters provided by the manufacturer
	      is much better than required by track reconstruction.
	\item The intrinsic bending of the SSD sensors is assumed to be negligible.
	\item The radius of the SFT is assumed to be independent of $\phi$ angle
	      and $z$ coordinate.
\end{itemize}

\begin{figure}[b]
  \begin{center}
    \includegraphics[width=0.495\textwidth]{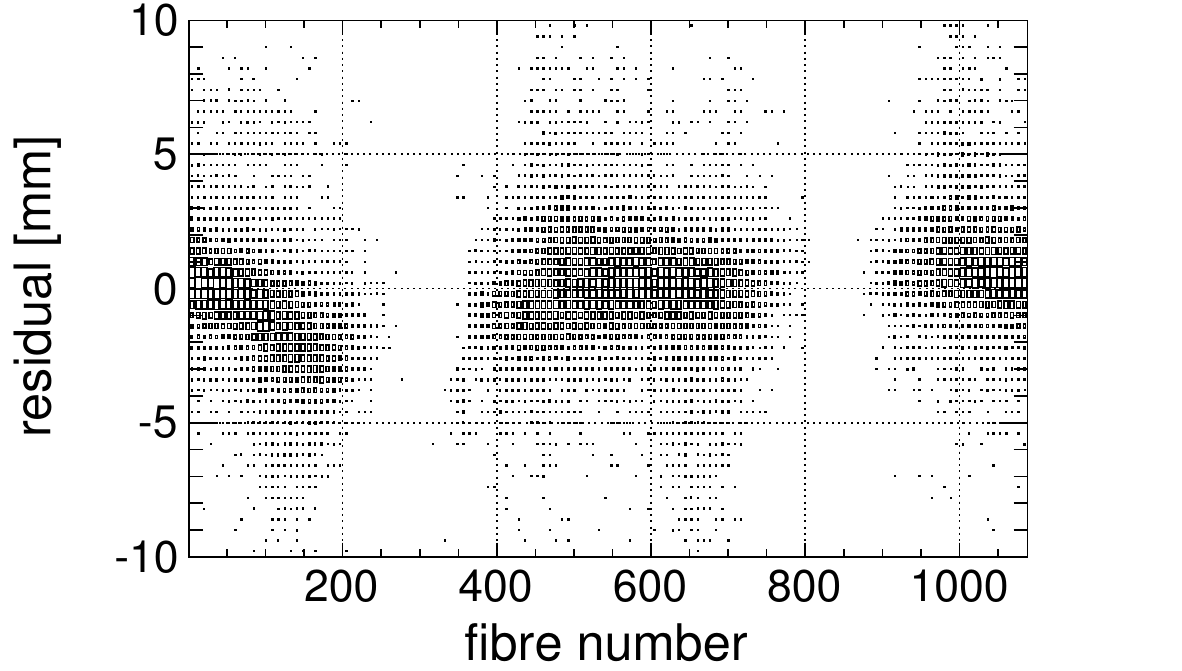}
    \includegraphics[width=0.495\textwidth]{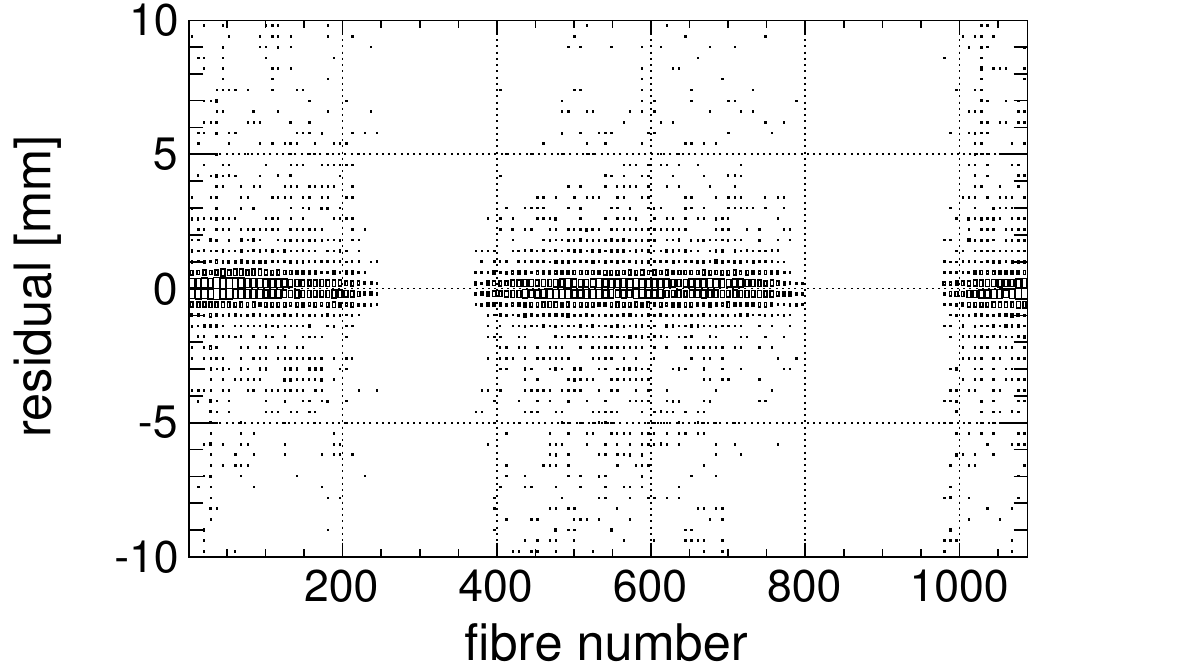}
    \caption{\mhl{Differences between the hit position in the outer SFT layer calculated
             from the track parameters and the measured position before (left) and
             after (right) alignment obtained from a cosmic ray data sample.}}
    \label{fig:OuterSFTalign}
  \end{center}
  \begin{center}
    \includegraphics[width=0.495\textwidth]{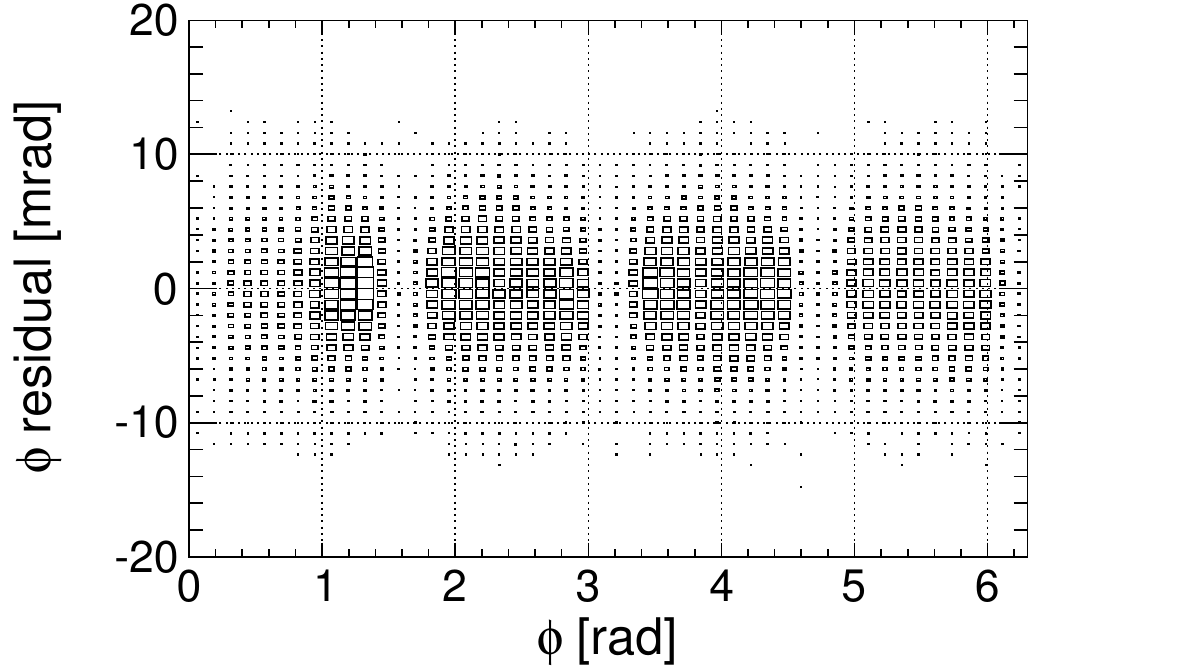}
    \includegraphics[width=0.495\textwidth]{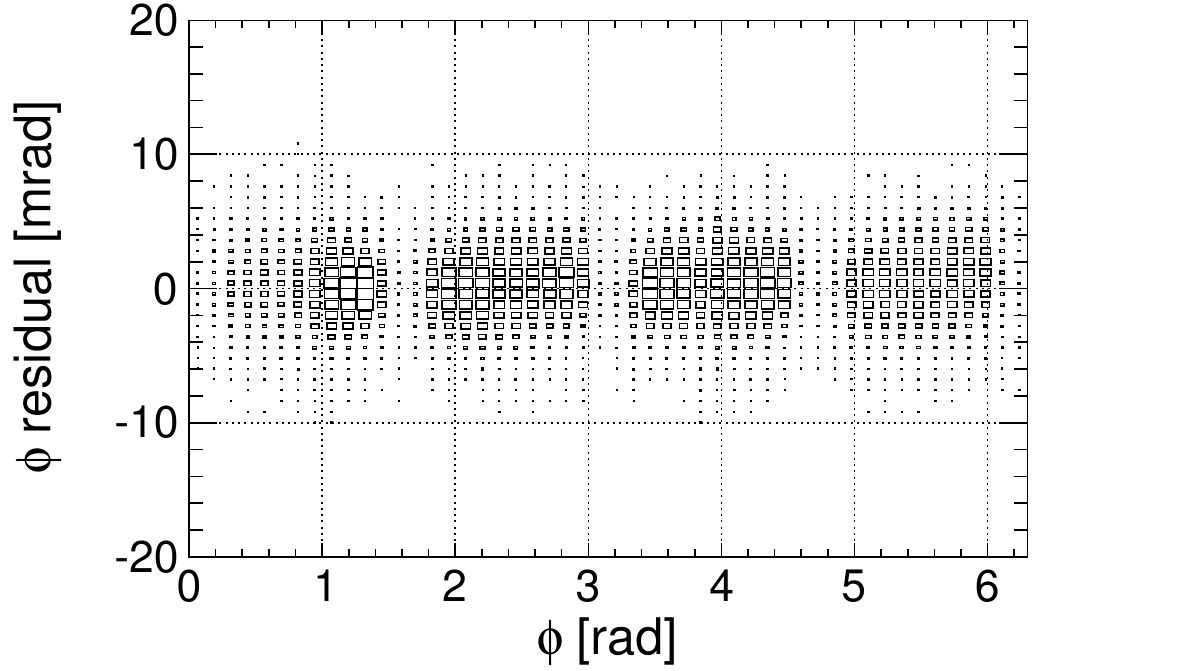}
    \caption{Differences between the $\phi$ angle of the hit in the inner (left) and outer (right)
             SSD layers calculated from the track parameters and the measured $\phi$ angle.}
    \label{fig:SSDres}
  \end{center}
\end{figure}

\begin{figure}[t]
  \begin{center}
    \includegraphics[width=0.495\textwidth]{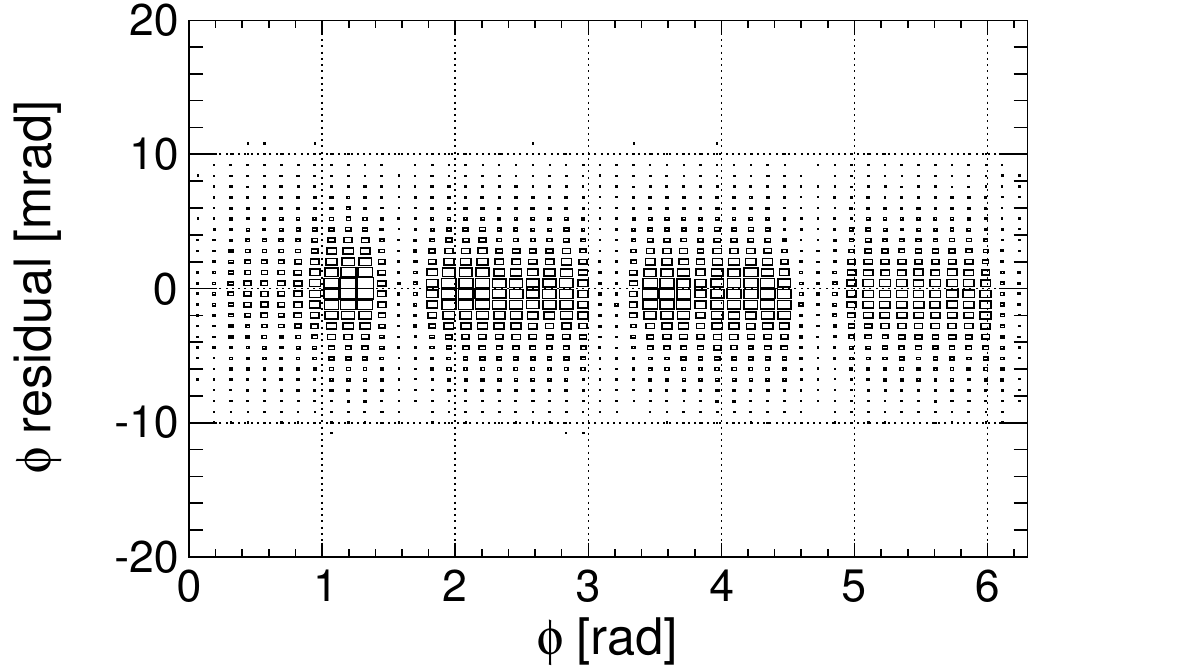}
    \includegraphics[width=0.495\textwidth]{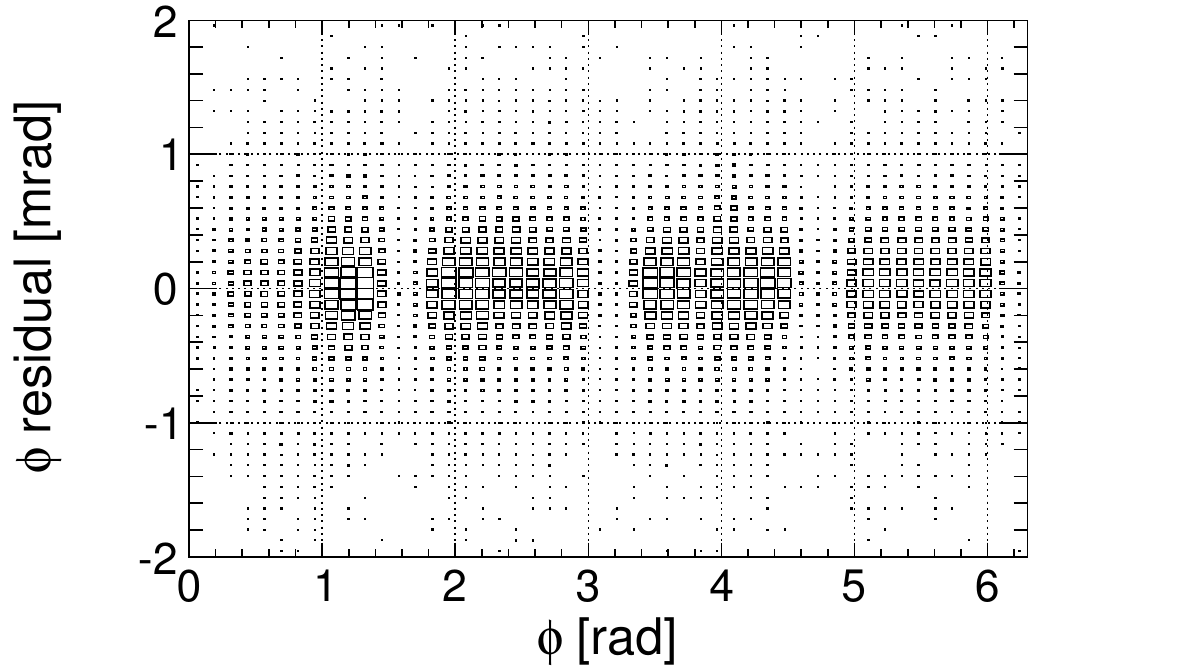}
    \caption{Differences between the $\phi$ angle of the hit in the inner (left) and outer (right)
             SFT layers calculated from the track parameters and the measured $\phi$ angle.}
    \label{fig:SFTres}
  \end{center}
\end{figure}

The first step in the alignment procedure is to determine the SFT fibre positions
based on data obtained at the DESY22 test beam exploiting $5\unit{GeV}$ electrons.
The measurement procedure is described in reference~\cite{Perez:PhD}. Straight-track
data was used to extract the fibre position along the detector and a lookup table
consisting of the $\phi$ angle and $z$ position of all possible intersections of
parallel and stereo fibres was produced.

As a next step, the SSD sensors are positioned relative to each other. Shifts and
rotations with respect to the nominal values of each sensor were fitted using an
event sample with selected tracks from a cosmic-ray data set collected without
magnetic field and reconstructed with a straight-line assumption. \mhl{The minimization
of alignment parameters is performed iteratively by refitting all tracks during each
iteration.}

With knowledge of the internal alignment of SSD and SFT, both detector components
are aligned relative to each other in a final step, using a cosmic-ray event sample
with straight tracks. In parallel also small correction of the SFT fibre positions in
comparison to the positions obtained from the DESY22 test-beam data are extracted.

The quality of the alignment can be tested using 'residuals' defined as the
differences between the hit position in a detection layer calculated from the
track parameters and the measured position. \mhl{As an example, the residuals obtained
from a cosmic ray data sample before - using the nominal fibre positions - and after
alignment are shown for the outer SFT layer in} figure~\ref{fig:OuterSFTalign}.
In figures~\ref{fig:SSDres} and \ref{fig:SFTres}, the azimuthal residuals obtained
from experimental data after the alignment are shown for the SSD and SFT layers.
\mhl{In all cases, all four detection layers were included in the track fit.
The outer layer of the SFT mainly defines the curvature in the magnetic field and
as expected the observed residual is much smaller than for the other layers.
A small residual miss-alignment is observed at $\phi \approx 6\unit{rad}$ even after
the iterative alignment procedure that can not be explained and further improved.}

\subsubsection*{Alignment of the PD}

The PD is aligned with respect to the SSD and SFT using reconstructed tracks
from pions originating from the interaction of the positron beam with the hydrogen
target. First, the orientation of each strip of the $B$ and $C$ layers was measured 
with respect to the beam line and averaged over all strips from the same layer.
Secondly, the position along the $x$ and $y$ axis and the rotation in azimuthal
angle $\phi$ of each of the layers was determined, and finally, the measurement
of the strip orientation was repeated to check for a possible correlation between
the two distinct alignment procedures. Within the uncertainty, no correlation
between both procedures was found.

The design value of the radius of each of the PD layers was assumed to be sufficiently
accurate by construction. The PD was also assumed not to be inclined with respect to
the beam axis, which is a reasonable assumption considering the low $\theta$ resolution
of the photon detector and results from survey measurements. Finally, the magnetic field
was assumed to be homogeneous.

\begin{figure}[t]
  \begin{center}
    \includegraphics[width=0.95\textwidth]{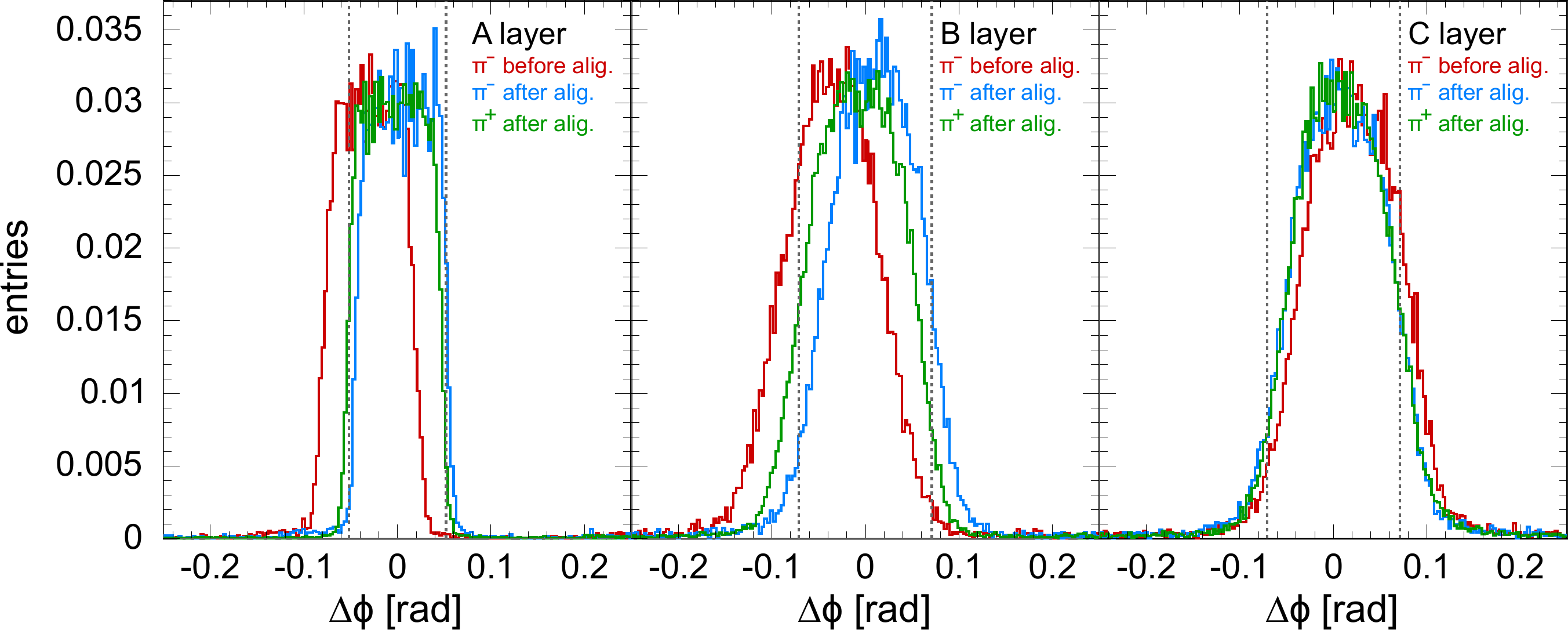}
    \caption{Distributions in $\Delta\phi$ before and after the alignment for the
             $A$ layer (left), $B$ layer (center), and $C$ layer (right) of the
             photon detector.}
    \label{fig:PD_AlignmentResults}
  \end{center}
\end{figure}

The effect of the PD alignment procedure is shown in figure~\ref{fig:PD_AlignmentResults}
by means of the $\Delta\phi$ distribution, which represents the difference between the
azimuthal angle of the strip center and the azimuthal angle of the track intercept
translated along the strip orientation to the upstream end of the detector layer. 
\mhl{The vertical dashed lines in the figure delimit one strip pitch. Before the alignment,
the mean of the $\Delta\phi$  distribution for $\pi^{-}$ amounts to $-30\unit{mrad}$,
$-35\unit{mrad}$, and $20\unit{mrad}$ for the $A$, $B$, and $C$ layer respectively.
After the alignment, the mean of the distribution for $\pi^{-}$ ($\pi^{+}$) 
amounts to $4.2\unit{mrad}$ ($-2.8\unit{mrad}$), $9.6\unit{mrad}$ ($-8.7\unit{mrad}$), 
and $8.7\unit{mrad}$ ($9.8\unit{mrad}$) for the $A$, $B$, and $C$ layer. The deviation
from $0\unit{rad}$ is largest for the two outer layers. As the additional knowledge
of the $z$-coordinate of the track intercept with the layer is needed for the $B$ and
$C$ layer, and the track's polar angle and $z$-vertex position are known with less
precision than the track's azimuthal angle, an additional bias can be introduced for
the two outer layers. Moreover, Monte-Carlo studies show that the strip orientation
can not be determined better than $3.5\unit{mrad}$/$7.0\unit{mrad}$ for a perfectly aligned
$B$/$C$ layer, while the layer alignment is limited to $2.8\unit{mrad}$ in $\phi$ and
$0.03\unit{cm}$ in $x$ and $y$ for a precisely known strip orientation. A combination of
these various factors can explain the magnitude of the observed shift for each of the
three detector layers. This has to be supplemented with a small misalignment in $\theta$
of the detector with respect to the beam line of $4.608\pm0.017\unit{mrad}$, a value far
below the $\theta$-resolution of the photon detector, and with the non-homogeneity of the
magnetic field, which is not taken into account. Although the above given arguments can
explain the magnitude of the shift, they can not explain the difference in shift of the
$\Delta\phi$ distribution between negatively and positively charged pions for the $A$ and
$B$ layer. It can be understood, however, if the radii of both layers are underestimated. 
The construction precision of the photon detector is not better than $0.5\unit{mm}$ for
the radius of each layer. The underestimation of the radii of course also contributes to 
the magnitude of the shift. The inclusion of the layers' radii as free parameters in the
layer-alignment procedure can reduce this shift. However, as the magnitude of the shift
in either direction does not exceed the shift observed in the $C$ layer, it was decided
not to elaborate on this. Further details on the PD alignment procedure can be found in}
reference~\cite{VanHulse:PhD}.

\section{Detector Performance}
%

\subsection{Momentum Reconstruction}

\begin{figure}[b]
  \begin{center}
    \includegraphics[width=0.495\textwidth]{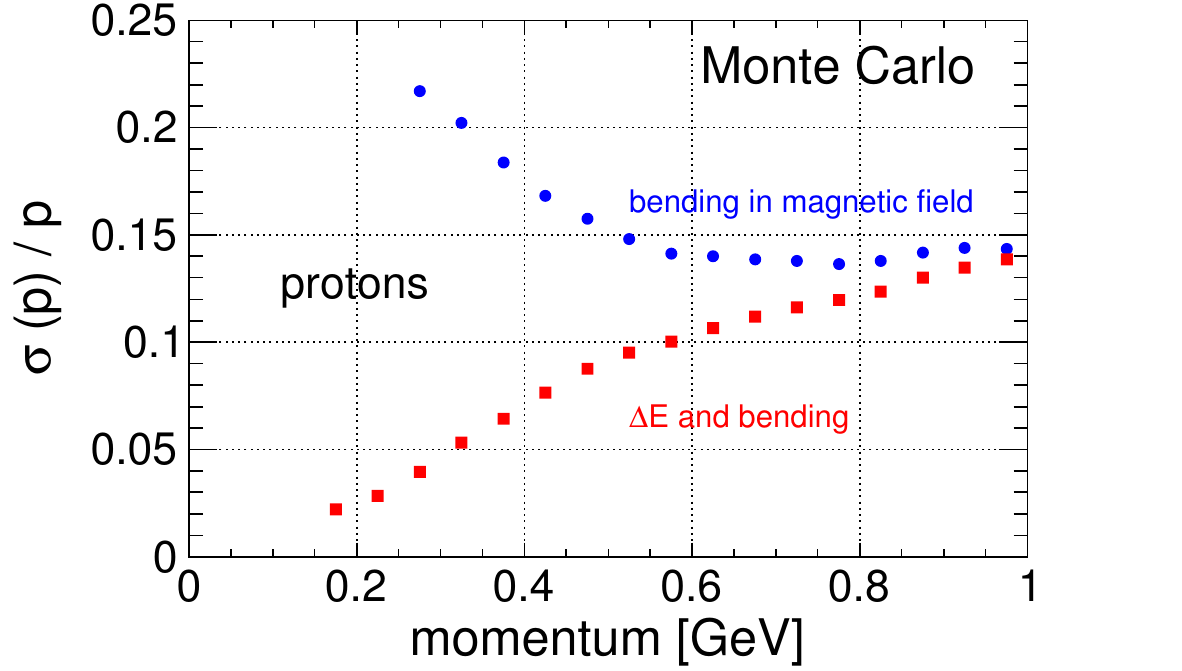}
    \caption{\mhl{Fractional momentum resolution versus momentum for proton reconstruction
             by coordinates only (circles) and by energy deposition in the SSD (squares).}}
    \label{fig:MomRes}
  \end{center}
\end{figure}

\begin{figure}[t]
  \begin{center}
    \includegraphics[width=0.495\textwidth]{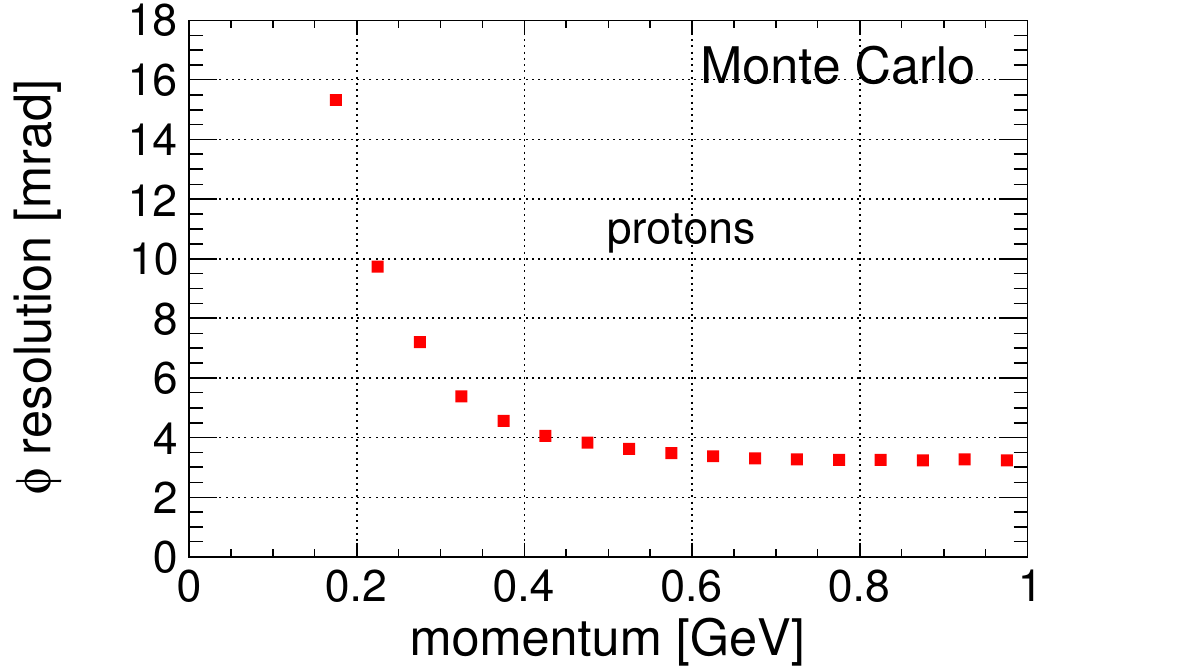}
    \includegraphics[width=0.495\textwidth]{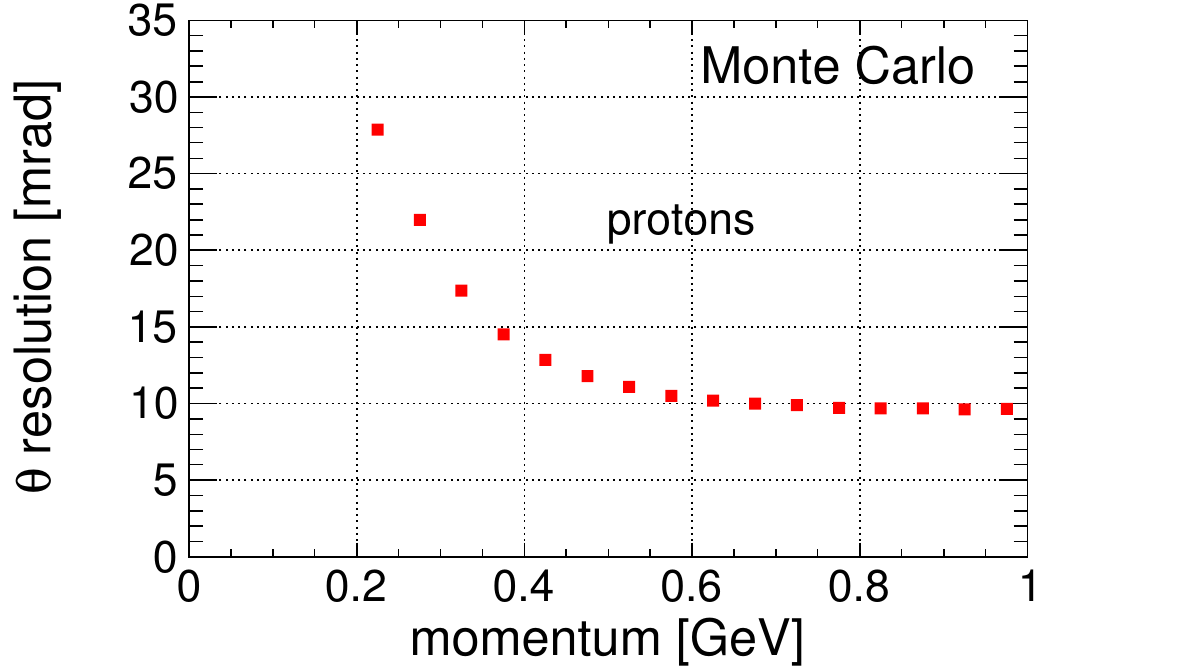}
    \caption{Azimuthal (left) and polar (right) angle resolution versus momentum for protons.}
    \label{fig:PhiThetaRes}
  \end{center}
\end{figure}

Momentum-resolution studies were performed based on Monte Carlo data samples containing
protons and pions. In figures~\ref{fig:MomRes} and~\ref{fig:PhiThetaRes}, the accuracy
of the momentum and angular reconstruction is presented for protons. Figure~\ref{fig:MomRes}
shows the momentum reconstruction accuracy using coordinate information only, in comparison
with the accuracy of the momentum reconstruction taking into account additionally energy
deposits in the SSD. Figure~\ref{fig:PhiThetaRes} shows azimuthal (left) and polar (right)
angle resolutions as calculated using both energy deposits and coordinates. For protons,
the angular resolutions deteriorate for momenta below $0.5\unit{GeV}$ because of multiple
scattering. For pions, the momentum resolution is about $12\%$ and the azimuthal (polar)
angle resolution is about $4\unit{mrad}$ ($10\unit{mrad}$), with no pronounced dependence
on momentum.

\subsection{Particle Identification}
\label{sec:PID}

For each reconstructed track, the energy deposited along the particle's
passage through the active detector components can be used to
determine the particle type. In the case of tracks that incorporate
coordinate information from the four detection layers, a total of six
independent energy-deposition measurements are available, coming from the
inner and outer SSD layers, and the parallel and stereo layers of the inner
and outer SFT barrel. As the protons and pions constitute the predominant
statistics, only the separation of these two particle types is considered.
Due to the absence of anti-protons, negatively charged tracks are always
assumed to be negatively charged pions.

For each energy-deposition measurement and hence each detection layer $l$,
a particle-iden\-ti\-fi\-ca\-tion probability $\mathrm{PID}_{l}$ depending
on the energy deposition $dE$ and the reconstructed momentum $p$ is calculated
according to

\begin{equation}
\mathrm{PID}_{l}(dE; p) =
\log_{10}\frac{P_{l}(dE; \beta\gamma = \frac{p}{m_{p}}) }
{P_{l}(dE; \beta\gamma = \frac{p}{m_{\pi}}) }.
\label{eq:PID_prob}
\end{equation}

\noindent
Here, $P_{l}$ are the so-called parent distributions, which are
energy-deposition distributions normalized to unity, and which depend on the momentum
and the particle type. The particle-type dependence in the distributions is
eliminated by using a $\beta\gamma = \frac{p}{m_{p/\pi}}$ binning for the energy depositions,
namely $\log_{10}\frac{p}{m_{p/\pi}}$.


The combined particle-identification probability is the sum of the
probabilities from the individual layers. A cut on the total probability is
used to discriminate between charged pions and protons. The actual value of the
cut depends on the specific application, i.e., the desired discrimination
efficiency and contamination of the other particle type. The particle-identification
efficiencies are covered in a subsequent paragraph.

\subsubsection*{Extraction of Parent Distributions}

\begin{figure}[t]
  \begin{center}
    \includegraphics[width=0.495\textwidth]{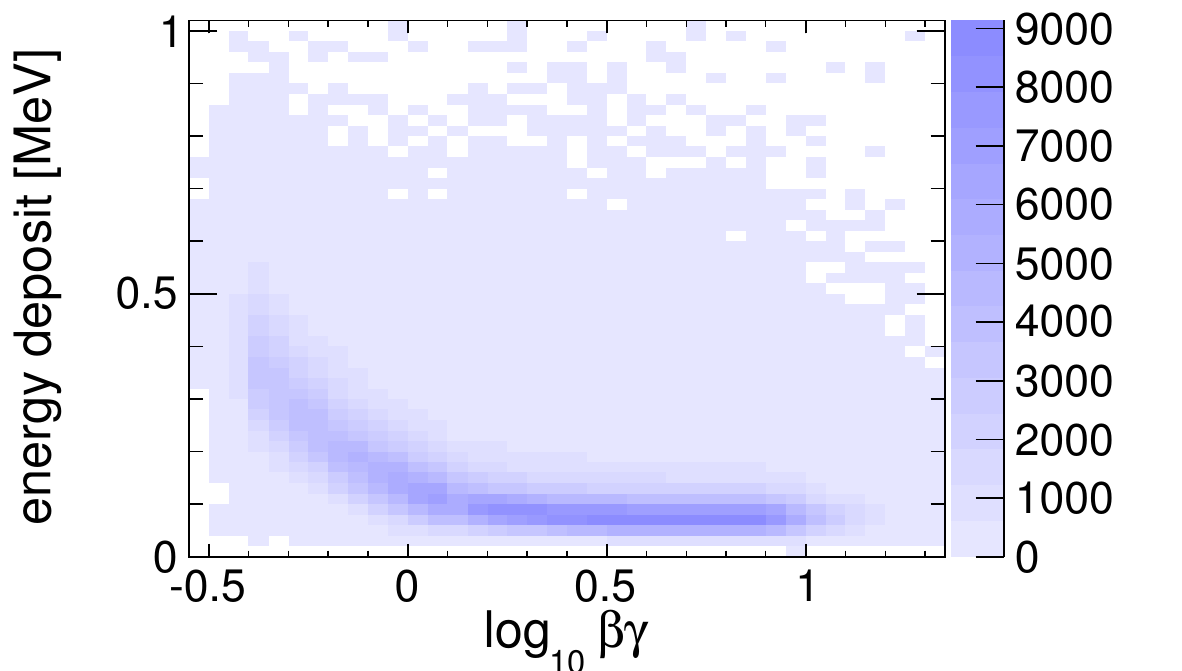}
    \includegraphics[width=0.495\textwidth]{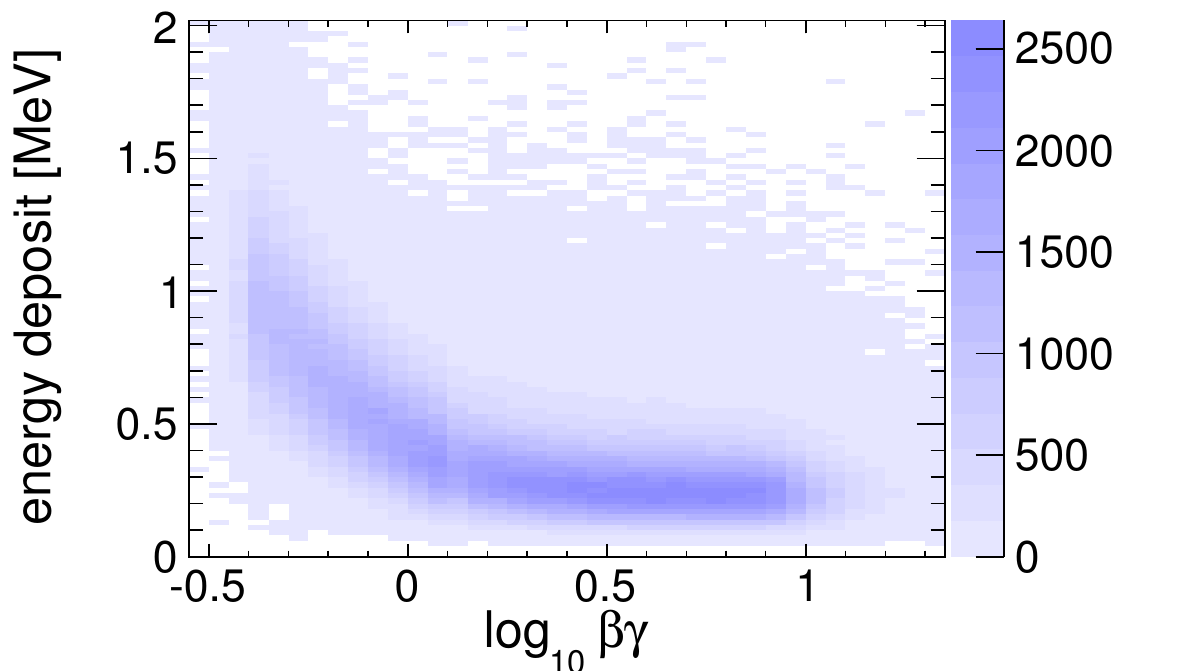}
    \caption{Energy deposit in the inner SSD layer (left) and the inner SFT
             layer (right) versus $\log_{10}(\beta\gamma)$ for the clean proton
             and charged pion track sample.}
    \label{fig:PID_dEvsBetaGamma}
  \end{center}
\end{figure}

The quality of the parent distributions $P_{l}$ is crucial to the
performance of the particle iden\-ti\-fi\-ca\-tion. In order to avoid effects originating
from any remaining miscalibration of the sub-detectors, the
distributions were extracted from experimental data instead of Monte Carlo data.
This procedure requires clean proton and pion track samples. Events have been
chosen with only a single reconstructed track consisting of hits in all
four tracking layers. Restrictive cuts on the energy deposit in the other five
layers are applied in order to eliminate remaining background when extracting
the energy-deposit distributions for a certain detection layer. The energy-deposit
distributions versus $\log_{10}(\beta\gamma)$ for both the inner SSD and inner
SFT layers are shown in figure~\ref{fig:PID_dEvsBetaGamma}. Protons with
momenta below $1\unit{GeV}$ populate the region
$\log_{10}(\beta\gamma) \leq 0.1$, whereas protons with larger momenta and
pions with momenta above $0.2\unit{GeV}$ populate the region with
$\log_{10}(\beta\gamma)$ larger than $0.1$.

For each $\log_{10}(\beta\gamma)$ bin a Landau distribution convoluted with a
Gaussian is fit to the energy deposit distribution. \mhl{The obtained
fit parameters are then parameterised as functions of $\log_{10}(\beta\gamma)$.}
The final parent distributions $P_{l}$ are obtained from the Landau-Gauss
parameterisations. This method ensures also reasonable parent distributions
in the transition region at $\log_{10}(\beta\gamma) \approx 0.1$ where a lack
of statistics is observed.


\subsubsection*{Particle Identification Efficiency and Contamination}

\begin{figure}[t]
  \begin{center}
    \includegraphics[width=0.495\textwidth]{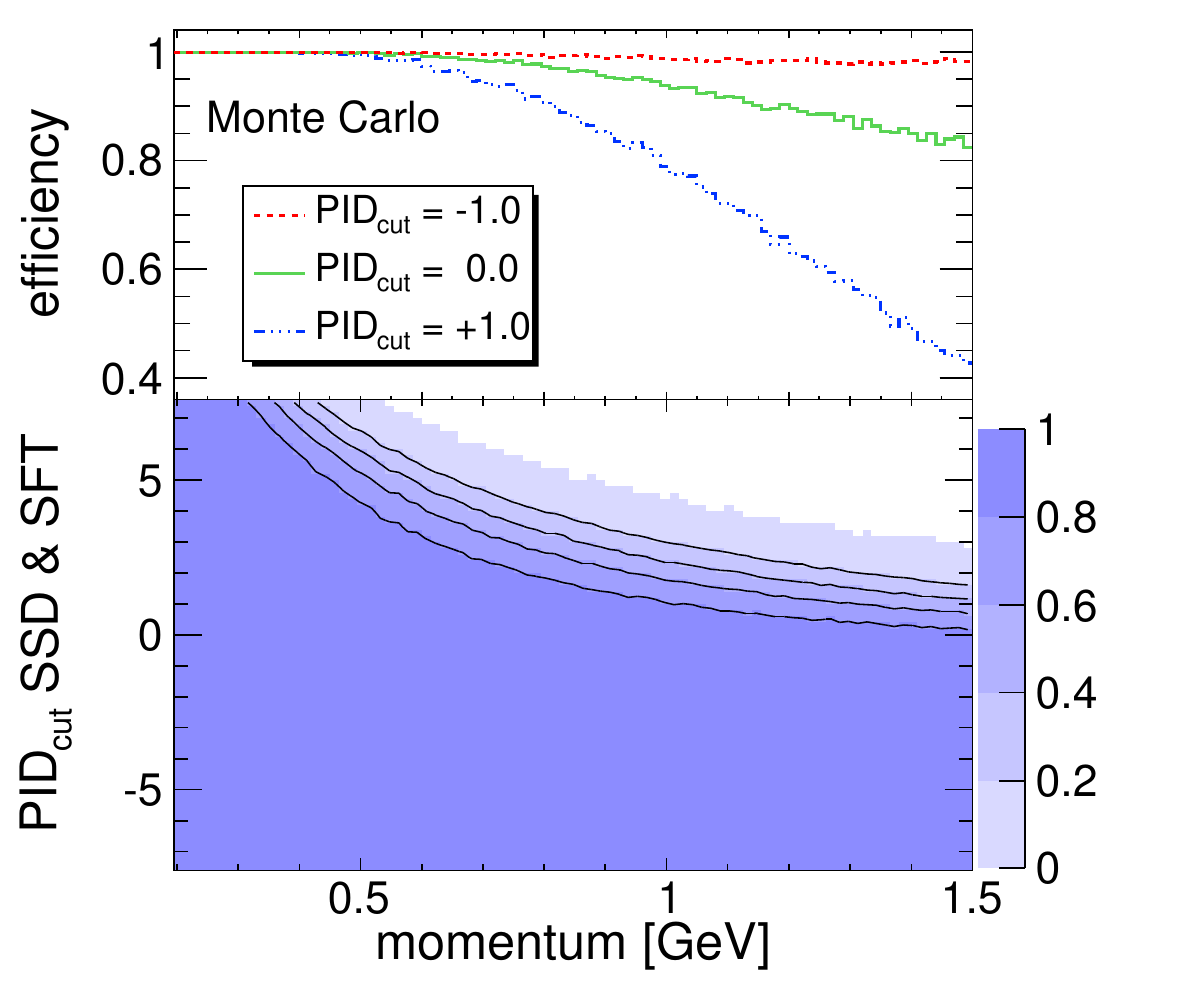}
    \includegraphics[width=0.495\textwidth]{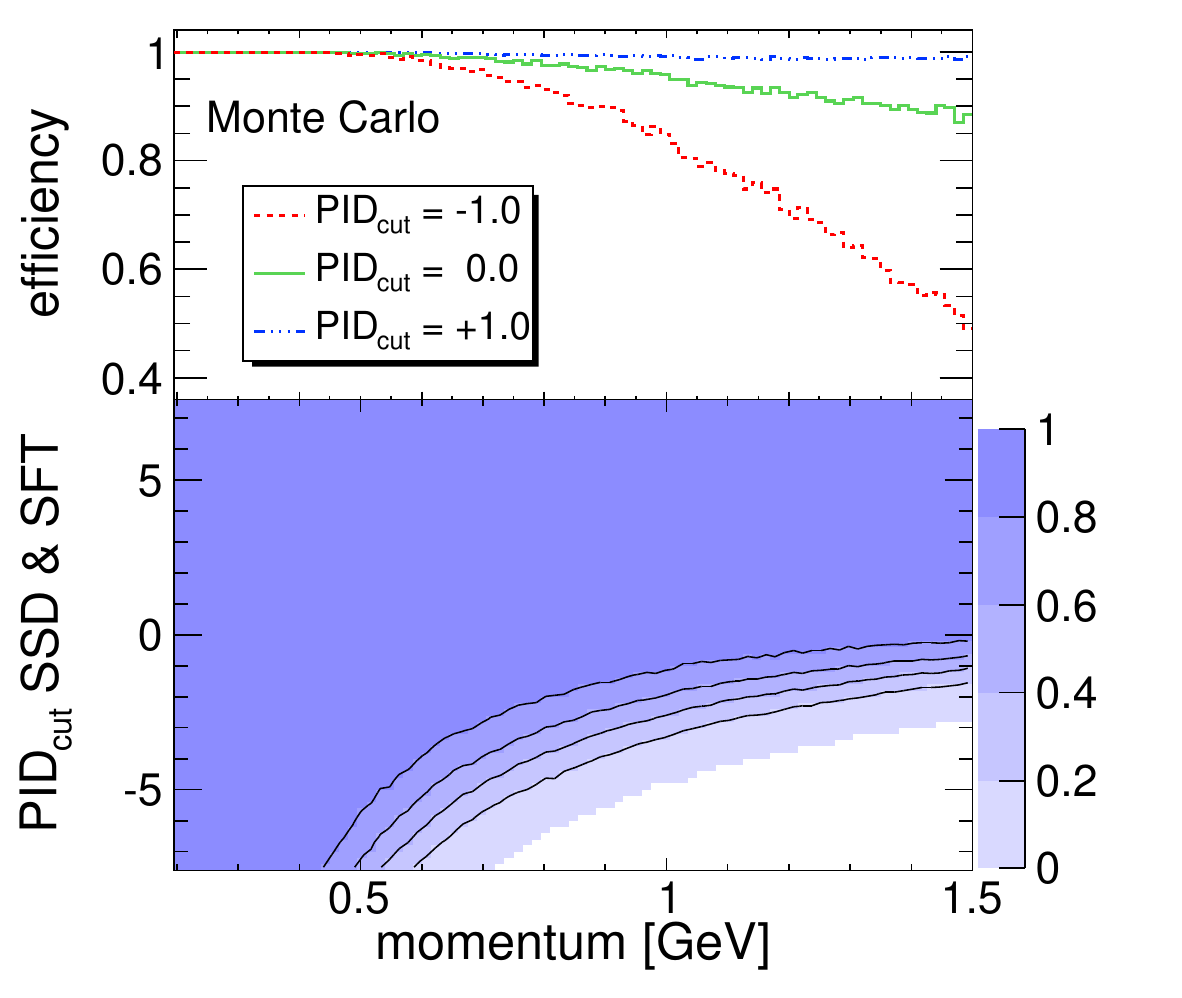} \\
    \includegraphics[width=0.495\textwidth]{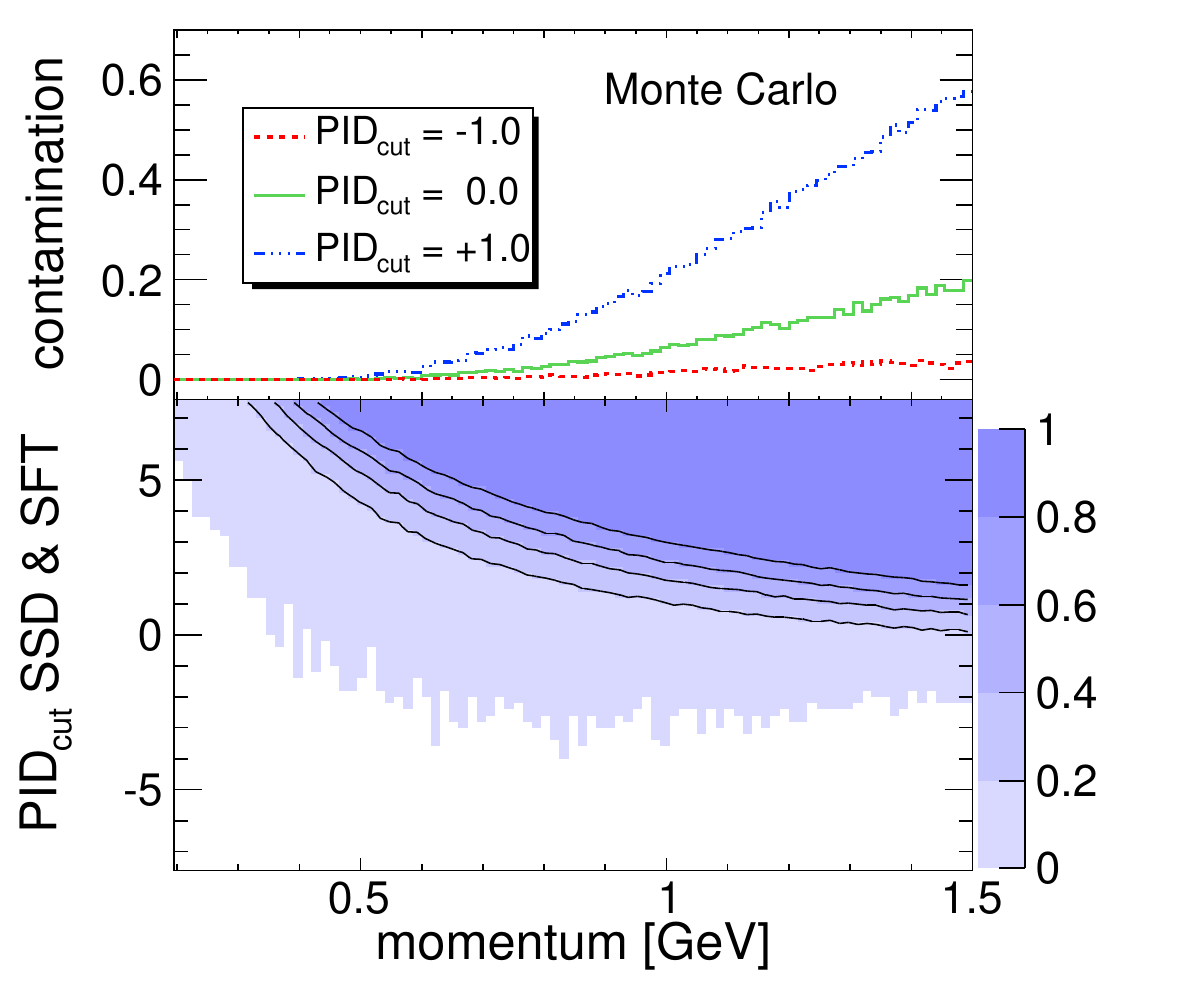}
    \includegraphics[width=0.495\textwidth]{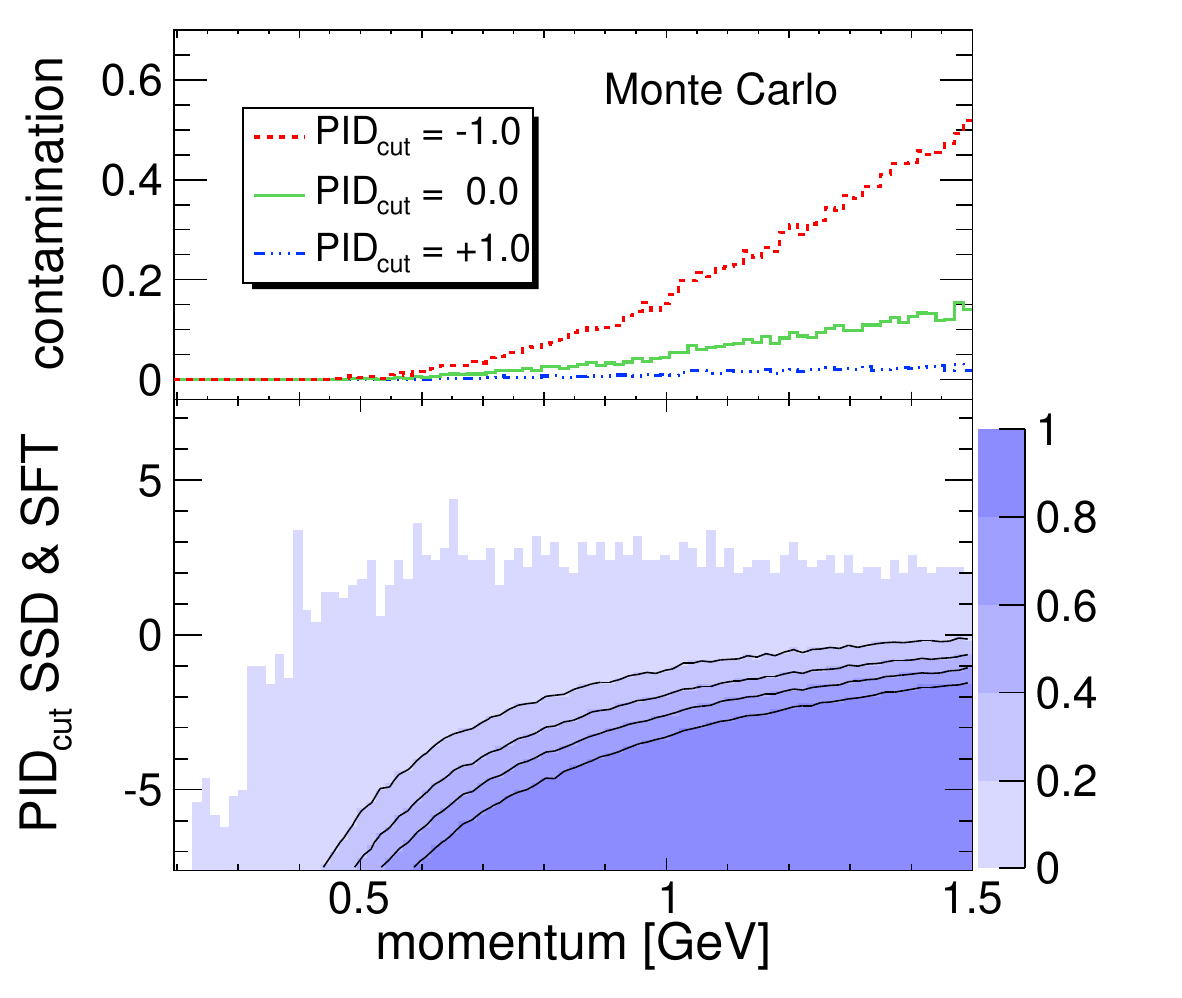}
    \caption{Top-left: Efficiency of the proton identification for tracks
             consisting of at least three spacepoints. The bottom sub-panel
             shows the efficiency as a function of particle momentum (x-axis)
             and a selection cut ($PID_{cut}$) applied to the combined $PID$ value
             (y-axis), i.e. the sum of the $PID$ values from
             the individual detection layers. The top sub-panel
             depicts the efficiency for three different values of the selection
             cut above which a particle is deemed to be a proton.
             Top-right: Efficiency of the pion identification. Here,
             particles are deemed to be a pion if the combined $PID$ value
             is below the $PID_{cut}$ value. Bottom row: Contamination of a
             selected particle sample with protons (left) and pions (right)
             as a function of momentum and combined $PID$ value. The upper
             sub-panels show the contamination for three different
             $PID_{cut}$ values.}
    \label{fig:PID_Performance}
  \end{center}
\end{figure}

For each reconstructed track, with either three or four spacepoints, the combined
particle-iden\-ti\-fi\-ca\-tion probability can be calculated from the energy
deposits in the layers, the reconstructed momentum and the parent distributions.
The combined $PID$ value is used to select protons (charged pions) by
requiring the value to be above (below) a certain threshold $PID_{cut}$.
Depending on the actual value of $PID_{cut}$, selected particle samples contain
a certain contamination by the other particle type. In addition, the efficiency of the
selection strongly depends on the threshold. Therefore the threshold used to select
certain particle types has to be tuned to the specific requirements.
In order to obtain values for the efficiency and contamination levels as a function of
the value of $PID_{cut}$, extensive Monte Carlo studies were performed.
The top row of panels in figure~\ref{fig:PID_Performance} shows the efficiency of
the proton (left) and pion (right) identification as a function of the reconstructed
particle momentum and $PID_{cut}$ value applied to the combined $PID$ (bottom sub-panel)
or versus momentum for three different $PID_{cut}$ values (top sub-panel). The bottom
row of panels in the figure shows the contamination of a selected track sample with
protons (left) and pions (right) as a function of momentum and $PID_{cut}$ value applied
to the combined $PID$ (bottom sub-panel) or versus momentum for three different
$PID_{cut}$ values (top sub-panel).

For the standard $PID_{cut}$ value of zero, the selection efficiency for protons (top-left
panel in figure~\ref{fig:PID_Performance}) is close to $1.0$ up to a momentum of
$0.7\unit{GeV}$ and then it drops to $0.85$ at $p \approx 1.5\unit{GeV}$. The corresponding
contamination with pions can be extracted from the bottom-right panel in the figure.
Up to $0.7\unit{GeV}$ the contamination is very low and increases to a value of $0.14$
for higher momenta. For pions the situation is very similar. The efficiency is close to
$1.0$ up to a momentum of $0.6\unit{GeV}$ and drops with increasing momentum to about $0.9$
at $p \approx 1.5\unit{GeV}$. The respective contamination by protons (bottom-left panel)
reaches a value of $0.2$ at a momentum of $p \approx 1.5\unit{GeV}$. Further details can
be found in reference~\cite{XLu:Master}.


\subsection{Detector Efficiencies}
\label{sec:DetEffy}

The detection efficiencies of the individual layers are extracted by
excluding the layer under study from the track search and reconstruction,
and by comparing the number of expected hits with the number
of detected hits in a narrow special window in that layer. As the SSD
was not designed for high efficiency for MIPs but rather for low energy
recoiling protons, the efficiency shows a strong dependence on the particle
momentum and type. In order to eliminate the particle-type dependence,
a binning of the extracted efficiency values in $\log_{10}(\beta\gamma)$ instead
of momentum and particle type is used for SSD and SFT.

\subsubsection*{Efficiency of the SSD}
\label{sec:DetEffySSD}

\begin{figure}[t]
  \begin{center}
    \includegraphics[width=1.0\textwidth]{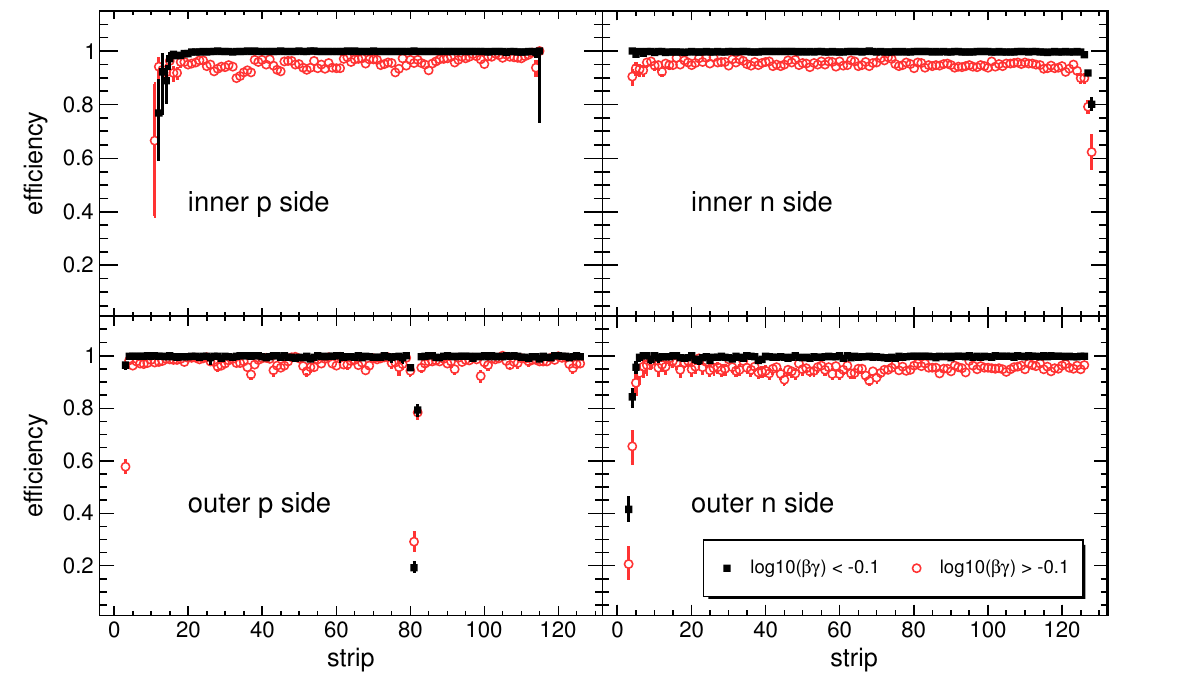}
    \caption{Cluster efficiency for the p- and n-sides of one of the inner
             SSD sensors (top) and one of the outer sensors (bottom) for two
             ranges in $\log_{10}(\beta\gamma)$. The range
             $\log_{10}(\beta\gamma) < -0.1$ constitutes the bulk of the proton
             statistics.}
    \label{fig:ssd_efficiencies}
  \end{center}
\end{figure}

\mhl{The efficiency of the SSD is extracted for hits, clusters and spacepoints. In
the case of the inner (outer) sensors the hit predictions are obtained from an
interpolation of the tracks reconstructed from the hits in the outer (inner)
SSD and inner SFT, and the beam constraint to the layer under study.}
Figure~\ref{fig:ssd_efficiencies} shows the cluster efficiency for both
p- and n-sides of one inner and one outer sensor for two $\log_{10}(\beta\gamma)$
ranges. The solid markers for which $\log_{10}(\beta\gamma) < -0.1$ constitute
the bulk of the proton statistics. In this region the cluster efficiency
for all sensors is above $99\unit{\%}$. The panel for the inner p-side efficiency
misses data points for low and high strip numbers. The reason for this is the
smaller $\phi$~acceptance of the corresponding outer sensor and therefore
absence of statistics in these strip regions in the inner sensors. For
$\log_{10}(\beta\gamma) > -0.1$, which corresponds to protons with momenta
above $800\unit{MeV}$ and most of the pion statistics, the cluster
efficiency is lower. The results shown in figure~\ref{fig:ssd_efficiencies}
represent sensors with rather high cluster efficiencies for pions. Due to
the fact that the hardware threshold was set to $50\unit{\%}$ of the most
probable energy deposit of a MIP, the cluster efficiency is strongly
affected by small variations in the gain and linearities of the high-gain
readout channel. For some sensors the MIP efficiency was as low as $30\unit{\%}$.
The lower left panel of figure~\ref{fig:ssd_efficiencies} also shows a drop
in efficiency for both $\log_{10}(\beta\gamma)$ ranges at around strip~80.
This is caused by a dead strip with zero efficiency which is smeared over
a few strips due to the limited position resolution.

\subsubsection*{Efficiency of the SFT}
\label{sec:DetEffySFT}

\begin{figure}[t]
  \begin{center}
    \includegraphics[width=1.0\textwidth]{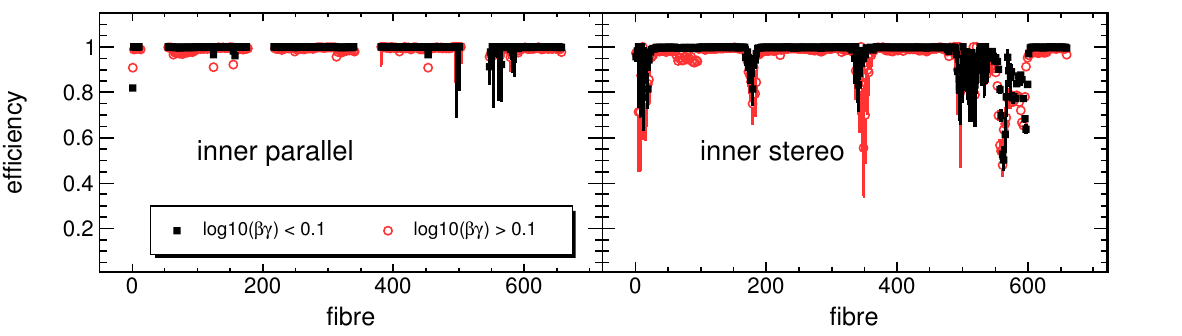}
    \includegraphics[width=1.0\textwidth]{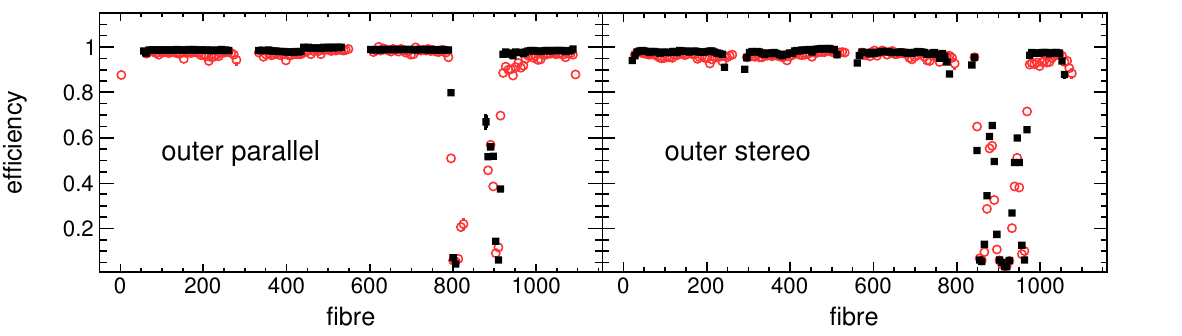}
    \caption{Efficiency of the SFT layers for two regions of $\log_{10}(\beta\gamma)$.
             The different panels show the individual layers of the SFT.}
    \label{fig:sft_efficiencies}
  \end{center}
\end{figure}

Very similar to the case of the SSD, the efficiency of the SFT is determined for
individual fibres. The cluster efficiencies as a function of fibre number are shown
in figure~\ref{fig:sft_efficiencies} for two different $\log_{10}(\beta\gamma)$
regions and the four layers of the SFT: inner parallel, inner stereo, outer parallel
and outer stereo. The gaps with undefined efficiency are caused by lack of
statistics due to shadowing by the SSD holding structure.
The inner SFT layers have an efficiency of above $95\unit{\%}$ in the region
$\log_{10}(\beta\gamma) < 0.1$ and above $90\unit{\%}$ for the high-$\log_{10}(\beta\gamma)$
region. In general, the outer layers show a very similar behavior. \mhl{However, the
outer parallel layers have a very low efficiency in the region between fibres $850$ and
$950$ due to two not properly functioning MAPMTs.} As a consequence, also the determination
of the efficiency of the other layers suffers from this effect by lack of statistics
and lower quality tracks. Moreover, the coverage in the angle $\phi$ of the
fibres connected to the non working MAPMTs has an overlap with the $\phi$ acceptance
of the non-fully functioning outer SSD module (see section~\ref{sec:DAQPerformance})
which causes an even larger decrease in track quality when neglecting the layer
under study during track search and fit. The combined effect on the efficiencies can
be seen for the parallel layer of the outer SFT at fibres $800$ and $900$, as well
as in the inner layers in the region of fibres $550$ through $600$. Details on the
efficiency of the SFT can be found in references~\cite{Perez:PhD} and~\cite{XLu:Master}.

\subsubsection*{Efficiency of the PD}

\begin{figure}[t]
  \begin{center}
    \includegraphics[width=0.495\textwidth]{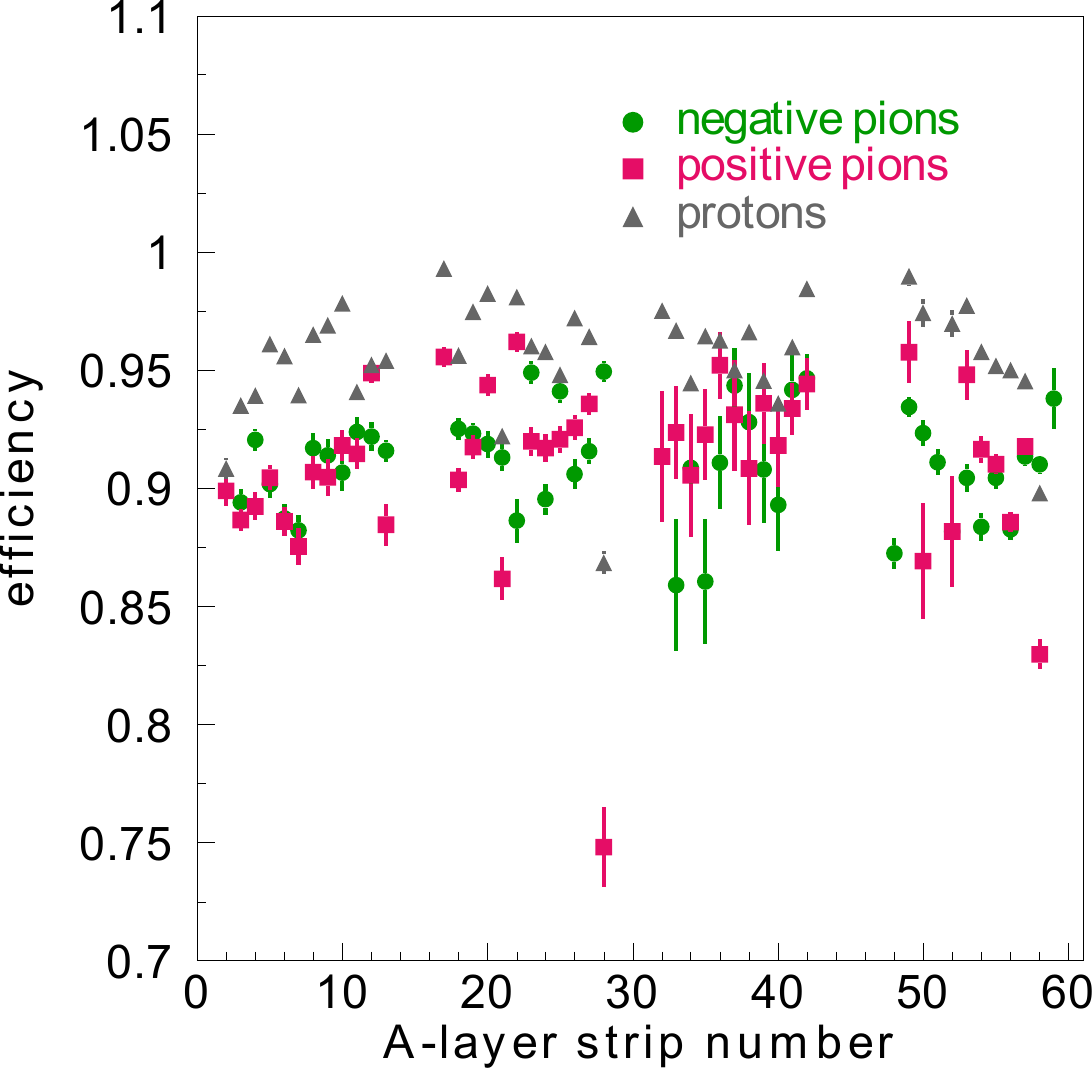}
    \includegraphics[width=0.495\textwidth]{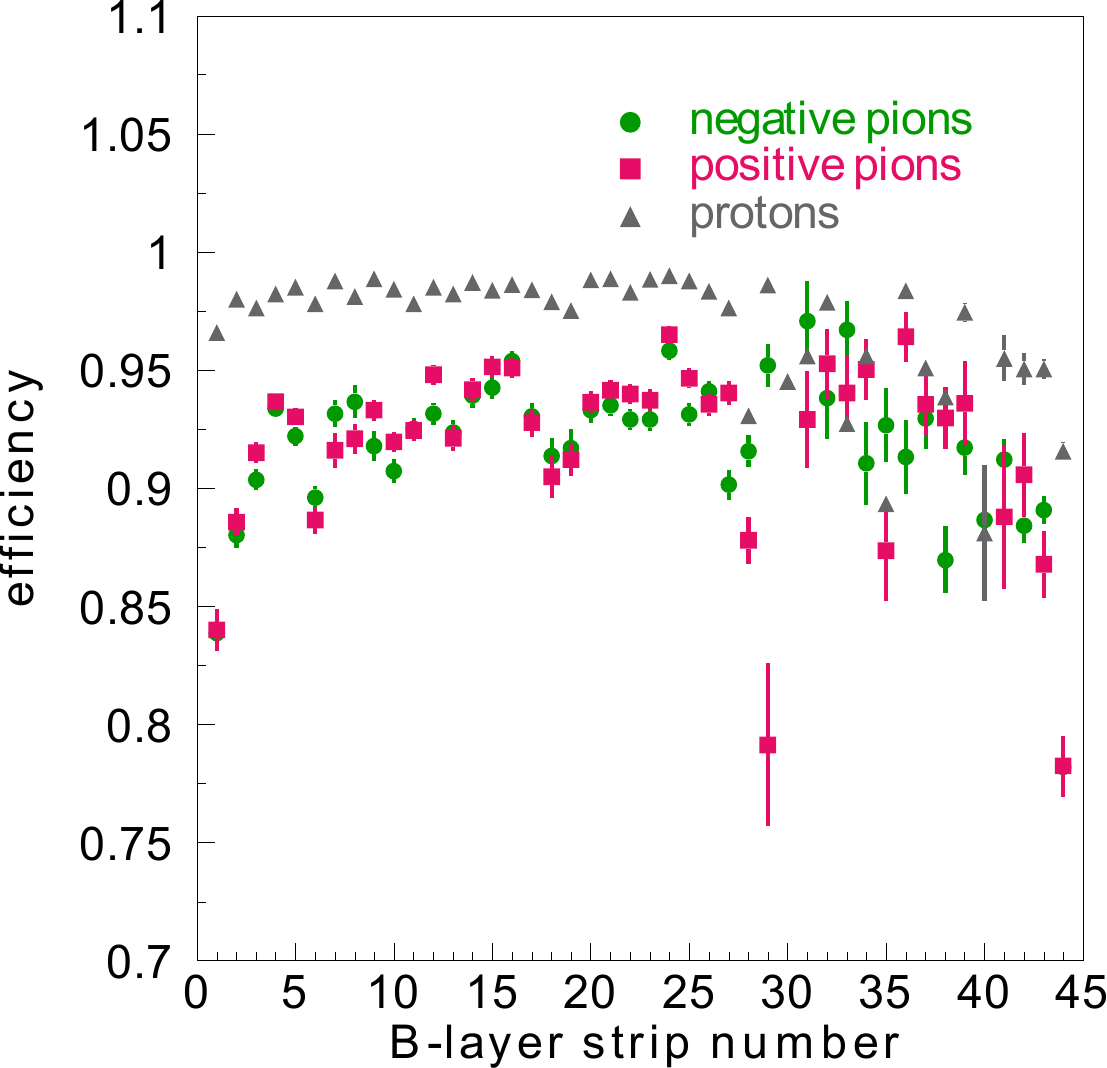}
    \caption{Efficiency of the $A$ layer (left) and $B$ layer (right) of the PD
    	         for charged pions and protons with momenta between $0.6$~GeV and
	         $0.7$~GeV.}
    \label{fig:PD_Efficiency}
  \end{center}
\end{figure}

The detection efficiencies of the individual PD layers are studied using
tracks reconstructed by both the SSD and SFT, and by requiring a signal in
the PD layers not under investigation. The efficiencies are presented
in figure~\ref{fig:PD_Efficiency} for signals from positively and
negatively charged pions and from protons in the $A$ layer and $B$ layer.
Since the $C$ layer is the outermost active detector component, it is not possible
to distinguish between particles that passed through the $C$ layer and those
that stopped before the $C$ layer. Therefore, the efficiency for the $C$ layer is
not determined to the same extent. In addition, the track selection limits the
acceptance for the extraction of the PD efficiency.

As can be seen in figure~\ref{fig:PD_Efficiency}, on average no difference
in detection efficiency for positively and negatively charged pions is
observed. For strips located around the gap in between two silicon strip
modules, the bending in opposite direction of the positively and negatively
charged particles in the magnetic field together with the track selection
leads to small differences. Apart from these edge effects and not considering
the problematic region with the not fully functioning SSD module (see
section~\ref{sec:DAQPerformance}) and the non-working SFT MAPMTs (see
section~\ref{sec:DetEffySFT}), which affects strips $46$--$50$ of the $A$-layer 
and strips $1$--$3$ and $38$--$44$ of the $B$-layer, the detection efficiency
shows no dependence on either the momentum or the charge of the pions. It is
about $92\%$ for both layers. 

As expected, the detection efficiency for protons lies well above the detection
efficiency for pions. \mhl{A higher proton-detection efficiency in the $B$ layer than
in the $A$ layer is observed, because of the requirement of a signal in the $A$
and $C$ layer or in the $B$ and $C$ layer, respectively. This requirement,
introduced in order to reject noise hits, results in a larger average energy
deposition in the $B$ layer than in the $A$ layer.}

%
\section{Event Selection with the Recoil Detector}
%

In the following chapter, details about the selection of DVCS events are
presented. In a similar manner, events from other exclusive processes can be selected
with the use of RD information.

DVCS event candidates are selected requiring the detection of exactly
one scattered electron or positron and exactly one photon measured in the forward spectrometer.
In addition, information on all tracks reconstructed in the recoil
detector is used in the analysis.

\begin{figure}[t]
\begin{center}
\includegraphics[width=0.495\textwidth]{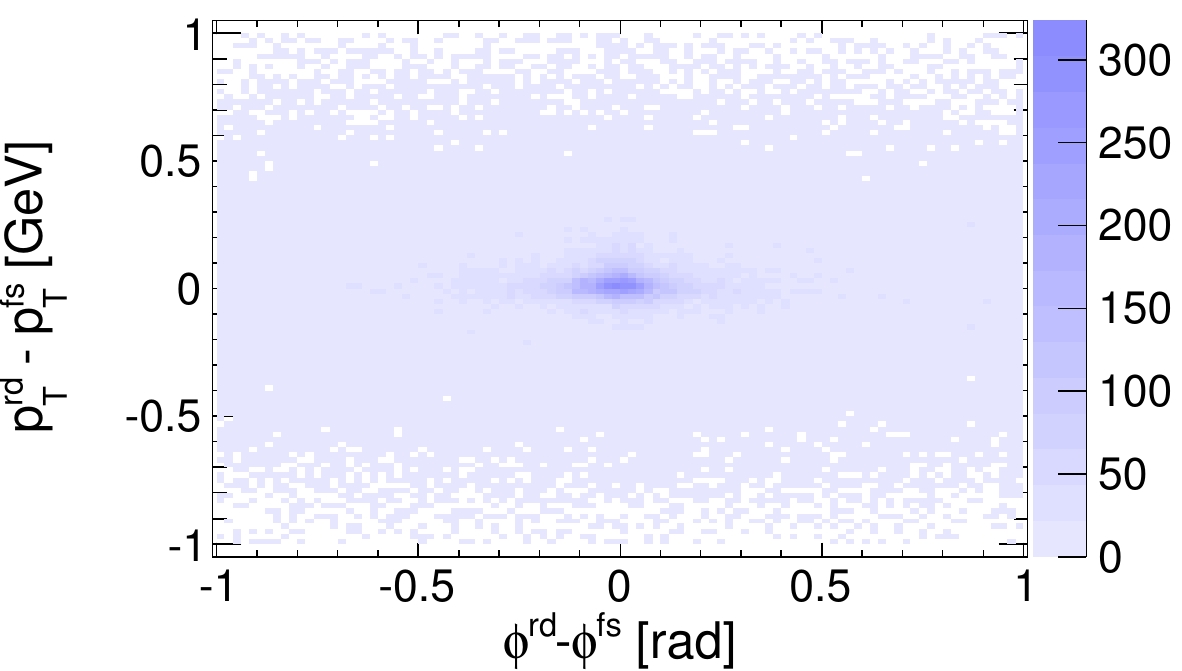}
\includegraphics[width=0.495\textwidth]{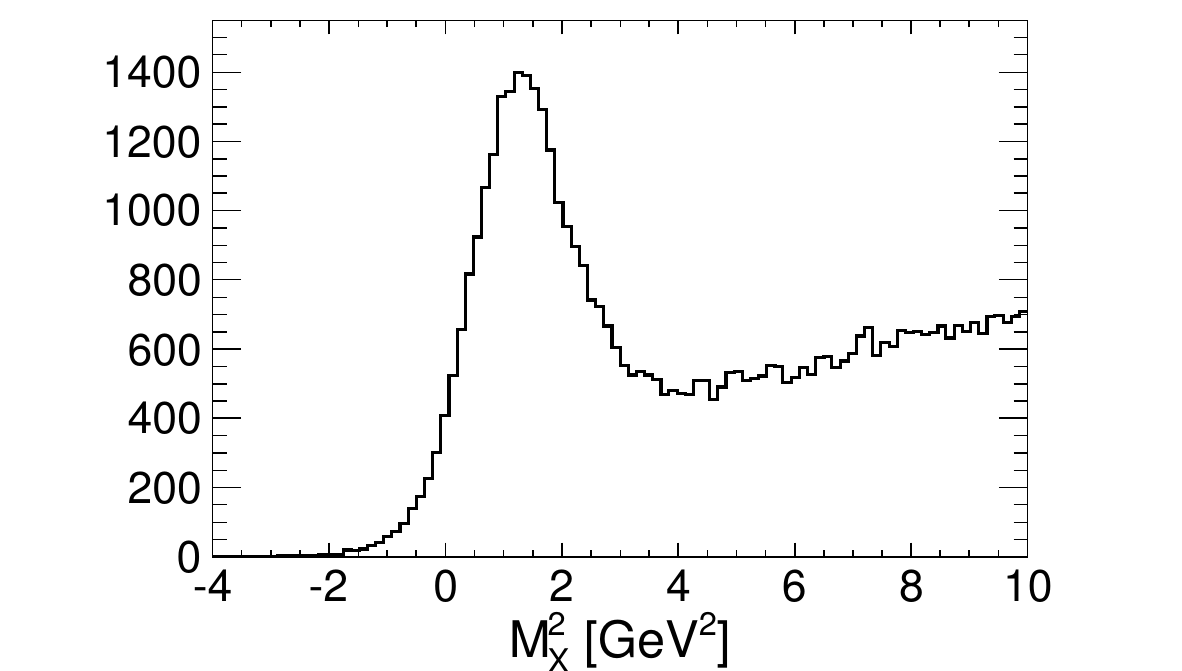}
\caption{Left panel: The difference between the transverse momentum of the proton
         reconstructed in the RD and the transverse momentum calculated from the
         measured 3-momenta of the scattered lepton and real photon versus the
         difference between the reconstructed and calculated azimuthal angles.
         Right panel: Squared missing mass calculated from the measured 3-momenta
         of the scattered lepton and real photon.}
\label{fig:dvcs_dphi_dpt_dvcs_mx2}
\end{center}
\end{figure}

In the left panel of figure~\ref{fig:dvcs_dphi_dpt_dvcs_mx2}, the difference between
the proton transverse momentum reconstructed with the RD and the missing transverse
momentum calculated using information from only the forward spectrometer is presented
versus the difference between reconstructed and calculated azimuthal angles. The
right panel of figure~\ref{fig:dvcs_dphi_dpt_dvcs_mx2} shows the distribution of
the squared missing mass

\begin{equation}
M_X^2=(k+p-k^\prime-q^\prime)^2,
\label{eq:mx2}
\end{equation}

\noindent
calculated using the 3-momenta of the scattered lepton and the real photon ($q^\prime$).
By applying a cut on these distributions one can select DVCS events and suppress
background events. However, for the selection of "pure" DVCS events, i.e., DVCS events
where the proton stays in its ground state, this simple approach is not optimal and an
alternative approach, kinematic event fitting, is used instead.

In the kinematic event fitting approach, all available information from the detected particles
is combined under a certain hypothesis on the event kinematics.
In the case of DVCS, four kinematic constraints can be constructed from the
energy-momentum conservation assuming electron, photon and proton as particle
types and absence of other particles in the final state.
Technically (and in the case that the measured kinematic parameters are not
correlated) a $\chi^2_{\mathrm{kin}}$ probability distribution function


\begin{equation}
\chi^2_{\mathrm{kin}} = \sum_{i = 1}^{n}(r_i^{fit}-r_i^{meas})^2/\sigma_i^2 
\end{equation}

\noindent is constructed under the conditions that

\begin{eqnarray}
f_j & = & f_j(r_1^{fit}, r_2^{fit}, ... r_{n}^{fit}) = 0, \ \ j = 1, ..., m.
\end{eqnarray}

\noindent Here $f_j$ are (usually nonlinear) functions of kinematic parameters,
$r_i^{meas}$ and $r_i^{fit}$ are the measured and fit kinematic parameters,
respectively, $\sigma_i$ are measurement uncertainties, $n$ is the number
of kinematic parameters and $m$ is the number of kinematic constraints.

In the case of DVCS, the following nine kinematic parameters are chosen:

$$r = \left\{\frac{p_x^e}{p_z^e}, \frac{p_y^e}{p_z^e}, \frac{1}{p^e}, \frac{p_x^{\gamma}}{p_z^{\gamma}}, \frac{p_y^{\gamma}}{p_z^{\gamma}}, \frac{1}{p^{\gamma}}, \phi^p, \theta^p, \frac{1}{p^p\cdot\sin(\theta^p)}\right\}.$$ 

\noindent
Here $p_x^e$, $p_y^e$, $p_z^e$, $p^e$ are the $x, y, z$
components and absolute value of the momentum of the scattered positron, 
$p_x^{\gamma}$, $p_y^{\gamma}$, $p_z^{\gamma}$, $p^{\gamma}$ the $x, y, z$ components and absolute
value of the momentum of the real photon and $\phi^p$, $\theta^p$ and $p^p$
the azimuthal and polar angles and the absolute value of the momentum
of the recoiling proton.

The minimisation of the $\chi^2_{\mathrm{kin}}$ function with constraints can be achieved
by the method of penalty functions (see, e.g, reference~\cite{Dymov:2000} for details).
In this method, penalty terms are added to the chi-square function

\begin{equation}
\chi^2_{\mathrm{pen}} = \chi^2_{\mathrm{kin}} + T\cdot \sum_{j=1}^{m} f_j^2/(\sigma_{j}^{f})^2,
\end{equation}

\noindent
where $\sigma_{j}^{f}$ is the error of the $j$-th constraint and $T$ is a constant penalty number.
For sufficiently large $T$, the constraints are automatically satisfied after convergence of 
the minimization procedure.

The measurement uncertainties are extracted from data sets generated by Monte
Carlo. The momentum dependence of the measurement errors is parametrized for
electrons and photons measured in the forward spectrometer and protons 
registered in the RD. Small corrections of the measurement uncertainties are
applied for experimental data based on the observed difference between data and
Monte Carlo.

\begin{figure}[t]
\begin{center}
\includegraphics[width=0.495\textwidth]{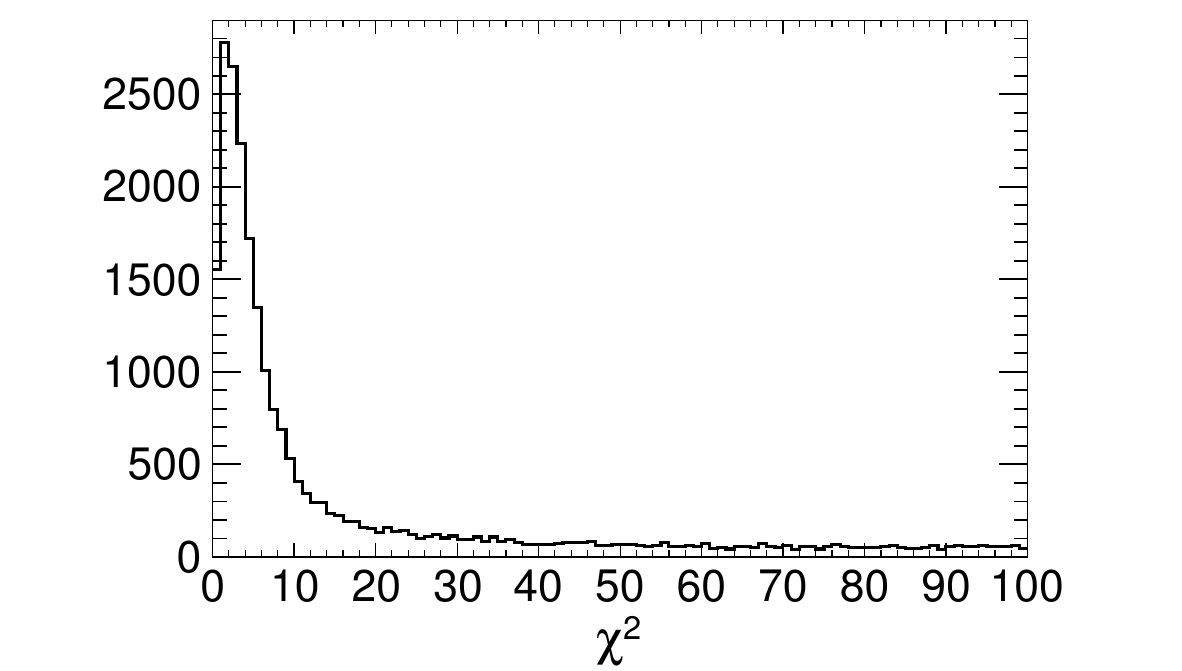}
\includegraphics[width=0.495\textwidth]{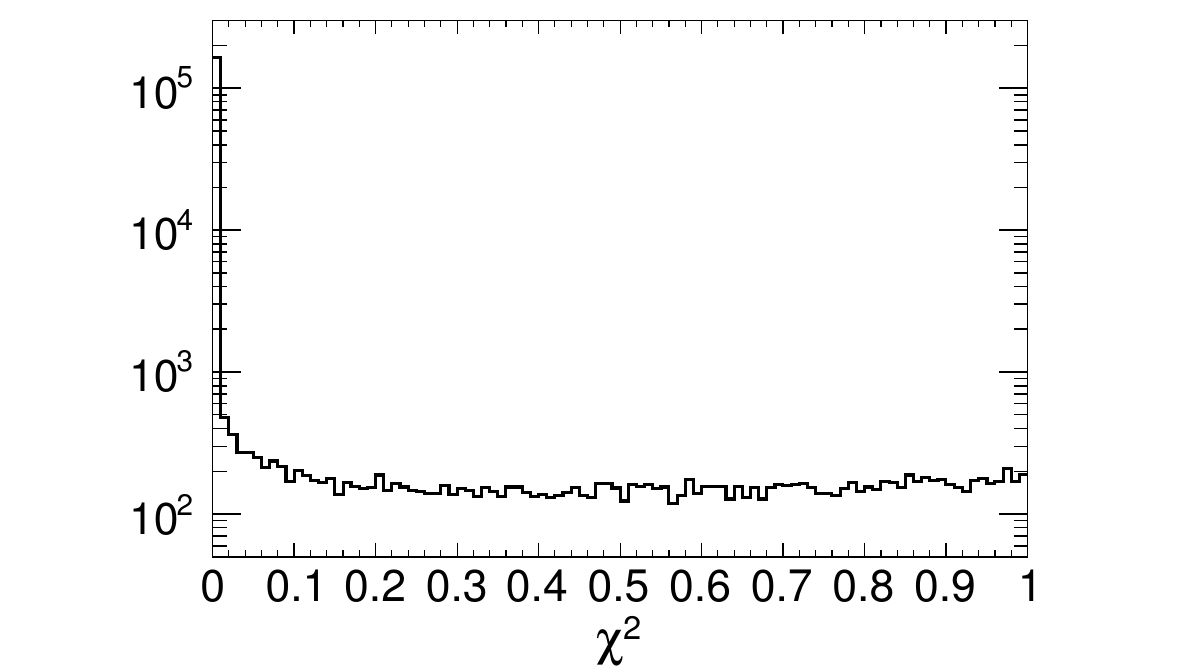}
\caption{Left panel: Distribution of $\chi^2_{\mathrm{kin}}$ from kinematic event fitting
         for DVCS event candidates. Right panel: Probability distribution
         corresponding to the $\chi^2$ distribution.}
\label{fig:chi2kinfit}
\end{center}
\end{figure}

The resulting $\chi^2_{\mathrm{kin}}$-distribution from kinematic event fitting is displayed in
the left panel of figure~\ref{fig:chi2kinfit} for DVCS event candidates
with one lepton and one photon measured in the forward spectrometer
and a proton candidate registered in the RD. In the right
panel the fit probability distribution is presented, which corresponds to the
$\chi^2$-distribution for four degrees of freedom. In the ideal
case of Gaussian distributions of the measurement uncertainties $\sigma_i$ and
absence of background, this probability is expected to be uniformly distributed
between zero and unity.  

In the case that there are multiple proton track candidates reconstructed in
combination with the scattered electron and the real photon, the proton candidate
that resulted in the smallest $\chi^2_{\mathrm{kin}}$-value is selected.
The probability calculated from  $\chi^2_{\mathrm{kin}}$ that a particular event
satisfied the DVCS hypothesis is required to be larger than 0.01, a value that is
adequate to ensure negligible background contamination. The performance of this
event selection is studied using an appropriate mixture of simulated signal and
background events~\cite{lepto, gmcDVCS}. Events satisfying all other previously
mentioned constraints are found to be selected with high efficiency (83\%) and
background contamination less than 0.2\%.

\begin{figure}[t]
\begin{center}
\includegraphics[width=0.495\textwidth]{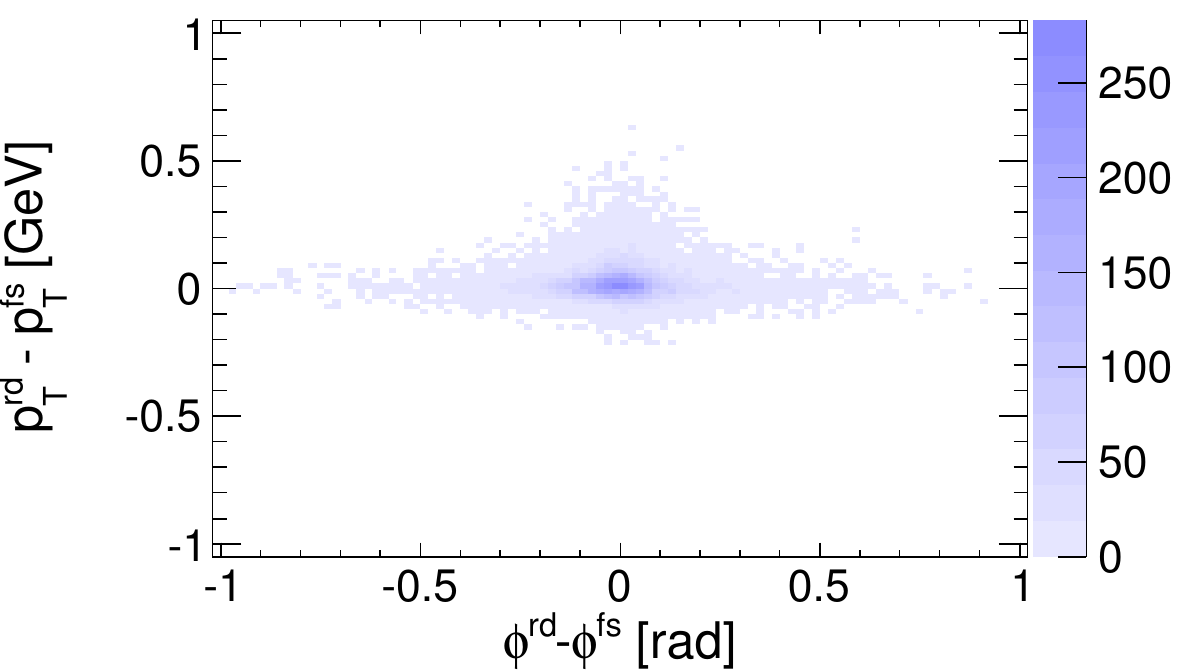}
\includegraphics[width=0.495\textwidth]{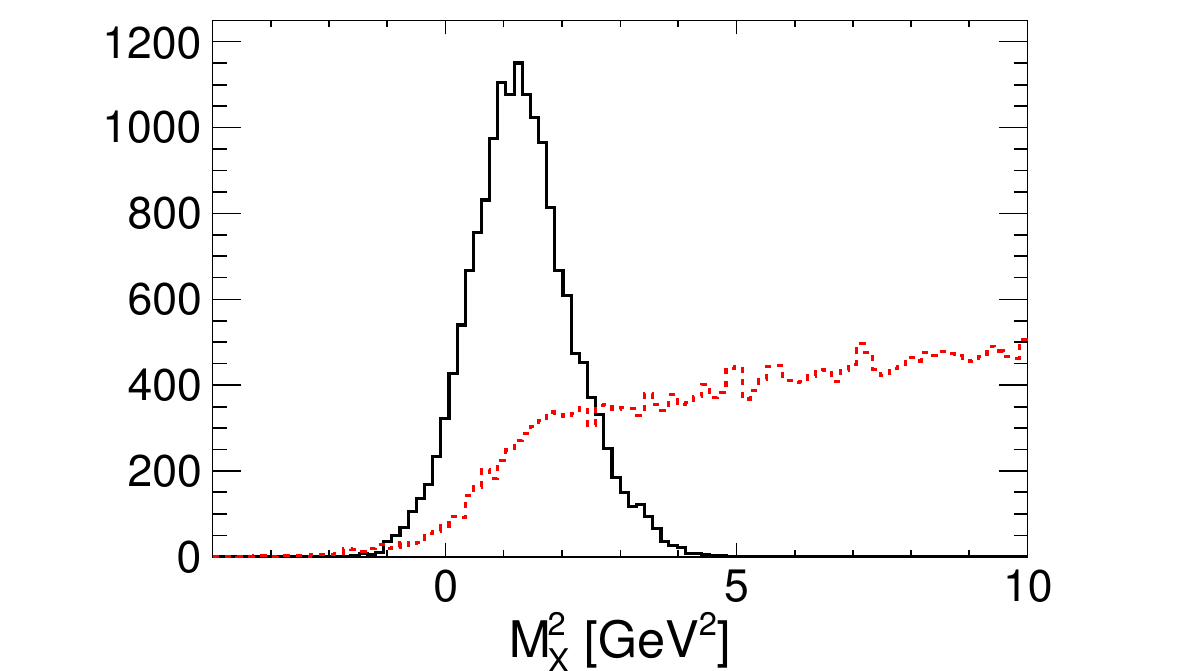}
\caption{Left panel: The difference between the transverse momentum of the proton
         reconstructed in the RD and the one calculated using only
         information from the forward spectrometer versus the difference between
         reconstructed and calculated azimuthal angles for DVCS events selected
         using kinematic event fitting with the fit probability above 0.01.
         Right panel: Squared missing mass calculated from the measured 3-momenta
         of the scattered lepton and real photon for selected DVCS events with
         the fit probability above 0.01 (solid line) and rejected events with the
         fit probability below 0.01 (dashed line).}
\label{fig:dvcs_dphi_dpt_dvcs_mx2_ch}
\end{center}
\end{figure}
\begin{figure}[t]
\begin{center}
\includegraphics[width=0.495\textwidth]{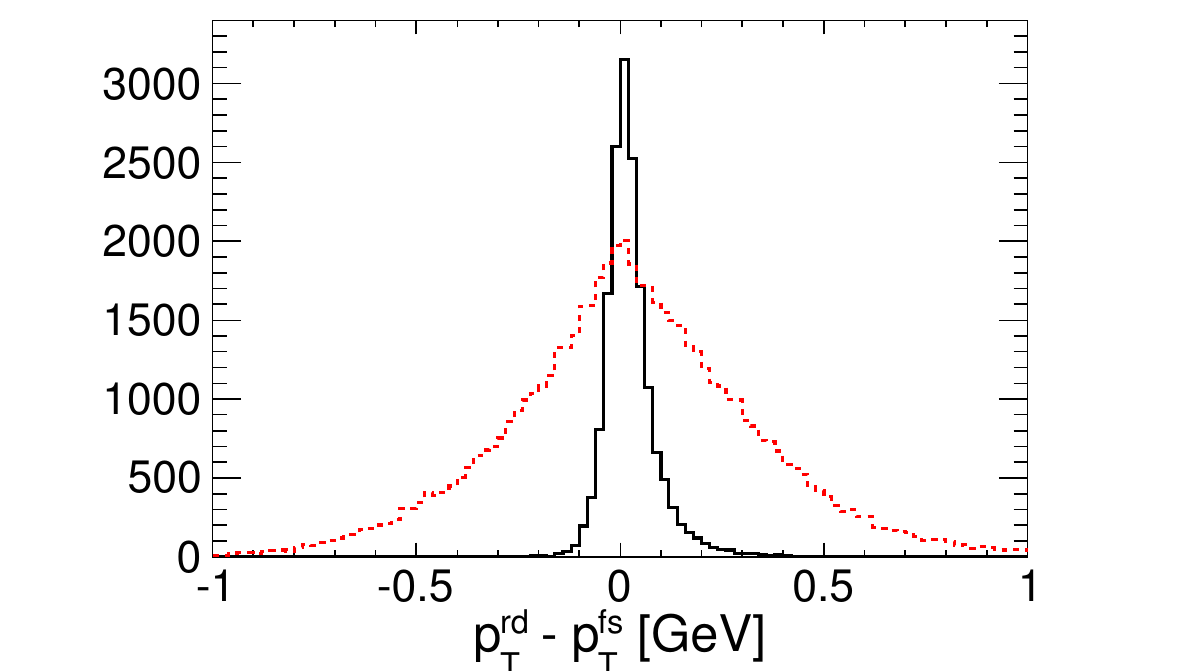}
\includegraphics[width=0.495\textwidth]{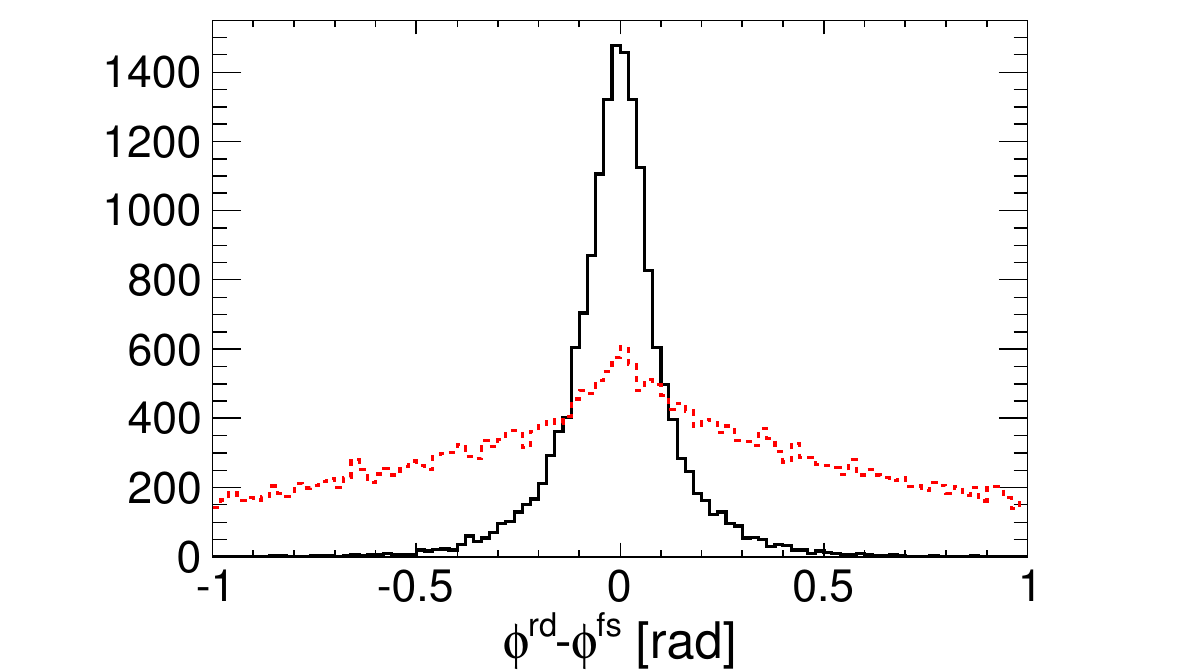}
\caption{Left panel: The difference between the transverse momentum
of the proton reconstructed in the RD and the one calculated
using only information from the forward spectrometer for selected DVCS
events with the fit probability above 0.01 (solid line)
and rejected events with the fit probability below 0.01 (dashed line).
Right panel: The difference between reconstructed and
calculated azimuthal angles for selected DVCS events with the fit
probability above 0.01 (solid line) and rejected events with 
the fit probability below 0.01 (dashed line).
}
\label{fig:dvcs_dpt_dvcs_dphi}
\end{center}
\end{figure}

In figures~\ref{fig:dvcs_dphi_dpt_dvcs_mx2_ch} and \ref{fig:dvcs_dpt_dvcs_dphi},
results of the event selection using kinematic event fitting are presented.
A clear peak in the difference between the transverse momentum of the proton
reconstructed in the RD and the transverse missing momentum calculated from the
measured 3-momenta of the scattered lepton and real photon as well as in the
difference between the reconstructed and calculated azimuthal angles is observed
(left panel of figure~\ref{fig:dvcs_dphi_dpt_dvcs_mx2_ch} and both panels of
figure~\ref{fig:dvcs_dpt_dvcs_dphi}). From the missing-mass distributions shown in
the right panel of figure~\ref{fig:dvcs_dphi_dpt_dvcs_mx2_ch}, only the one with
the fit probability above $0.01$ (solid line) shows Gaussian behavior,
confirming a clear and background-free selection of the pure DVCS sample.
Such a distinction performance is not possible with applying simple constraints
on kinematic distributions. 

%
\section{Conclusion}
%

The main purpose of the RD was the registration of protons and charged pions with momenta
from $0.125\unit{GeV}$ to $1.4\unit{GeV}$ and large polar angles in order to measure the
complete kinematics of hard exclusive processes in electroproduction. For this aim, a set
of silicon strip detectors situated inside the {\sc Hera} beam vacuum, a scintillating-fibre
tracker and a photon detector, \mhl{all surrounded by a superconducting magnet with a
field strength of $1\unit{T}$, were installed} in the {\sc Hermes} target region.
Commissioning of the detector started in spring
2006 when part of the detector was already operational and finished in fall 2006 resulting in
stable running of the detector until the {\sc Hera} shutdown in June of 2007.
All detector components were calibrated, detector efficiencies were studied and the alignment
of all subdetectors was performed based on cosmic ray and experimental data from
normal detector running periods. The tracking and momentum-reconstruction algorithms developed
allow for a momentum reconstruction in a wide momentum range. For low-momentum protons,
in addition to coordinate information from the SSD and SFT, energy deposits in the SSD
are used to improve the momentum reconstruction. The main purpose of the PD was the detection
of photons, but it was also capable of detecting high-energetic protons and charged pions.
The particle-identification techniques, which use energy deposits in the SSD, SFT and PD,
allow for a reliable selection of protons and charged pions with momenta below $0.7\unit{GeV}$.

First results on single-charge beam-helicity asymmetry in deeply virtual Compton scattering
using the RD were recently obtained \cite{DC92}. As intended, the use of the RD allows
a suppression of the background to a negligible level.

%
\section*{Acknowledgments}
%

We gratefully acknowledge the {\sc Desy} management for its support and the staff
at {\sc Desy} and the collaborating institutions for their significant effort.
This work was supported by 
the Ministry of Economy and the Ministry of Education and Science of Armenia;
the FWO-Flanders and IWT, Belgium;
the Natural Sciences and Engineering Research Council of Canada;
the National Natural Science Foundation of China;
the Alexander von Humboldt Stiftung,
the German Bundesministerium f\"ur Bildung und Forschung (BMBF), and
the Deutsche Forschungsgemeinschaft (DFG);
the Italian Istituto Nazionale di Fisica Nucleare (INFN);
the MEXT, JSPS, and G-COE of Japan;
the Dutch Foundation for Fundamenteel Onderzoek der Materie (FOM);
the Russian Academy of Science and the Russian Federal Agency for 
Science and Innovations;
the Basque Foundation for Science (IKERBASQUE) and the UPV/EHU under program UFI 11/55;
the U.K.~Engineering and Physical Sciences Research Council, 
the Science and Technology Facilities Council,
and the Scottish Universities Physics Alliance;
the U.S.~Department of Energy (DOE) and the National Science Foundation (NSF);
as well as the European Community Research Infrastructure Integrating Activity
under the FP7 "Study of strongly interacting matter (HadronPhysics2, Grant
Agreement number 227431)".






\end{document}